\documentclass[twocolumn,nolinenumbers]{aastex7}
\usepackage{amsmath,array,graphicx}
\usepackage{tabularx}
\usepackage{float}

\usepackage{amsmath}
\usepackage{breqn}

\def\lapprox{\hbox{\lower .8ex\hbox{$\,\buildrel < \over\sim\,$}}}
\def\gapprox{\hbox{\lower .8ex\hboxeening{$\,\buildrel > \over\sim\,$}}}

\def\aap{\hbox{A\&A}}

\begin{document}

 \title
    {''SNe Ia twins'' in the Hubble flow,
      and the determination of H$_{0}$}
    \author{P. Ruiz-Lapuente}
\email{pilar@icc.ub.edu}
\affiliation{Instituto de F\'{\i}sica Fundamental, Consejo Superior de 
Investigaciones Cient\'{\i}ficas, c/. Serrano 121, E-28006, Madrid, Spain}
\affiliation{Institut de Ci\`encies del Cosmos (UB--IEEC),  c/. Mart\'{\i}
i Franqu\'es 1, E--08028, Barcelona, Spain}

\bigskip

  \author{A.Quintana--Estellés}
\email{antonio.quintana@iff.csic.es}
\affiliation{Instituto de F\'{\i}sica Fundamental, Consejo Superior de 
Investigaciones Cient\'{\i}ficas, c/. Serrano 121, E-28006, Madrid, Spain}
\affiliation{Institut de Ci\`encies del Cosmos (UB--IEEC),  c/. Mart\'{\i}
i Franqu\'es 1, E--08028, Barcelona, Spain}

\bigskip

\author{J.I. Gonz\'alez Hern\'andez}
\email{jonay@iac.es}
\affiliation{Instituto de Astrof\'{\i}sica de Canarias, E-38200 La Laguna, 
Tenerife, Spain}
\affiliation{Universidad de La Laguna, Dept. Astrof\'{\i}sica, E38206 
La Laguna, Tenerife, Spain}

\bigskip

\author{A. Pastorello}
\email{andrea.pastorello@inaf.it}
\affiliation{INAF - Osservatorio Astronomico di Padova, Vicolo dell'Osservatorio 5, 35122 Padova, Italy}

\bigskip

\begin{abstract}

\noindent
  We have applied our approach of using ''SNe Ia twins''in the Hubble flow to
  obtain distances to SNe Ia at z $>$ 0.015 and  derive H$_{0}$.
  Our results, taking a single step between the low z domain
  and the Hubble flow,
  validate the three rung classical method.
    We find, however, that the full compilation of distances, both 
    in Pantheon+ and in the Carnegie-Chicago Hubble Program (CCHP),
    contain some inaccurate values in the 
   colors due to an  undersestimate of reddening
  by dust, or  due to the adoption of not well--defined 
  light curve declines. This produces  odd individual values for H$_{0}$
  from single SNe Ia.
  On the average, those erroneous estimates 
   do not affect the mean value of H$_{0}$, which is characterized by
   the bulk of well--modeled SNe Ia. 
  Our sample of carefully
  addressed SNe Ia in the
  Hubble flow contains a dozen supernovae, for which
  the distances are  determined with high accuracy.
  Three of these SNe Ia are of the  Broad Line subtype 
  and can be compared with SN 1989B in M66, a host galaxy  with a unique
  convergence of the Cepheid distance determination and the Tip of the Red Giant Branch  stars (TRGB)  determination
  by the CCHP group. 
  There is as well a very good agreement on the distances to NGC 7250 and
  NGC 5643 between those derived with Cepheids by SH0ES and
  those derived with the use of J-Asymptotic Giant Branch stars
  (JAGB stars) by the CCHP, which makes them very good anchors.
  The sample of 12 SNe Ia gives a value
  of H$_{0}$ $=$  72.56 $\pm$ 1.54 (stat) $\pm$ 1.33(sys)
  km s$^{-1}$ Mpc $^{-1}$, 
  when anchored in Cepheids, 
  and 
  of H$_{0}$ $=$ 72.20 $\pm$ 1.53 (stat) $\pm$ 1.33 (sys)
  km s$^{-1}$ Mpc $^{-1}$,
  when anchored in JAGBs by the CCHP.
   We take a mean of the two values of H$_{0}$ as derived by the Cepheids
   and by JAGB (from the CCHP) and obtain 
   H$_{0}$ $=$  72.38 $\pm$ 1.54(stat) $\pm$ 1.33(sys)
   km s$^{-1}$ Mpc $^{-1}$.  
    Our findings confirm that the Hubble tension is real.

 \end{abstract}


\section{Introduction}

\noindent
Type Ia supernovae (SNe Ia) at high $z$ led to the discovery of the acceleration
of the Universe (Riess et al. 1998; Perlmutter et al. 1999) and dark energy.
Large compilations of SNe Ia have recently suggested a tentative evolution
in time of such  unknown major component of the energy density of the Universe
(Rubin et al. 2025; DESyr5, DESI\footnote{SH0ES stands for
''Supernova H0 for the Equation of State of Dark Energy''. DESyr5 for
''Dark Energy Survey year 5'', DESI
for ''Dark Energy Spectroscopic Instrument''.}).

\bigskip

\noindent
Apart from this use in cosmology, a puzzle emerged some  years ago and still
requires an explanation: the expansion rate
of the Universe today as obtained with
SNe Ia calibrated with Cepheids  gives a value of H$_{0}$ higher than that
measured
from the Cosmic Microwave Background (CMB).
The value of H$_{0}$ derived from the CMB is
  67.4$\pm$0.5 km s$^{-1}$ Mpc$^{-1}$ (Planck Collaboration 2020) and the latest
 {\it SH0ES}
 value of  H$_{0}$ = 73.3 $\pm$0.9 km s$^{-1}$ Mpc$^{-1}$
 (Murakami et al et. 2023) have now a discrepancy at the 5.7$\sigma$ level
 or even  at the 6$\sigma$ level (Riess et al. 2025). 
 The {\it SH0ES}  group  has now addressed three additional methods: their own
 Tip of the Red Giant Branch stars (TRGB) measurements, to be compared
 with that of the Carnegie Chicago Supernova Project (CCHP), their 
 own J-Asymptotic Giant Branch stars (JAGB) and the
 use of Mira stars to compare
  with Cepheids, in conjunction with SNe Ia in the Hubble flow from 
  the Pantheon+ SN catalog (Scolnic et al. 2022). They
  obtain values of 72.1-73.3 $\pm$ 1.8 km s$^{-1}$ Mpc$^{-1}$
  (depending on methodology) (Li et al. 2025).

  \bigskip

  \noindent
  Such
  difference in H$_{0}$, when comparing with the H$_{0}$ provided by the CMB,
  if confirmed, would stand as a challenge to the 
$\Lambda$CDM  model (Di Valentino et al. 2025a,b).
 A number of suggestions have been made, including 
the need of new physics in the early Universe to make the
{\it Planck} value compatible with that derived by methods
involving low-$z$ astrophysical distance indicators such as Cepheids
(see Di Valentino
et al. 2021a,b, 2025a,b; Kamionkowski \& Riess 2023;  Sch\"oneberg et al.
2021; M\"ortsell \&
Dhawan 2018, Poulin et al. 2025, for reviews).

\bigskip

\noindent
Nowadays, several other  methods are giving values of H$_{0}$ around
74--72 km s$^{-1}$ Mpc$^{-1}$ : 
the sample of distances obtained by the {\it Surface
Brightness Fluctuations (SBF)} method, dominated by early-type galaxies as 
required by the method, calibrated with TRGB, gives rise to an H$_{0}$ $=$ 
73.8 $\pm$ 0.7 (stat)  $\pm$ 2.3 (syst)
km s$^{-1}$ Mpc$^{-1}$ (Jensen et al. 2025). Garnavich et al. (2023) had  used
as well the SBF method and found H$_{0}$ $=$ 74.6 $\pm$ 0.9 (stat)
$\pm$ 2.7 (syst) km s$^{-1}$ Mpc$^{-1}$. Galbany et al.(2023) used the
peak of brightness in the infrared bands J and H from a sample  of SNe Ia
in the Hubble flow together with Cepheids by SH0ES as nearby calibrators
and obtained  H$_{0}$ of 72.3 $\pm$ 1.4 (stat) $\pm$ 1.4 (syst)
km s$^{-1}$ Mpc$^{-1}$ when using the J-band and
72.3 $\pm$ 1.3 (stat) $\pm$ 1.4 (syst)
km s$^{-1}$ Mpc$^{-1}$ when using the H-band.
  The Expanding Photosphere method, using SNe II, which does not rely in any
  nearby calibrator, gives 74.9 $\pm$ 1.9 (stat)
km s$^{-1}$ Mpc$^{-1}$ (Vogl et al. 2024).

\bigskip

  \noindent
  While new  methods are now providing  the above quoted values for
   H$_{0}$, the debate on
obtaining H$_{0}$ using Cepheids, TRGB  
or JAGB stars distances to calibrate Type Ia supernovae (SN Ia)
 persists. 
The SNe Ia calibration depends on three rungs: Rung
1 is the calibration of TRGB/Cepheids/JAGB stars with local
anchors using geometrical methods,
rung 2 refers to the calibration of the distance to SNe
Ia host galaxies with  TRGB/Cepheids/JAGB stars, and rung 3
is the overall calibration of SNe Ia in the Hubble ﬂow, which is
done in different ways but usually involves the derivation of
a fiducial value for the absolute magnitude of SNe Ia
and  the final dependence of each particular SN Ia magnitude from such
absolute magnitude
and from factors such as rate of decline (hereafter called stretch)
of the light curve,
color of the SN Ia at maximum and mass of its host galaxy.
The two collaborations are making this step in different ways:
Pantheon+ uses the SALT2  relation (Brout et al. 2022) while the CCHP uses a
polynomial derivation (Freedman et al. 2019).

\bigskip

\noindent
    When using the TRGB and JAGB methods the two collaborations get different
      values of distances to the same galaxies and, therefore,
  different H$_{0}$ values
  from their use of these methods (see for instance Pantos \& Perivolaropoulos, 2026; Ruiz-Lapuente \& Gonzalez Hern\'andez 2024: hereafter RLGH24). 

\bigskip

\noindent
Given  the discrepancies between the two groups on H$_{0}$, we decided  to
introduce a new approach (RLGH24).
We address the Hubble tension issue with a method  which uses
the spectral evolution of SNe Ia that have not only similar stretch but
identical spectral features along their lifetimes, i.e. ''SNe Ia twins
for life''. The precision of the
distance provided by those twins should be higher than for  SNe Ia with just
similar stretch. The total reddening could also be obtained.

\bigskip

\noindent
The selection of 
twins is made of SNe Ia with a similar stretch, being then of similar luminosities,
but in addition the ''twinness factor'' can make more precise the relative distance estimate, with a modulus error of 0.04 mag (RLGH24). 
 As described in RLGH24, once the distance is determined by the MCMC, we position the twins at the same distance. 
We pass the spectrum through the B, V, R, and I filters and calculate the magnitude at B, V, R, and I. 
We do this for both SNe Ia twins. Then, we observe what the intrinsic dispersion of the twins might be, based on the magnitude difference. 
This yields a typical error of 0.04 mag at the filters.

\bigskip

\noindent
    In this work, we add another estimate of the accuracy of the method
    with an internal error, $\sigma_{int}$, from our MCMC calculations (see the Methodology section). It confirms an error  $\leq$ 0.055 mag. 

\bigskip

\noindent
The anchors (/calibrators)\footnote{We will call anchors or calibrators to the host galaxies in this first step of our method.} used to obtain
absolute distances should be galaxies where there
is a confluence on the
distance estimates obtained by various methods and groups. 

\bigskip

\noindent
Our method does not rely on many host galaxies of SNe Ia, but
on a sample of 4 nearby host galaxies for which
there is a  very good agreement in the
distance from the SH0ES and the CCHP collaborations.
We use  the distances to
M101, NGC 7250, NGC 5643, M66 to anchor the twins.
The uncertainty in the anchor is a part of the systematic error in
the  obtained H$_{0}$. A difference of 0 mag indicates 
  that the SH0ES and the CCHP collaborations agree in
  the anchor distances. 
  However, an error of 0 mag is not reflecting the real uncertainty in the calibrators. We estimate
  it to be of 0.04 mag, as quoted typically in the most recent papers of the collaborations.
  Therefore, we have taken it to be 0.04 mag from anchoring of the calibrators for all SNe Ia.

\bigskip

\noindent
The advantage to select this group of well measured galaxies is to
avoid the whole range of discrepancies introduced by the present
samples in rung 2
 when applying different methods (Cepheids, TRGB, JAGB stars)   
by the two collaborations. 
 The choice of these galaxies has been driven, as well, 
 by  the availability in number
 of SNe Ia twins in the Hubble flow of the kind found in these low-z
 galaxies.

\bigskip

\noindent
The three--rung method
(see Figure 6 un Riess et al. 2024) presents
a  sample of calibrators in rung 2 which  looks far from being ideal:
the calibrators point to individual H$_{0}$ values
from 60 to 86 km s$^{-1}$ Mpc$^{-1}$. 
The calibrators in rung 2 are taken together with the sample of SNe Ia
in the Hubble flow to define a fiducial absolute magnitude M$_{B}$ for
SNe Ia and the Hubble diagram. 
It seems clear that selecting particular subsamples of these galaxies
for calibrating the SNe Ia would produce biases in H$_{0}$. This fact
has been identified as one of the causes of the  divergent results for H$_{0}$
(Riess et al. 2024).
In our method, the representation of  the types of SNe Ia in the Hubble flow
with
a given intrinsic luminosity of SN Ia matches 
 the luminosity  of its twin SN Ia at low z. The uncertainty will come 
from the distance of the low z galaxies of each SN Ia type. We will avoid
the selection effects arising from modeling a {\bf diverse} SNe Ia family.   

\bigskip

\noindent
A crucial task  now is  to verify if the
three-rung process, which culminates with  distances given
to the SNe Ia in the Hubble flow, is in accordance with the
distances that we found by our more direct method. Our procedure is
then a test of what is the most important delivery of the three rung
method: the distances of SNe Ia in the Hubble flow and the value of
H$_{0}$.

\bigskip

\noindent
The paper is organized as follows: 
In section 2, we describe the sample of nearby SNe Ia 
which will have twins in the Hubble flow. In section 3, we present the
reasons to use those SNe Ia and their host galaxies as anchors. In section 4,
we present our sample of SNe Ia in the Hubble flow.
In section 5, we describe our method. In section 6, we
present the results on the distances of the SNe Ia.
In this section, we discuss the distance moduli of SNe Ia
and the H$_{0}$ obtained. We compare our results to the distances obtained
by the Pantheon+ sample and by the CCHP. 
In section 7, we present a summary and give our conclusions.

\section{The use of twin SNe Ia}

\bigskip

\noindent
In our previous paper (RLGH24), we tested
the method in two SNe Ia from the same galaxy. Instead of $\mu$, we were interested
in $\Delta\mu$, the relative distance between the 2 twin SNe Ia, 
SN 2013aa and
SN 2017cbv in NGC 5643. The
results point to $\Delta E(B - V)$ = 0.0$\pm$0.0 mag,
as it should have been
expected from the two SNe being in the same galaxy  with similar
 conditions of negligible reddening in the host galaxy
  and same reddening by dust in our Galaxy, and a $\Delta\mu$ $=$ 
0.004 $\pm$ 0.005 mag in an early spectrum and $\Delta\mu$ =--0.023$^{+0.008}_{-0.007}$ mag
in a nebular one. Those correspond
to a precision of 23 kpc from the early spectrum and around 100 kpc from
the late
one. The late phase has a larger error. The joint result from the two phases gives a distance difference
compatible with 0 with
$\Delta\mu$ $=$ -0.005 $\pm$ 0.004 mag. 

\bigskip

\noindent
The preceding approach
demonstrated  the possibility to combine early
and late phase information to obtain relative distances or differences in
distance moduli between the hosts of SN Ia twins.

\bigskip

\noindent
     To estimate the magnitude accuracy that twins have
      in their spectral
  comparison, we calculated the difference in filters B, V, R, I filters 
  of the twin spectra at the same phase once they are shifted to the
  same distance
 (this distance is calculated given the information provided by our emcee
  computation). The error estimate is of  $\sigma$ $=$ 0.04 mag
  in all filters
 and all phases. This confirms the power of this method to disentangle
 errors in SNe Ia host galaxies distances provided by TRGB, Cepheids, JAGB.

\bigskip

\noindent
Twin spectra of SNe Ia near maximum should have a similar
shape of the pseudo--continuum and similar pseudo--equivalent widths
(pWs) of the different lines. All that is required because otherwise
it is not possible to have a difference  of
$\sigma$ $=$ 0.04 mag in magnitude in all filters.

\bigskip

\noindent
Other approaches using 
the spectral diversity of SNe Ia twins 
in the context of improving the  determination of the nature of dark energy
 were presented by Fakhouri et al. (2015) and  Boone  et al. (2016).
In our case, we want to use the ''twins'' to determine the
Hubble constant. 

\bigskip

\noindent
We follow the subclassification of SNe Ia according the
{\it Branch classification} scheme (Branch et al. 2006),
which has been set in more quantitative terms
by the Carnegie Supernova Project. The {\it Branch classification}
subdivides SNe Ia in four types: Core Normal (CN), Broad Line (BL),
Shallow Silicon (SS) and Cool  (CL). The main factor to be taken into account
is the pseudo--equivalent widths (pWs) of the lines in the spectrum.
Burrow et al.(2020)
show how these classes are well separated in  the pWs diagrams.
The physical explanation of these subclasses is linked to the
velocities and chemical composition of the SN Ia core 
which is revealed as the supernova photosphere recedes showing the inner
ejecta. Those
differences can be seen as four different subtypes with typical
pseudo--equivalent widths (pWs) at various phases.

\bigskip

\noindent
The spectra of the
twins have very  similar pseudo-equivalent widths (pWs) in the list
of classifying  lines of the subtypes (see Morrell et all. 2024). Those lines
are 
pW1 (Ca H \& K), pW2 (Si II $\lambda$ 4130 \AA \ ), pW3
(Mg $\lambda$ 4481; blended with Fe II),
pW4 (Fe II at $\sim$ 4600 \AA \, blended SII), pW5
(S II ''W'' $\sim$ 5400 \AA \ ), pW6 (Si II $\lambda$ 5972 \AA \ ),
pW7 (Si II $\lambda$ 6355 \AA \ ), pW8 (Ca II IR \AA \ ).
In particular, the line associated with Si II $\lambda$ 5972 \AA \
should be very similar in twin SNe Ia, with a discrepancy of less than
5$\%$. This line correlates with stretch ($\Delta m(B)_{15}$, s$_{BV}$), and
amongst SNe Ia of similar stretch, has to be almost the same for twins. 
In the pW6 over pW7 diagram, the twins fall in very close positions. 
If the equivalent widths are similar, the ratios of the two Si II
at $\lambda$ 5972 \AA \ and $\lambda$ 6355 \AA \ , $\Re_{Si II}$,
should coincide. Thus, for the application of this method, both the
  pWs of the lines and the pseudo-continuum of the spectra have to be very
  similar between the twin SNe Ia. In particular, pW6 (Si II $\lambda$ 5972 \AA \ ) and pW7 (Si II $\lambda$ 6355 \AA \ ) should be similar within 5$\%$
  (see RLGH24). The  diagram of those pWs  define the subtypes. In addition, 
  pW6 has correlations with the stretch s$_{BV}$ and $\Delta m(B)_{15}$.

\bigskip

\noindent
We looked,  in our method paper
(RLGH24), at  the abovementioned
signs for the classification as twins of nearby SNe Ia. We need
at least two phases to obtain a reliable result. (One phase is not enough).
\bigskip

\noindent
In that previous paper (RLGH24), 
we used a single subtype of SNe Ia: 
the Core Normal (CN) subtype. In the present work we have extended it 
to the Broad Line (BL) subtype. We have found a good anchor of BL 
SNe Ia: SN 1989B in M66.

\bigskip

\noindent
Previous research done within the core normal subtype started with a
comparison between the twins SN 2013aa/SN 2017cbv, 
SN 2013aa and SN 2017cbv appeared 
 in the outskirts of the same host galaxy, NGC 5643, so the reddening
E(B-V) by the host should be negligible. The reddening
in our Galaxy in the direction of NGC 5643 is $(E-B)_{MW}$ = 0.15 mag 
(Schlafly \& Finkbeiner, 2011). 
SN 2013aa and SN 2017cbv  have
similar decline rates and B-peak magnitudes.
The studies of 
these two SNe Ia also reveal similar characteristics in other aspects. Table
1 specifies the decline
rates of those SNe Ia not only by $\Delta m_{15} (B)$  but also by s$_{BV}^{D}$
(Burns et al. 2020).
Fitting the B-V
color curves with cubic splines, Burns et al (2020)
ﬁnd identical color-stretch s$_{BV}^{D}$ = 1.11 for SN 2013aa and SN 2017cbv.
This value is smaller 
 (slower decline) than in typical SNe Ia. They also found similar values in the
 {\it Branch classification}, concerning the  pseudo equivalent widths (pWs) of
 the lines of different elements. This proves that their interior was
 very alike in chemical composition. 
The
use as anchor of the host galaxy of both SN 2013aa and SN 2017cbv
NGC 5643 is justified given the converging values of distance moduli to
this galaxy with the use of Cepheids by SH0ES and with JAGB stars by
the CCHP. 

\bigskip

\noindent
In the previous paper, we found that 
SN 2013dy is practically equal at late phases to SN 2017cbv (there are
no spectra of SN 2013aa at that phase to allow a comparison). A
comparison shows that the light curve at late phases is
slightly more luminous than in SN 2013aa, but coincides with SN 2017cbv.
At very early phases
it is similar to both SN 2017cbv and SN 2013aa though slightly redder.
After maximum brightness, it is also identical to SN 2013aa. 
SN 2013dy has a the Galactic reddening 
E(B-V)$_{MW}$ $=$0.14 mag (Schlafly \& Finkbeiner 2011), though there are
indications that the total reddening might be 0.15 mag. 
There is the possibility, then, to use SN 2013dy as an anchor.
The fact that this supernova  was discovered a few hours after explosion
and had a  very intense photometric and
spectroscopic follow up (Pan et al. 2015) enables to have many phases
for comparison with their SNe Ia twins in the Hubble flow.  
There are, as well, converging values
of distance moduli to
the host galaxy of this SNIa, NGC 7250, 
with the use of Cepheids by SH0ES and with JAGB stars by
the CCHP.

\bigskip

\noindent
SN 2011fe, of the same {\it Branch subtype} as SN 2013aa,
SN 2017cbv, and SN 2013dy, is also identified as a very good reference
 as both  SH0ES and CCHP find similar distances to the
 host galaxy M101. 
SN 2011fe was discovered in its very early 
phase and  was classified as a normal SN Ia (Nugent et al. 2011).  
At maximum light, the color was 
(B$_{max}$ - V$_{max}$) = –0.07 $\pm$ 0.02 mag. 

\bigskip

\noindent
Concerning the luminosity decline parameter of SN 2011fe,
Burns et al (2023, private communication) measured a
$\Delta m_{15} (B)$ = 1.07$\pm$0.006 mag. 

\bigskip

\noindent
SN 2011fe is a faster and more underluminous SN Ia than SN 2017cbv and
SN 2013aa. 
The difference between them is seen at both early and late phases.
In fact,
SN 2013aa and SN 2017cbv synthesized about 0.23 M$_{\odot}$ more  $^{56}$Ni than
 SN 2011fe. Studies by Jacobson-Gal\'an et al. (2018) suggest
 that SN 2013aa and SN 2017cbv
 synthezised 
 0.732 $\pm$ 0.151 M$_{\odot}$ of $^{56}$Ni. The mass of $^{56}$Ni 
   of SN 2011fe, as found by many authors, quoted in RLGH24, is
   $\sim$ 0.5 M$_{\odot}$.

\bigskip
  
\begin{table}
  \scriptsize
  \centering
  \caption{SN 2013aa, SN 2017cbv, SN 2013dy SN 2011fe and SN 1989B}
  \begin{tabular}{lcc}
    \hline
    \hline
        {\bf SN 2013aa} &      &     \\
        RA, DEC$^{a}$ &  14:32:33.881 & -44:13:27.80 \\
        Discovery date$^{a}$ & $\dots$ &  2013-02-13 \\
        Phase ( referred to maximum light)$^{b}$& $\dots$ & -7 days \\
        Redshift$^{c}$ & $\dots$ & 0.004 \\
        E(B-V)$_{MW}$$^{d}$ & $\dots$ & 0.15$\pm$0.06 mag \\
        $m_{B}^{max}$ $^{b}$ & $\dots$ & 11.094$\pm$0.003 mag \\
        $\Delta m(B)_{15}$$^{b}$ & $\dots$ & 0.95$\pm$0.01 \\
        Stretch factor $s_{BV}^{D}$$^{b}$ & $\dots$ & 1.11$\pm$0.02 \\
        Phases of the spectra used & $\dots$ & -2, 34, 42.07 days \\
        \hline
            {\bf SN 2017cbv} &     &    \\
            RA, DEC$^{e}$ & 14:32:34.420 & -44:08:02.74 \\
            Discovery date$^{e}$ & $\dots$ & 2017-03-10 \\
            Phase (referred to maximum light)$^{b}$ & $\dots$ & -18 days \\
            Redshift$^{c}$ & $\dots$ & 0.004 \\
            E(B-V)$_{MW}$$^{d}$ & $\dots$ &  0.15$\pm$0.06 mag \\
            $m_{B}^{max}$$^{b}$ & $\dots$ & 11.11$\pm$0.03 mag \\
            $\Delta m(B)_{15}$$^{b}$ & $\dots$ & 0.96$\pm$0.02 \\
            Stretch factor $s_{BV}^{D}$$^{b}$ & $\dots$
            & 1.11$\pm$0.03 \\
            Phases of the spectra used & $\dots$ & 13.14 days \\
            \hline
       {\bf SN 2013dy} &      &     \\
        RA, DEC$^{f}$ & 22:18:17.599 & +40:34:09.59 \\
        Discovery date$^{f}$ & $\dots$ & 2013-07-10 \\
        Phase (referred to maximum light)$^{g}$ & $\dots$ & -18 days \\
        Redshift$^{h}$ & $\dots$ & 0.0039 \\
        E(B-V)$_{MW}$$^{d}$ & $\dots$ & 0.15 mag \\
        $m_{B}^{max}$$^{i}$ & $\dots$ & 13.229$\pm$0.010 mag \\
        $\Delta m(B)_{15}$$^{j}$ & $\dots$ & 0.92$\pm$0.006 mag \\
         Stretch factor $s_{BV}^{D}$$^{j}$ & $\dots$
            & 1.091$\pm$0.03 \\
        Phases of the spectra used & $\dots$ & 3.7, 5 days \\
        \hline
      {\bf SN 2011fe} &      &     \\
        RA, DEC$^{a}$ & 14:03:05.810 & +54:16:25.39 \\
        Discovery date$^{k}$ & $\dots$ & 2011-08-24 \\
        Phase (referred to maximum light)$^{l}$ & $\dots$ & -20 days \\
        Redshift$^{m}$ & $\dots$ & 0.0012 \\
        E(B-V)$_{MW}$$^{d}$ & $\dots$ & 0.008 mag \\
        $m_{B}^{max}$$^{n}$ & $\dots$ & 9.983$\pm$0.015 mag \\
        $\Delta m(B)_{15}$ & $\dots$ & 1.07$\pm$0.06 \\
        Stretch factor $s_{BV}$  & $\dots$  &  0.97$\pm$ 0.002 \\
        Phases of the spectra used & $\dots$ & 4.65, 13 days \\
        \hline
          {\bf SN 1989B} &      &     \\
            RA, DEC$^{o}$ & 11:20:13.900 & +13:00:19.00 \\
            Discovery date$^{o}$ & $\dots$ & 1989-01-30 \\
            Phase (referred to maximum light)$^{p}$ & $\dots$ & -7 days \\
            Redshift$^{q}$ & $\dots$ & 0.00242 \\
            E(B-V)$_{MW}$$^{\bf d}$ & $\dots$ & {\bf 0.030 mag} \\
            $\Delta m(B)_{15}$$^{q}$ & $\dots$ & 1.053$\pm$0.11 mag \\
            Stretch factor $s_{BV}$$^{q}$ & $\dots$ & 0.954$\pm$0.034 mag \\
            Phases of the spectra used & $\dots$ & -1, 3, 7, 9, 11 days \\
            \hline
  \end{tabular}
  \begin{tabular}{ll}    
    $^{a}$Waagen (2013). & $^{b}$Burns et al. (2020). \\
    $^{c}$Parrent et al.
    (2013). &
    $^{d}$Schlafly \& Finkbeiner (2011). \\
    $^{e}$Tartaglia et al. (2017). &
    $^{f}$Casper et al. (2013). \\
    $^{g}$Zheng et al. (2013). &
      $^{h}$Schneider et al. (1992). \\
      $^{i}$Pan et al. (2015). &
      $^{j}$Uddin et al (2024) \\
    $^{k}$Richmond \& Smith (2012) &
    $^{l}$Cenko et al. (2011) \\
    $^{m}$Burns
    (private communication, 2023). &
    $^{n}$Yang et al. 2020. \\
    $^{p}$Evans (1989). &  
    $^{q}$Morrell et al. (2024). \\
    $^{r}$Wells et al. (1994). & \\
    \hline
    \end{tabular}

\end{table}

\noindent
These four SNe Ia are the anchors for the CN SNe Ia in the Hubble
flow. For BL  SNe Ia in the Hubble flow, we can use as
anchor SN 1989B. This supernova was studied in detail  by Wells et al. (1994).
We have been able to compare with our method significantly
reddened SNe Ia in the Hubble flow with SN 1989B.  One
of the  advantages of SN 1989B is that  the  distance
determinations coincide using the Cepheids (Saha et al. 1999)
and the TRGB by Hoyt et al. (2019) (see also Freedman et al 2019).

\bigskip

\noindent
We present these SNe Ia in Table 1.

\bigskip

\section{On the convergence of the distance to our anchors}

\subsection{M101}

\bigskip

\noindent
Recently, a convergence has been achieved on the distance to M101 by
the SH0ES group and the CCHP. This gives an extraordinary opportunity
to be safe on the class of  SN 2011fe--like supernovae.
The distance to M101 is  $\mu$  $=$ 29.188 $\pm$ 0.055 mag for the
{\it SH0ES}  group using Cepheids (Riess et al. 2024)
  and $\mu$ $=$ 29.18 $\pm$ 0.04 mag
also from Cepheids  by the CCHP in 2024 (Freedman et al. 2024) and
$\mu$ $=$  29.151 $\pm$ 0.04 mag using TRGBs. In the most
recent CCHP publication of the distance to this galaxy (Freedman
et al. 2025), it is only slightly different. 

\bigskip

\noindent
There are other approaches giving slightly smaller distances
to M101, like the Miras variables (Huang et al. 2024)
and the  Blue Supergiants
(Bresolin et al 2025). All are within 1$\sigma$ of the values from 
the SH0ES collaboration and the latest CCHP value quoted above. 
The JAGB stars distance to M101 from the CCHP (Freedman et al. 2025) is 
29.208 $\pm$ 0.045 mag. 

 \subsection{NGC 5643}

\bigskip

\noindent
In the most recent papers 
by the SH0ES collaboration (years 2024--2025), the distances to NGC 5643 are given
not only using the Cepheids method, but also the other methods
used by the CCHP. 

\bigskip

\noindent
The values from SH0ES in the Table 2 from Li et al. (2024)  include
TRGB distances obtained with the JWST and they are compared with their Cepheids
distance determination to the same galaxies. Their TRGB SH0ES values are not far
from the Cepheids ones (only differing  by $\pm$ 0.02 mag at most). 
The Cepheid distance modulus of reference for NGC 5643 is
$\mu$ $=$ 30.55 $\pm$ 0.063 mag, which is the same as in Riess et al. (2022).
Their TRGB values slightly change if one uses different values
for their smoothing s$=$ 0.10 or s$=$ 0.05. Fortunately, for NGC 5643
the difference is small  ($\Delta \mu$ $=$ 0.01): the distance is
30.57 $\pm$ 0.06 mag, when using a smoothing of s=0.10, or 30.58 $\pm$ 0.06
mag when using a smoothing of s=0.05. 
The latest value from the TRGB method for NGC 5643 in 2025 obtained by the CCHP collaboration
is $\mu$ $=$ 30.643 $\pm$ 0.066 mag  (Hoyt el al. 2025) but, in the
compilation of Freedman et al. (2025),  the error is of 0.071 mag.
This value has changed
often during the last year. 
The  distance has gone in the last months 
from  $\mu$ $=$ 30.61 $\pm$
0.07 mag (Freedman et al. 2024) to $\mu$ $=$ 30.643 $\pm$ 0.066 mag
 (Hoyt et al. 2025). In 2021, the TRGB distance to NGC 5643
  was significantly lower with a distance modulus of
  $\mu$ $=$ 30.48 $\pm$ 0.03 (stat) $\pm$ 0.07 (sys) (Hoyt et al. 2021).
There is a considerable difference in the TRGB values 
provided by SH0ES and by the CCHP, and even between the values provided
by the CCHP in the last years.

\bigskip

\noindent
Fortunately, there is an  agreement on the distance to NGC 5643 between
 Cepheids used by SH0ES (Riess et al. 2022)
and  JAGB stars used by the CCHP
(Freedman et al. 2025). As mentioned above the Cepheid distance
value by SH0ES is $\mu$ $=$ 30.55 $\pm$ 0.06 mag, and their TRGB value
$\mu$ $=$ 30.57/30.58 $\pm$ 0.06 mag. Whereas the TRGB value by the CCHP
has been changing often, their JAGB value seems  stable at 
$\mu$ $=$ 30.582 $\pm$ 0.038 mag. (The JAGB value by SH0ES is
30.54 $\pm$ 0.04, well in line with the previous values). 
We compare then our results using  as anchor for
this galaxy the Cepheids by SH0ES with the value anchored in the
JAGB stars by the CCHP in section 6.

\subsection{NGC 7250}

\noindent
In NGC 7250 there is good agreement between the distance obtained
by SH0ES and the distance obtained by the CCHP
with the JAGB stars. 

\bigskip

\noindent
The latest SH0ES measurement gives a distance  of
$\mu$ $=$  31.49 $\pm$ 0.05 mag using Cepheids measured with the JWST
(Appendix 1 in Riess et al. 2025). The distance to NGC 7250 with Cepheids
before the JWST had  a $~$ 0.125 mag error.
There is a very good agreement between the SH0ES collaboration
and the CCHP collaboration  using JAGB stars. The value by the SH0ES
project is 31.59 $\pm$ 0.04 mag  by Li et al. (2025) 
and  31.59 $\pm$ 0.02 (stat) $\pm$ 0.04 (sys) mag by
the the CCHP project  (Lee  et al.  2025). 
This makes of NGC 7250
a good anchor, as central values coincide. 
On the contrary, the value using TRGB by the 
CCHP is $\mu$ $=$  31.629 $\pm$ 0.034 and it is
different from the TRGB by SH0ES.

\bigskip

\subsection{M66}

\noindent
This is a crucial SN Ia host whose distance has been been measured
with TRGB by Freedman et al. (2019) giving a distance modulus of
$\mu$ $=$ 30.22 $\pm$ 0.04 mag. In fact, this measurement seems to be
based on Hoyt et al. (2019) TRGB  measurement of the distance to M66,
though the value quoted in this work is $\mu$ $=$
30.23 $\pm$ 0.04 (stat) $\pm$ 0.06 
(syst) mag. The CCHP distance to M66 coincides with the distance
modulus obtained with Cepheids by Saha et al. (1999) of 
$\mu$ $=$ 30.22 $\pm$ 0.12 mag.
The agreement between the distance obtained by Cepheids by Saha et al (1999) and the one quoted 
using TRGB by the CCHP is not a simple coincidence. The zero point adopted by Saha
et al. (1999) was an LMC distance of µ = 18.5 mag, while the CCHP adopted and LMC distance of
µ = 18.48 mag.
This agreement has been found despite the fact of the crowding of Cepheids in M66 (the WFPC2
fields are quite crowded and of lower quality than modern observations taken with ACS or WFC3).  Despite other possible systematic effects of the two
methods, the zero point to LMC has been crucial in their agreement.

\bigskip

\noindent
M66 looks  like a good anchor for
the BL SNe Ia in the Hubble flow. 
It is  ideal for getting distances to SN 2011cn in IC 4758,
SN 2008go in  2MASX J22104396-2047256, SN 1999ek in UGC 3329.
 Concerning the systematic error of the distance of the anchor,
  we assign an error of 0.04 mag to the distance of M66, as for the rest
  of the anchor galaxies.

\bigskip

\noindent
For the comparison of SN 1989B and the SNe Ia
which are BL but also high velocity (HV),  we need to
apply a shift to the reference SN 1989B
in accordance with the HV they have.
  SNe Ia from the BL subtype move faster than those
  of the CN subtype. A shift in the spectral features to the blue to
  compare
  the spectra between BL subtypes is a step expected
  within our twin to twin comparison. 
 The amount for which 
one SN Ia of the BL subtype is  moving faster than another one
from the BL subtype is not
a  parameter in the Branch et al. (2006) classification.

\section{Sample of SNe Ia in the  Hubble flow}

\bigskip

\noindent
The selection of twins has to follow the requirements of being within
the same subtype of SNe Ia (within a given Branch et al. 2006 class),
have a similar stretch within errors, and having various epochs for
the spectra.  A number of  SNe Ia from the Carnegie Supernova Project (CSP)
fullfil those criteria.

\bigskip

\noindent
Table 2 shows the references for the spectra and photometry used in this paper.
Some spectra come directly from the Carnegie Supernova Project archive,
and some others taken from  the {\it WISeREP}. All are
carefully calibrated in flux in agreement with the published photometry.
The error in the flux calibration is  taken into account. 

\bigskip

\noindent
The comparison of the light curves of the Hubble flow SNe Ia
with those nearby, is a first selection criterion in the use of the ''SNe Ia
twins for life'' method.
We have taken similar 
$\Delta m(B)_{15}$ and the stretch factor  $s^{D}_{BV}$ (see Tables 3 and 4).
Burns et al. (2014) describe a relation between the SNooPy stretch and
$\Delta m_{15}(B)$ as: 

\begin{equation}
 s_{BV}  = 0.955-0.458 ( \Delta m_{15}(B) -1.1) 
\end{equation}

\noindent
where  $\Delta m_{15}(B)$ is the decline rate parameter established by
Phillips (1993), which is at the core of the discovery of the correlation
of the brightness  of a SN Ia and
the rate of decline of the B light curve in the 15 days past maximum.
The color stretch $s_{BV}$ is the  rate of decline of the B-V color
in 30 days. One disadvantage of $s_{BV}$ is that one needs to have a
photometric B and V coverage from B maximum until approximately 40 days
past maximum. The CSP and Burns et al. (2020) quote,  instead of $s_{BV}$, the
stretch  $s^{D}_{BV}$, where  the ``D'' in the superindex refers to the
direct comparison in B, without having to take into account the {\it BVRI}
light curves,

\bigskip

\noindent
In this research, we have been very careful in selecting for the
comparison of SNe Ia in the Hubble flow with their twins, SNe Ia
with very similar $\Delta m_{15}(B)$ and stretch  $s^{D}_{BV}$ values.

\begin{table*}
    \scriptsize
  \centering
  \caption{Spectra and Photometry used for the SN fits}
  \begin{tabular}{lcccccc}
    \hline
    \hline
    Supernova & Galaxy & $z_{{\bf hel}}$ & Spectral range (\AA) & Phase & Date &  
    Reference \\
    \hline

SN 2012bo & NGC 4726 & 0.025431 & 3665-8847 & +14 &
    2012-04-18 & WISeREP \\
    $\dots$ $\dots$ $\dots$ & $\dots$ & $\dots$ & 3634-9602 & +26 &
    2012-04-30 & $\dots$ \\
SN 2008bq & ESO 308-G25 & 0.03400 & 3675-8931 & +5.57  &
    2008-04-12 & CSP-DR1 \\
    $\dots$ $\dots$ $\dots$ & $\dots$ &    &   & +33.44 &
    2008-05-10 & $\dots$ \\
SN 2008bz & A123857+1107 & 0.0603 & 3626-9433 & +4.65 &
    2008-04-26 & $\dots$ \\
    $\dots$ $\dots$ $\dots$ & $\dots$ & $\dots$ & 3800-9235 & +13 &
    2008-05-05 & $\dots$ \\
LSQ12fxd & ESO 487-G004 & 0.0312 & 3352-7475 & -2 & 2012-11-13 &
    CSP-DR1 \\
    $\dots$ $\dots$ $\dots$ & $\dots$ & $\dots$ & 3579-9615 & +3.7 &
    2012-11-19 & $\dots$ \\
SN 2008bf & NGC 4055/4057 & 0.0235 &  3715-9974  & -2.70 &
    2008-03-26 & $\dots$  \\
    $\dots$ $\dots$ $\dots$ & $\dots$ & $\dots$ & 3710-9018 & +42.07 &
    2008-05-11 & $\dots$  \\
SN 2007A & NGC 105 & 0.01764 & 3473-7409 & -2.86 &
    2007-01-10 &  WISeREP \\
    $\dots$ $\dots$ $\dots$ & $\dots$ & $\dots$ & 3378-1027 & +13.14 &
    2007-02-24 & $\dots$  \\
ASASSN-15db & NGC 5996  & 0.01099 & 3604-9164 & +3 & 2015-02-25 &$\dots$ \\
    $\dots$ & $\dots$ & $\dots$ & $\dots$ & +4 & 2015-02-26 & $\dots$\\
LSQ14gov & GALEXMSC J040601.67-160139.7  & 0.0896 & 3650-9250 & -6
     & 2013-12-30 & $\dots$ \\
$\dots$ & $\dots$ & $\dots$ & 3430-9100 & +1 & 2014-01-06 & $\dots$ \\
     SN 2007ca & MCG -02-34-61 & 0.01407 & 3455-9494 & +1.4 & 2007-05-08 &
     CSP-DR1 \\
     $\dots$ & $\dots$ & $\dots$ & 3297-9416 & +4.2 & 2007-05-11 & $\dots$ \\
SN 2001cn & IC 4758 & 0.0156 & 3827-7218 & +2.2 &
     2001-06-14 &  $\dots$ \\
     $\dots$ & $\dots$ & $\dots$ & 3715-7115  & +10  &
     2001-06-22 & $\dots$ \\
SN 2008go & 2MASX J22104396-2047256 & 0.0623 & 3800-9234  & +0.6 &
     2008-10-27 & $\dots$ \\
     $\dots$ & $\dots$ & $\dots$ & 3636-9450 & +7.5 &
     2008-11-04 & $\dots$ \\
SN 1999ek & UGC 3329 & 0.0177 & 3720-7540 & +3.3 &
     1999-11-03 & $\dots$ \\
     $\dots$ & $\dots$ & $\dots$ & 3720-7540 & +9.32  &
     1999-11-09 & $\dots$ \\
   
    \hline

    &  &  & Photometry &  &  &  \\

    \hline

    SN 2012bo & $\dots$ & $\dots$ & $\dots$ & $\dots$ & Morrell
    & $\dots$ \\
    SN 2008bq & $\dots$ & $\dots$ & $\dots$ & $\dots$ & CSP-DR1 & $\dots$ \\
    SN 2008bz & $\dots$ & $\dots$ & $\dots$ & $\dots$ & CSP-DR1 & $\dots$ \\
    LSQ12fxd  & $\dots$ & $\dots$ & $\dots$ & $\dots$ & Morrell & $\dots$ \\
    SN 2008bf & $\dots$ & $\dots$ & $\dots$ & $\dots$ & CSP-DR1 & $\dots$ \\
    SN 2007A  & $\dots$ & $\dots$ & $\dots$ & $\dots$ & CSP-DR1 & $\dots$ \\
    ASASSN-15db & $\dots$ & $\dots$ & $\dots$ & $\dots$ & Morrell & $\dots$ \\
    LSQ14gov  & $\dots$ & $\dots$ & $\dots$ & $\dots$ & Morrell & $\dots$ \\
    SN 2007ca & $\dots$ & $\dots$ & $\dots$ & $\dots$ & CSP-DR1 & $\dots$ \\
    SN 2001cn & $\dots$ & $\dots$ & $\dots$ & $\dots$ & Krisciunas & $\dots$ \\
    SN 2008go & $\dots$ & $\dots$ & $\dots$ & $\dots$ & CSP-DR1 & $\dots$ \\
    SN 1999ek & $\dots$ & $\dots$ & $\dots$ & $\dots$ & Krisciunas & $\dots$ \\
    \hline
     
  \end{tabular}
 
\end{table*}

\begin{table}
  \scriptsize
  \centering
  \caption{SNe 2012bo, 2008bq, 2008bz, LSQ12fxd, 2008bf, 2007A}
  \begin{tabular}{lcc}
    \hline
    \hline
        {\bf SN 2012bo} in NGC 4726  &      &     \\
        RA, DEC$^{a}$ & 12:50:45.23 & -14:16:08.5 \\
        Discovery date$^{a}$ & $\dots$ & 2012-03-27 \\
        Phase (referred to maximum light)$^{b}$ & $\dots$ & -8 days \\
        Redshift$^{b}$ & $\dots$ & 0.026548 \\ 
        E(B-V)$_{MW}$$^{c}$ & $\dots$ & 0.046 mag \\
        Stretch factor $s^{D}_{BV}$ $^{b}$ & $\dots$ & 1.160$\pm$0.0411 \\      
        $\Delta m(B)_{15}$ $^{b}$ & $\dots$ & 0.931$\pm$0.061 \\ 
        Phases of the spectra used & $\dots$ & 14, 26 days \\
        \hline
        \end{tabular}
\begin{tabular}{ll}
        $^{a}$Itagaki et al. (2012). &  $^{b}$Morrell et al. (2023). \\
  $^{c}$Schlafly \& Finkbeiner (2011). &
  \end{tabular}
\begin{tabular}{lcc}
  
        {\bf SN 2008bq} in ESO 308-G25 &      &     \\
        RA, DEC$^{a}$ & 06:41:02.51 & -38:02:19.0 \\
        Discovery date$^{a}$ & $\dots$ & 2008-04-02 \\
        Phase (referred to maximum light)$^{b}$ & $\dots$ & -4 days \\
        Redshift$^{b}$ & $\dots$ & 0.03446 \\ 
        E(B-V)$_{MW}$$^{c}$ & $\dots$ & 0.080 mag \\
        Stretch factor $s^{D}_{BV}$ $^{d}$ & $\dots$ & 1.056$\pm$0.014 \\
        $\Delta m(B)_{15}$ $^{d}$ & $\dots$ & 0.895$\pm$0.062 \\ 
        Phases of the spectra used & $\dots$ & 5.57, 33.44 days \\
        \hline
        \end{tabular}
\begin{tabular}{ll}
    $^{a}$Luckas et al. (2008). &  $^{b}$Folatelli et al. (2013). \\ 
    $^{c}$Schlafly \& Finkbeiner (2011). &
    $^{d}$Krisciunas et al. (2017).   
  \end{tabular}
\begin{tabular}{lcc}
  
        {\bf SN 2008bz} in A123957+1107 &      &     \\
        RA, DEC$^{a}$ & 12:38:57.74 & +11:07:46.2 \\
        Discovery date$^{a}$ & $\dots$ & 2008-04-22 \\
        Phase (referred to maximum light)$^{b}$ & $\dots$ & $\simeq$0 days \\
        Redshift$^{b}$ & $\dots$ & 0.0614 \\ 
        E(B-V)$_{MW}$$^{c}$ & $\dots$ & 0.023 mag \\
        Stretch factor $s^{D}_{BV}$ $^{d}$ & $\dots$ & 0.948$\pm$0.046 \\
        $\Delta m(B)_{15}$ $^{b}$ & $\dots$ & 1.093$\pm$0.079 \\ 
        Phases of the spectra used & $\dots$ & 4.65, 13 days \\
        \hline
\end{tabular}
\begin{tabular}{ll}
    $^{a}$Yuan et al. (2008). &  $^{b}$Krisciunas et al. (2020). \\ 
    $^{c}$Schlafly \& Finkbeiner (2011)t al. (2023). &
  $^{d}$Morrell et al. (2023).
  \end{tabular}
\begin{tabular}{lcc}
         
        {\bf LSQ12fxd} in ESO487-G004 &      &     \\
        RA, DEC$^{a}$ & 05:22:17.02 & -25:35:47.1 \\
        Discovery date$^{a}$ & $\dots$ & 2012-10-31 \\
        Phase (referred to maximum light)$^{b}$ & $\dots$ & -15 days \\
        Redshift$^{b}$ & $\dots$ & 0.031 \\ 
        E(B-V)$_{MW}$$^{c}$ & $\dots$ & 0.022 mag \\
        Stretch factor $s^{D}_{BV}$ $^{b}$ & $\dots$ & 1.161$\pm$0.040 \\       
        $\Delta m(B)_{15}$ $^{b}$ & $\dots$ & 0.940$\pm$0.060 \\ 
        Phases of the spectra used & $\dots$ & -2, 3.7 days \\
        \hline
        \end{tabular}
\begin{tabular}{ll}
    $^{a}$Bufano et al. (2012). &  $^{b}$Morrell et al. (2023). \\
    $^{c}$Schlafly \& Finkbeiner (2011). &   
  \end{tabular}
\begin{tabular}{lcc}
  
        {\bf SN 2008bf} in NGC 4055/4057  &      &     \\
        RA, DEC$^{a}$ & 12:04:02.90 & +20:14:42.6 \\
        Discovery date$^{a}$ & $\dots$ & 2008-03-18 \\
        Phase (referred to maximum light)$^{b}$ & $\dots$ & -11 days \\
        Redshift$^{b}$ & $\dots$ & 0.02510 \\ 
        E(B-V)$_{MW}$$^{c}$ & $\dots$ & 0.036 mag \\
        Stretch factor $s^{D}_{BV}$ $^{d}$ & $\dots$ & 1.058$\pm$0.012 \\
        $\Delta m(B)_{15}$ $^{d}$ & $\dots$ & 0.97$\pm$0.060 \\ 
        Phases of the spectra used & $\dots$ & -2.70, 42.07 days \\
        \hline
 \end{tabular}
\begin{tabular}{ll}
   $^{a}$Panski et al.(2008). &  $^{b}$Folatelli et al. (2013). \\
    $^{c}$Schlafly \& Finkbeiner (2011). &
    $^{d}$Krisciunas et al. (2017).
     \end{tabular}
\begin{tabular}{lcc}
  
        {\bf SN 2007A} in NGC 105 &      &     \\
        RA, DEC$^{a}$ & 00:25:16.66 & +12:53:12.5 \\
        Discovery date$^{a}$ & $\dots$ & 2007-01-02 \\
        Phase (referred to maximum light)$^{b}$ & $\dots$ & -10 days \\
        Redshift$^{b}$ & $\dots$ & 0.01648 \\ 
        E(B-V)$_{MW}$$^{c}$ & $\dots$ & 0.074 mag \\
        Stretch factor $s^{D}_{BV}$ $^{d}$ & $\dots$ & 1.012$\pm$0.061 \\
        $\Delta m(B)_{15}$ $^{d}$ & $\dots$ & 0.971$\pm$0.074 \\ 
        Phases of the spectra used & $\dots$ & -2.86, 13.14 days \\
        \hline
         \end{tabular}
\begin{tabular}{ll}
    $^{a}$Picket et al. (2007). & $^{b}$Folatelli et al. (2013). \\
    $^{c}$Schlafly \& Finkbeiner (2011). & 
    $^{d}$Krisciunas et al. (2017).
\end{tabular}

\end{table}

\begin{table}
  \scriptsize
  \centering
  \caption{SNe ASASSN-15db, LSQ14gov, 2007ca, 2001cn, 2008go, 1999ek}  
  \begin{tabular}{lcc}

    \hline
    \hline

        {\bf ASASSN-15db} in NGC 5996 &      &     \\
        RA, DEC$^{a}$ & 15:46:58.9 & 17:53:03 \\
        Discovery date$^{a}$ & $\dots$ & 2015-02-15 \\
        Phase (referred to maximum light)$^{b}$ & $\dots$ & -7 days \\
        Redshift$^{b}$ & $\dots$ & 0.0114 \\ 
        E(B-V)$_{MW}$$^{c}$ & $\dots$ & 0.03 mag \\
        Stretch factor $s^{D}_{BV}$ $^{b}$ & $\dots$ & 0.953$\pm$0.040 \\      
        $\Delta m(B)_{15}$ $^{b}$ & $\dots$ & 1.089$\pm$0.061 \\ 
        Phases of the spectra used & $\dots$ & 3, 4 days \\
        \hline
           \end{tabular}
\begin{tabular}{ll}      
    $^{a}$ Holoien et al. (2015). &  $^{b}$Morrell et al. (2023). \\
  $^{c}$Schlafly \& Finkbeiner (2011).
           \end{tabular}
\begin{tabular}{lcccccc}
  {\bf LSQ14gov} in GALEXMSC J040601.67-160149.7 & & & & & & \\
\end{tabular}
\begin{tabular}{lcc}
        RA, DEC$^{a}$ & 04:06:01.53 & -16:01:38.86 \\
        Discovery date$^{a}$ & $\dots$ & 2014-12-21 \\
        Phase (referred to maximum light)$^{b}$ & $\dots$ & -15 days \\
        Redshift$^{b}$ & $\dots$ & 0.0896 \\ 
        E(B-V)$_{MW}$$^{c}$ & $\dots$ & 0.0377 mag \\
        Stretch factor $s^{D}_{BV}$ $^{b}$ & $\dots$ & 1.141$\pm$0.041 \\      
        $\Delta m(B)_{15}$ $^{b}$ & $\dots$ & 0.979$\pm$0.061 \\ 
        Phases of the spectra used & $\dots$ & -6, 1 days \\
        \hline
            \end{tabular}
\begin{tabular}{ll}      
    $^{a}$Ducrof et al. (2014). &  $^{b}$ Morrell et al. (2023). \\
    $^{c}$Schlafly \& Finkbeiner (2011). &
           \end{tabular}
\begin{tabular}{lcc}
  
        {\bf SN 2007ca} in MGC-02-34-61 &      &     \\
        RA, DEC$^{a}$  & 13:31:05.81 & -15:06:06.6 \\
        Discovery date$^{a}$ & $\dots$ & 2007-04-25 \\        
        Phase (referred to maximum light)$^{b}$ & $\dots$ & -12 days \\
        Redshift$^{b}$ & $\dots$ & 0.01508 \\
        E(B-V)$_{MW}$$^{b}$ & $\dots$ & 0.067 \\
        Stretch factor $s^{b}_{BV}$ $^{b}$ & $\dots$ & 1.077$\pm$0.012 \\
        $\Delta m(B)_{15}$ $^{b}$ & $\dots$ & 0.904$\pm$0.032 \\
        Phases of the spectra used & $\dots$ & 1.4, 4.2 days \\
        \hline
            \end{tabular}
\begin{tabular}{ll}
    $^{a}$Itagaki et al. (2007). & $^{b}$Uddin et al. (2020). \\ 
           \end{tabular}
\begin{tabular}{lcc}
  
        {\bf SN 2001cn} in IC 4758 &      &     \\
        RA, DEC$^{a}$  & 18:46:17.84 & -65:45:41.8 \\
          Discovery date$^{a}$ & $\dots$ & 2001-06-12 \\
          Phase (referred to maximum light)$^{b}$ & $\dots$ & -4 days \\
          Redshift$^{b}$ & $\dots$ & 0.0156 \\
          E(B-V)$_{MW}$$^{c}$ & $\dots$ & 0.052 \\
          Stretch factor $s^{b}_{BV}$ $^{b}$ & $\dots$ & 0.927$\pm$0.020 \\ 
          $\Delta m(B)_{15}$ $^{b}$ & $\dots$ & 1.080$\pm$0.044 \\
          Phases of the spectra used & $\dots$ & 2.2, 10 days \\
          \hline
\end{tabular}
\begin{tabular}{ll}
    $^{a}$Chassagne (2001) & $^{b}$Morrell et al. (2024) \\
    $^{c}$Schlafly \& Finkbeiner (2011). &   \\
\end{tabular}
\begin{tabular}{lcccccccc}
        {\bf SN 2008go} in 2MASX J22104396-2047256 &  &  &  &  &  &  &  & \\
\end{tabular}
\begin{tabular}{lcc}
        RA, DEC$^{a}$ & 22:10:44.03 & -20:47:17.2 \\
        Discovery date$^{a}$ & $\dots$ & 2008-10-22 \\
        Phase (referred to maximum light)$^{b}$ & $\dots$ & -5 days \\  
        Redshift$^{c}$ & $\dots$ & 0.06 \\
        E(B-V)$_{MW}$$^{b}$ & $\dots$ & 0.033  \\
        Stretch factor $s^{b}_{BV}$ $^{b}$ & $\dots$ & 0.939$\pm$0.044 \\
        $\Delta m(B)_{15}$ $^{b}$ & $\dots$ & 1.093$\pm$0.089 \\
        Phases of the spectra used & $\dots$ & 0.6, 7.5 days \\
        \hline
\end{tabular}
\begin{tabular}{ll}
        $^{a}$Griffith et al. (2008). &
        $^{b}$Stahl et al. (2019). \\
        $^{c}$Schlafly \& Finkbeiner (2011). &   \\
 \end{tabular}
\begin{tabular}{lcc}

        {\bf SN 1999ek} in UGC 3329 &      &     \\
        RA, DEC$^{a}$  & 05:36:31.60 & +16:38:17.8 \\
        Discovery date$^{a}$ & $\dots$ & 1999-10-20 \\        
        Phase (referred to maximum light)$^{b}$ & $\dots$ -11 days \\
        Redshift$^{b}$ & $\dots$ & 0.0177 \\
        E(B-V)$_{MW}$$^{b}$ & $\dots$ & 0.502 \\
        Stretch factor $s^{b}_{BV}$ $^{b}$ & $\dots$ & 0.921$\pm$0.008 \\
        $\Delta m(B)_{15}$ $^{b}$ & $\dots$ &1.097$\pm$0.019 \\
        Phases of the spectra used & $\dots$ & 3.3, 9.32 days \\
        \hline
        \end{tabular}
  \begin{tabular}{ll}
    $^{a}$Johnson \& Li (1999. & $^{b}$Morrell et al. (2024). \\
 \end{tabular}
  
\end{table}  

\bigskip

\subsection{The SN 2011fe--like in the Hubble flow}

\bigskip

\noindent
We have thus M101 and its supernova SN 2011fe at a distance of
d $=$ 6.880 $\pm$ 0.174 Mpc ($\mu$ $=$ 29.188 $\pm$ 0.055).
In the Hubble flow we have found an excellent twin
in SN 2008bz. It has a
$\Delta m(B)_{15}$ of 1.093 $\pm$ 0.079. The $\Delta m(B)_{15}$
of SN 2011fe is 1.07 $\pm$ 0.006.\footnote{The two SN 2011fe-like SNe Ia
have the following
stretches: SN 2008bz, $s^{D}_{BV}$ = 0.948$\pm$0.948$\pm$0.046;
ASASSN-15db, $s^{D}_{BV}$ = 0.953$\pm$0.04. Both compare well with
SN 2011fe, with $s^{D}_{BV}$ = 0.97$\pm$0.033.}
SN 2008bz's heliocentric
redshift is 0.06, thus the full expansion of the modulus
in q$_{0}$ is needed. The result can be seen in Figure 1. It
is remarkable that with this method we can go from 6.8 Mpc to
 $\sim$ 270 Mpc in a single step. We will come back to the result
in distance in section 6.

\bigskip

\noindent
Another SN Ia in the Hubble flow of the same type as SN 2011fe is
ASASSN-15db in NGC 5996. It has a nearly identical light--curve decline
value as SN 2011fe, with
$\Delta m(B)_{15}$ $=$ 1.089 $\pm$ 0.061
for ASSASN-15db and
$\Delta m(B)_{15}$ $=$ 1.07 $\pm$ 0.006 for SN 2011fe.
Despite the fact that we do not have the full light curve of this
SN Ia (which belongs to the DR of CSP II soon to come), we can calibrate
the spectra with the photometry for the two phases.

\subsection{SNe Ia in the Hubble flow like SN 2013aa/2017cbv}

\noindent
 SNe Ia that can be considered twins of SN 2013aa are
SN 2007A, SN 2008bf and LSQ14gov. 

\bigskip

\noindent
SN  2007A has  $\Delta m(B)_{15}$ $=$
0.971$\pm$0.074 which compares well with  SN 2013aa and SN 2017cbv,
since $\Delta m(B)_{15}$ $=$ 0.95$\pm$0.01 for SN 2013aa.\footnote{The stretches of
the three SNe Ia of this group are: for SN 2007A,
$s^{D}_{BV}$ = 1.025$\pm$0.061; for SN 2008bf, $s^{D}_{BV}$ = 1.058$\pm$0.012;
for LSQ14gov, $s^{D}_{BV}$ = 1.141$\pm$0.041. All of them compare well with
the stretch of the anchor SN 2013aa: $s^{D}_{BV}$ = 1.11$\pm$0.02.}

\bigskip

\noindent
This supernova has an internal reddening of 0.2 magnitudes. Such a
high extinction by dust has been well captured by  the SNooPY light curve
fitter. It was not well captured by SALT2 and, therefore, the value of the
distance derived by Pantheon+ to this supernova is too large.  

\bigskip

\noindent
SN 2008bf compares very well with SN 2013aa. The first one
has a $\Delta m(B)_{15}$ $=$ 0.97 $\pm$ 0.06
and the second one  $\Delta m(B)_{15}$ $=$ 0.95$\pm$0.01.
For the case of doubt about their similarity, a direct comparison
of the light curves can be seen  for this one and for all
SNe Ia from CSP I (see Appendix A). SN 2008bf as a
perfect twin of SN 2013aa.

\bigskip 

 \noindent
 For the last comparison, we have LSQ14gov
which has $\Delta m(B)_{15}$ $=$ 0.979$\pm$0.061 to compare with
$\Delta m(B)_{15}$ 0.95 $\pm$ 0.01 of SN 2013aa.

\subsection{SNe Ia in the Hubble flow like SN 2013dy}

\bigskip

\noindent
The advantage of these  SNe Ia is that there is a pretty good
agreement between the distances given by the two groups
for this anchor.

\bigskip

\noindent
Four SNe Ia from the SN 2013dy class are SN 2012bo and SN 2008bq,
SN 2007ca and LSQ12fxd.

\bigskip
 
 \noindent
 SN 2012bo has a  $\Delta m(B)_{15}$ $=$ 0.931$\pm$0.061 
 which compares  
 well with the $\Delta m(B)_{15}$ $=$ 0.92 $\pm$ 0.006
 from SN 2013dy.\footnote{The stretches of
the four SNe Ia of this group are: for SN 2012bo,
$s^{D}_{BV}$ = 1.160$\pm$0.0411; for SN 2008bq, $s^{D}_{BV}$ = 1.056$\pm$0.014;
for SN 2007ca, $s^{D}_{BV}$ =1.077$\pm$0.012; for LSQ12fxd,
 $s^{D}_{BV}$ = 1.161$\pm$0.040. All of them compare well with
the stretch of the anchor SN 2013dy: $s^{D}_{BV}$ = 1.091$\pm$0.03.}

\bigskip

\noindent
We have paired SN 2008bq with SN 2013dy. It has
a similar $\Delta m(B)_{15}$ of 0.90 $\pm$ 0.062 versus 0.92$\pm$0.006
for SN 2013dy.

\bigskip

\noindent
SN 2007ca has
 $\Delta m(B)_{15}$ $=$ 0.904$\pm$0.032  while
$\Delta m(B)_{15}$ is 0.92 $\pm$ 0.006 for SN 2013dy.
 They are very good twins.

\bigskip
\noindent
LSQ12fxd has  $\Delta m(B)_{15}$ $=$ 0.940$\pm$0.06,
versus $\Delta m(B)_{15}$ $=$ 0.92$\pm$0.006 of
SN 2013dy.

\subsection{SNe Ia in the Hubble flow like SN 1989B}

\noindent
SN 1989B is a broadline SNe Ia that appeared in M66, a nearby
galaxy for which there is a coincident distance by TRGBs 
(Hoyt et al. 2019; Freedman et al. 2019) and Cepheids (Saha et al. 1999).
There is no published distance by SH0ES
  because SN 1989B shows too much host reddening
  to be in the fiducial SH0ES samples.

\bigskip

\noindent
Three SNe Ia in the CSP I  are broad line and
have a  rate of decline in the light curve similar to SN 1989B.
Usually broadline supernovae have also
high velocity ejecta. This produces a blueshift in the lines
of the supernova. 
SN 1989B is not a high velocity supernova.  Two of the broadline
SNe Ia at high--z have high velocity features: SN 2001cn and
SN 2008go. In contrast, SN 1999ek is a  broad line  that does not
exhibit high velocity in  the its spectral features. 

\bigskip

\noindent
We have paired SN 2001cn, SN 2008go and SN 1999ek with
SN 1989B. There will be a blueshift applied to SN 1989B to compare
with SN 2001cn and SN 2008go, whereas in SN 1999ek it is not
necessary. 

\bigskip

\noindent
SN 2001cn has  $\Delta m(B)_{15}$ = 1.080$\pm$0.044, which fits well with
$\Delta m(B)_{15}$ = 1.053$\pm$0.11 of SN 1989B.

\bigskip

\noindent
SN 2008go has $\Delta m(B)_{15}$ = 1.093$\pm$0.089, also concordant with
the $\Delta m(B)_{15}$ = 1.053$\pm$0.11 of SN 1989B.

\bigskip

\noindent
SN 1999ek has $\Delta m(B)_{15}$ = 1.097$\pm$0.019, again compatible with
the $\Delta m(B)_{15}$ = 1.053$\pm$0.11 of the anchor\footnote{Concerning the
stretch, SN 1989B has $s^{D}_{BV}$ = 0.954$\pm$0.034. The corresponding values
for its distant twins are: for SN 2001cn, $s^{D}_{BV}$ = 0.927$\pm$0.020; for
SN 2008go, $s^{D}_{BV}$ = 0.939$\pm$0.044; for SN 1999ek,
$s^{D}_{BV}$ = 0.921$\pm$0.008. So, the stretches  of the three distant SNe Ia
 compare well with that of the anchor SN 1989B.}

\section{Methodology}

\noindent
We have used eight SNe Ia from Carnegie I included in the DR1.
Their characteristics  are described in Folatelli et al. (2012)
and four SNe Ia in  Carnegie II (Morrell et al. 2024).
We have been able to calibrate the spectra,
which are uploaded to
the {\it WISeREP}.

\bigskip

\noindent
All the spectra have been corrected from Galactic reddening.
This is done with the purpose of allowing to compare Galactic dust-free
spectra. It also helps to learn about dust extinction in the host.
The spectra are compared in the frame of the Hubble flow supernova.
For that purpose  nearby spectra are shifted in wavelength and 
their flux is diluted in accordance with the (1 + z) law. The shift
brings  the nearby SN Ia to the redshift of the Hubble flow supernova taking
into account the heliocentric redshift of the nearby SNIa (shifting
by (1 + z$_{flow}$)/(1+z$_{nearby}$), though the redshift of nearby
SNe Ia is always low). 

\bigskip

\noindent
The method primarly determines  the  distance between  the nearby 
SN Ia and that in the Hubble flow. It will also give the relative
$\Delta E(B - V)$ reddening.
  The $\Delta E(B - V)$ value will indicate whether $E(B–V)$ is
  larger or smaller for one of the two SNe with respect to its twin, 
  taken as reference. Concerning the distance modulus, in reality we are calculating distances
  relative to the SNe Ia of reference. So, we have a distance scale factor
  between the reference and the SN in the Hubble flow. For the sake of
   producing a final result, we give the absolute distance modulus $\mu$.

\bigskip

\noindent
To find the best values and the uncertainties of all variables, we explore the
parameter space with Markov Chain Monte Carlo (MCMC) techniques after
converting the $\chi^{2}$ into a log likelihood function (the probability of
a dataset given the model parameters) that we aim to optimize:

\begin{equation}
{\rm log} \mathcal{L} = -0.5 * \sum{(f_{\lambda}^{obs} -
  f_{\lambda}^{ref})^{2}/\sigma_{\lambda}^{2}} + {\rm log} \sigma_{\lambda}
\end{equation}

\bigskip

\noindent
where {\it obs} corresponds to a given supernova and {\it ref} to its  reference twin, which represents a whole class. 
$f_{\lambda}^{obs}$ and $f_{\lambda}^{ref}$ correspond to the fluxes of the two SNe at different wavelengths, 
  and $\sigma_{\lambda}$, defined as $\sigma_{\lambda}^2 = \epsilon_{obs}^2 + \epsilon_{ref}^2 + \epsilon_{int}^2$ includes the quadratic addition of the uncertainty on the fluxes of both SNe and the intrinsic uncertainty in the flux comparison. This uncertainty can be described as the maximum flux of the observed SNIa multiplied by a factor, $\eta_{obs}$, as $\epsilon_{int} = max(f_{\lambda}^{obs}) * \eta_{obs}$ . This intrinsic
  uncertainty absorbes the remaining uncertainty to help to maximize the likelihood. We note that in RLGH24, this intrinsic uncertainty was already included with a slightly different definition using the $f_{\lambda}^{obs}$ multiply by the factor $\eta_{obs}$. This new definition of $\epsilon_{int}$ provides an estimate of the intrinsic magnitude scatter, as $\sigma_{int} = (2.5 / \ln(10)) \times ( \epsilon_{int} / max(f_{\lambda}^{obs}) )$.   

The MCMC allows to get the maximum likelihood values of each parameter from the posterior distribution of samples, together with their
uncertainties, including the propagation of all the parameter uncertainties on each parameter.

\bigskip

\noindent
In fact, as we are using different phases, we have a total likelihood function.
For two phases, $n$ corresponds to 1 and 2 below. Thus the total likelihood
function is written as

\begin{equation}
{\rm log} \mathcal{L}_{n} = {\rm log} \mathcal{L}_{1} +
{\rm log} \mathcal{L}_{2}
\end{equation}
  
  \noindent
  i.e. just the addition of all individual SN phases.

\noindent
We use a Python package, EMCEE (Foreman–Mackey et al. 2013)\footnote{https://emcee.readthedocs.io/} to explore the
likelihood of each variable. EMCEE utilizes an affine invariant MCMC ensemble
sampler proposed by Goodman et al. (2010). This sampler
tunes only two parameters to get the desired output: number of
walkers and number of steps.
The run starts by assigning initial maximum likelihood values of the
variables to the walkers. The walkers then start wandering and explore the
full posterior distribution.
After an initial run, we inspect the samplers for their performance. We
do this by looking at the time series of variables in the chain and computing
the autocorrelation time, $\tau$\footnote{https://emcee.readthedocs.io
/en/stable/tutorials/autocorr/}. In our case, the maximum autocorrelation
time among the different variables is about 50.
When the chains are sufficiently burnt-in, they
forget their initial start point), we can safely throw away some
 steps that are a few times higher than the burnt-in steps. In our case,
 we run EMCEE with 32 walkers and 10,000 steps and throw
away the first  $\sim$ 250 samples, equivalent to $\sim$ 4 times
the maximum autocorrelation time. 
Thus, our burn--in value in this computation is equal to 4 times the maximum autocorrelation time. 
  A criterion
of good sampling is the acceptance fraction, $a_{f}$. This is the fraction of
steps that are accepted after the sampling is done. The suggested value of
$a_{f}$ is between 0.2 - 0.5 (Gelman et al. 1996).
In each run, we typically  obtain $a_{f} \sim$ 0.25.

\bigskip

\noindent
      The priors in the difference between E(B-V) of the host galaxies
      go from  [--0.4, + 0.4].  As we said before, we correct from 
      Galactic reddening the SNe Ia spectra before running the MCMC.
      So our intrinsic $\Delta (E(B-V))$ is enterely attributted to
      the host galaxies.
      In relation to the priors in distance for each supernova, they vary
      depending of the particular
      SN Ia.\footnote{ Distance priors are (in Mpc):
      [95.0, 115.0]
(SN 2012bo); [130.0, 155.0] (SN 2008bq); [200.0, 350.0] (SN 2008bz);
[110.0, 140.0] (LSQ12fxd); [101.0, 110.0] (SN 2008bf); [62.0; 82.0]
(SN 2007A); [35.0, 65.0] (ASASSN-15db); [370.0, 410.0] (LSQ14gov);
[40.0, 80.0] (SN 2007ca); [50.0, 80.0] (SN 2001cn); [230.0, 300.0]
(SN 2008go); [60.0, 80.0] (SN 1999ek).}

\bigskip

\noindent
One can visualize the output of
two-dimensional and one-dimensional posterior probability distributions in a
corner plot corresponding to 1$\sigma$, 2$\sigma$, 3$\sigma$. This corner
plot shows one and two dimensional projections of the posterior
probability distributions of our parameters.


\noindent
We present  the results of the MCMC for all the SNe Ia
in the Hubble flow that have been included in this paper.
We give the results for each phase and the combined phase
(see Figures 1 to 12). 

\section{Results and discussion}

\subsection{The test of the three rung method}

\noindent 
The present work can serve to test
the three rung method for the determination of
H$_{0}$. In this respect, 
the results obtained by the SH0ES-Pantheon+ collaboration and by
 our method using only one step are the same.
That could in principle not have been the case, as
the 3 rung method goes through the calibration of a fiducial
absolute magnitude of SNe Ia which is not required by our method.

\bigskip

 \noindent
     We include in Table 5
      the distance moduli using the ''SNe Ia twins for life''
      method as well as those provided by Pantheon+ with SH0ES and
      by the Carnegie Supernova Project. Those last two methods are
      calibrated with the Cepheid distances.
      The SNe Ia moduli in the Hubble flow  from the Pantheon+ and SH0ES
      are quite close to those from the CCHP.

\bigskip
      
\noindent
 The Pantheon+ data come with their
errors in distance moduli in their release and the CCHP distance moduli
and errors are generated as in Uddin et al.
(2024) code given the entries in the light curve parameterization.
We see similar distances moduli in Table 5 by our method when compared
with those from Pantheon+ and with Cepheids by the CCHP.
There are some
differences (in 3 SNe Ia) mostly
due to the underestimate of reddening (see Figure 13), as we will
discuss in the next subsection.

\bigskip

\noindent
To obtain the error in the ''SNe Ia  twins''
distance modulus we add the various uncertainty terms. 
 We take a
 systematic error due to anchors to be of 0.04 mag.

\bigskip

\begin{figure}[H]
\includegraphics[width=0.45\textwidth]{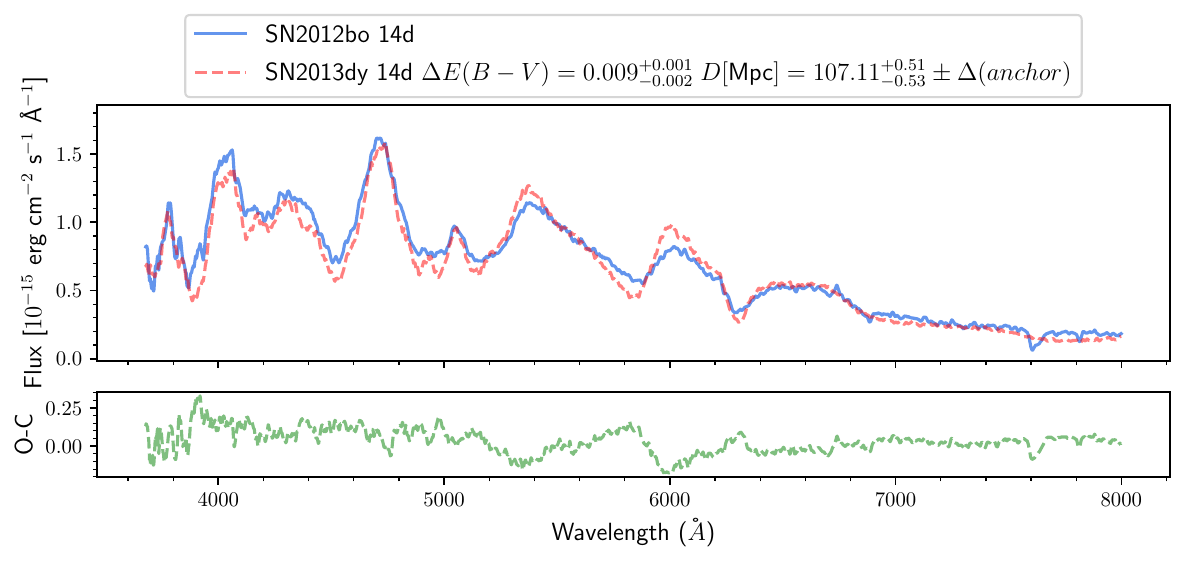}
\includegraphics[width=0.45\textwidth]{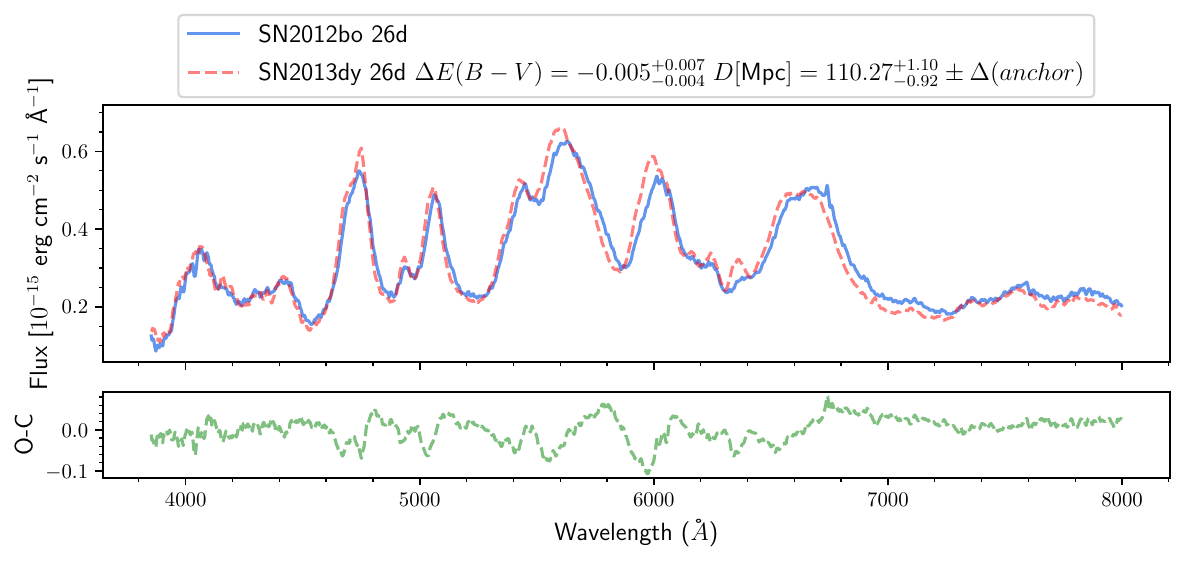}
\includegraphics[width=0.3\textwidth]{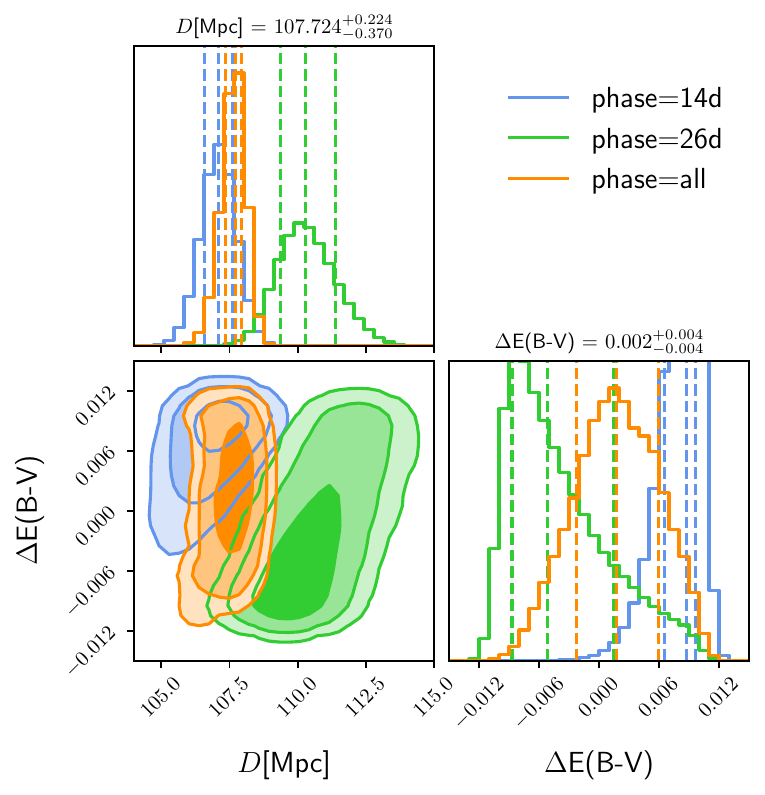}
\includegraphics[width=0.3\textwidth]{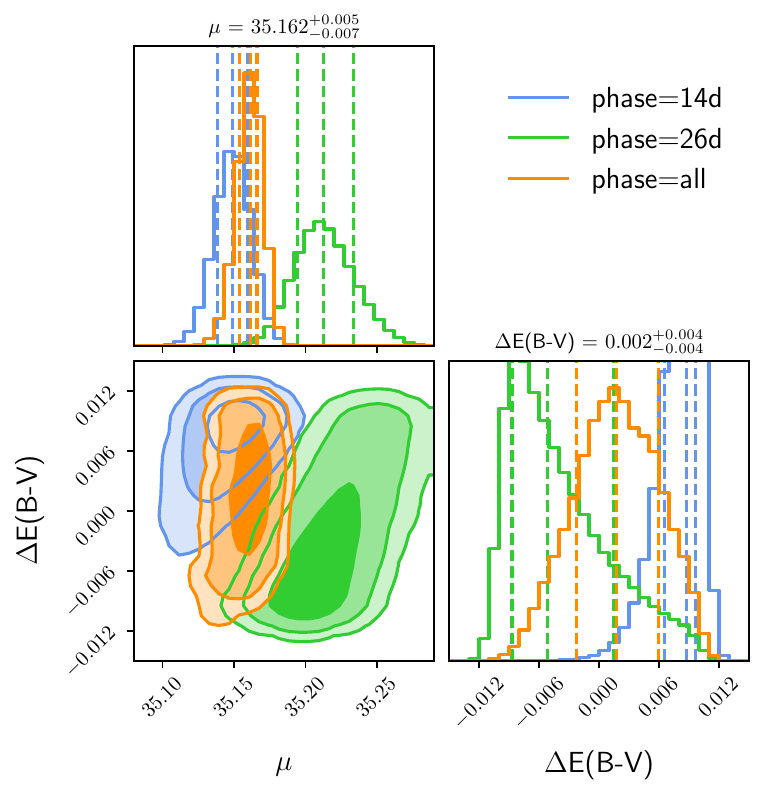}
\caption{ Top: Comparison of the spectrum at 14 days past maximum
  of SN 2012bo with that of SN 2013dy at the same phase.
  Middle below top: Comparison of the spectrum of SN 2012bo
  at 26 days past maximum with that of SN 2013dy at the same phase.
  Middle above bottom: Corner plot with the posterior probability at 1$\sigma$, 2$\sigma$,
  3$\sigma$ of distance and relative intrinsic reddening of 
 SN 2012bo in relation to SN 2013dy. Bottom: Corner plot with the posterior probability at 1$\sigma$, 2$\sigma$,
  3$\sigma$ of distance moduli and relative intrinsic reddening of
 SN 2012bo in relation to SN 2013dy. }
\end{figure}

\begin{figure}[H]
\includegraphics[width=0.45\textwidth]{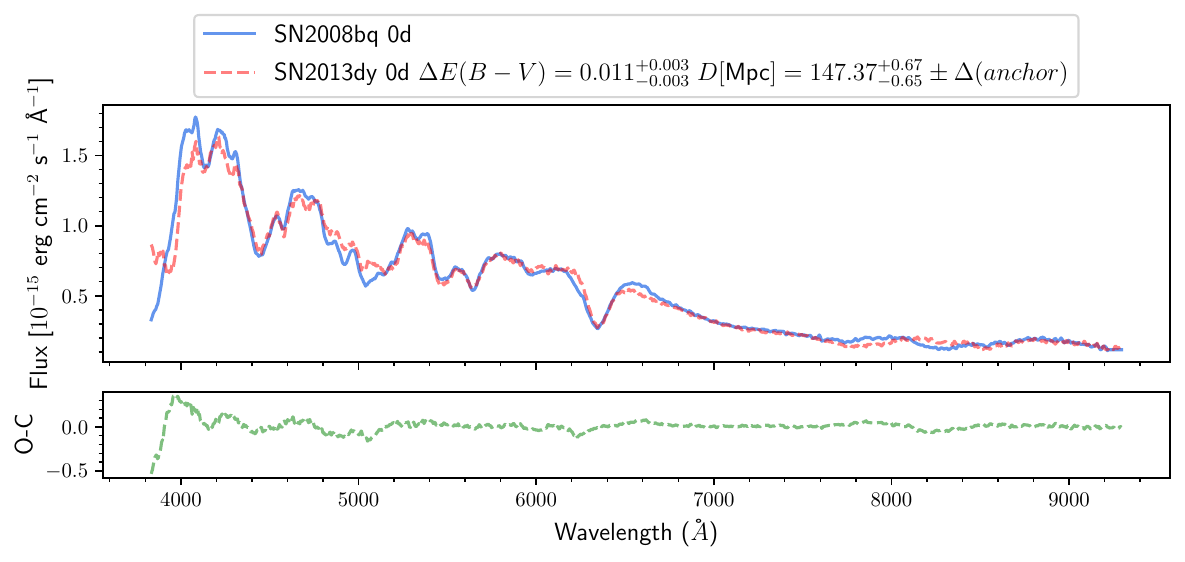}
\includegraphics[width=0.45\textwidth]{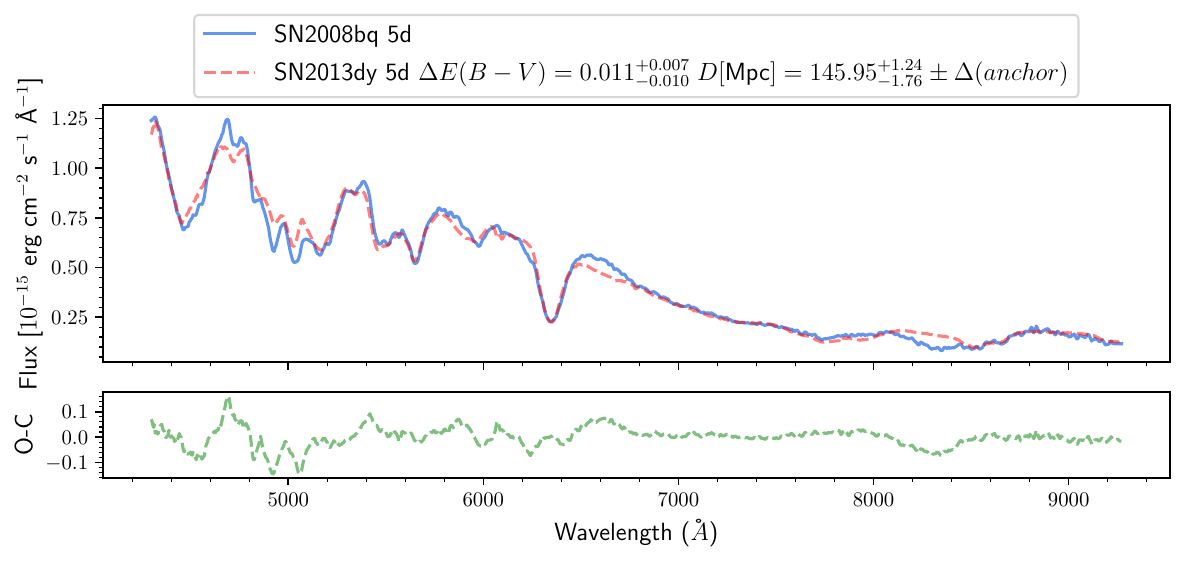}
\includegraphics[width=0.3\textwidth]{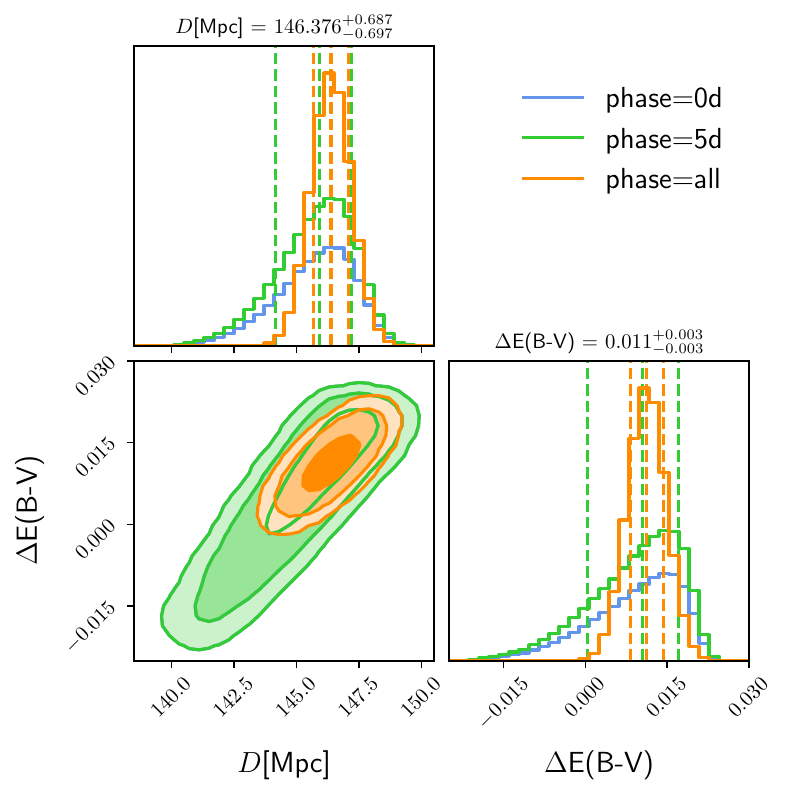}
\includegraphics[width=0.3\textwidth]{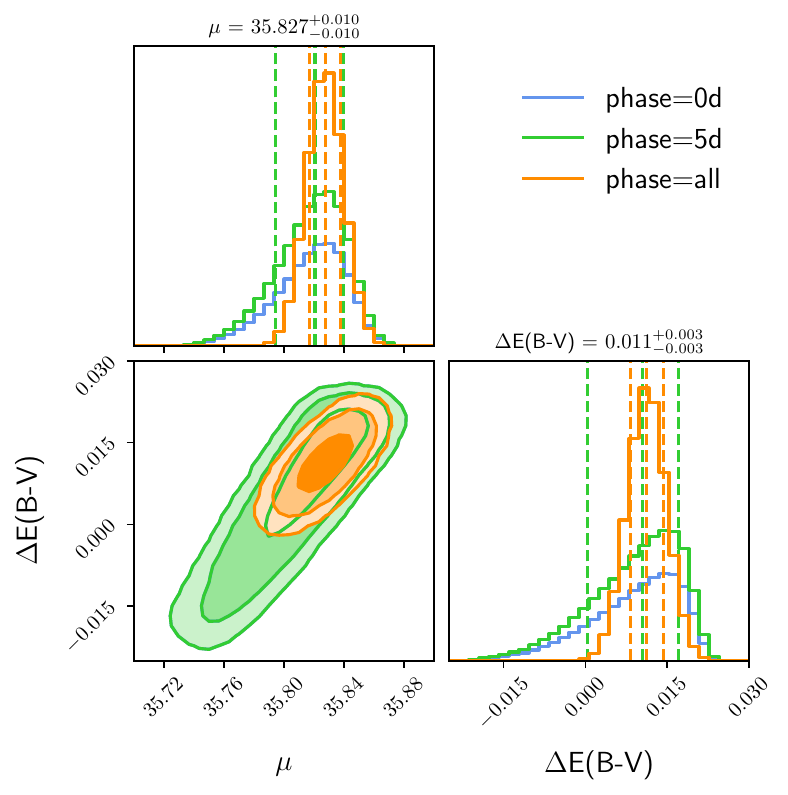}
\caption{ Top: Comparison of the spectrum at 0 days past maximum
  of SN 2008bq with that of SN 2013dy at the same phase.
  Middle below top: Comparison of the spectrum at 5 days past maximum
  with that of SN 2013dy at the same phase.
   Middle above bottom: Corner plot with the posterior probability at 1$\sigma$, 2$\sigma$,
  3$\sigma$ of distance and relative intrinsic reddening of 
  SN 2008bq in relation to SN 2013dy. Bottom:
  Corner plot with the posterior probability at 1$\sigma$, 2$\sigma$,
  3$\sigma$ of distance moduli and relative intrinsic reddening of 
  SN 2008bq in relation to SN 2013dy.}
\end{figure}

\begin{figure}[H]
  
  \centering
\includegraphics[width=0.45\textwidth]{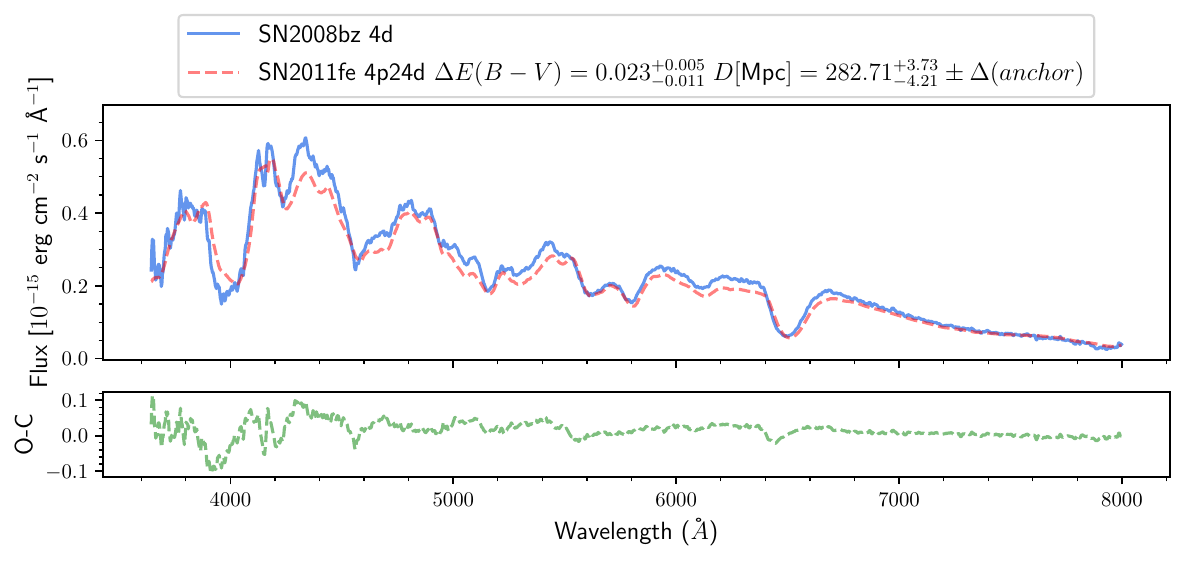}
\includegraphics[width=0.45\textwidth]{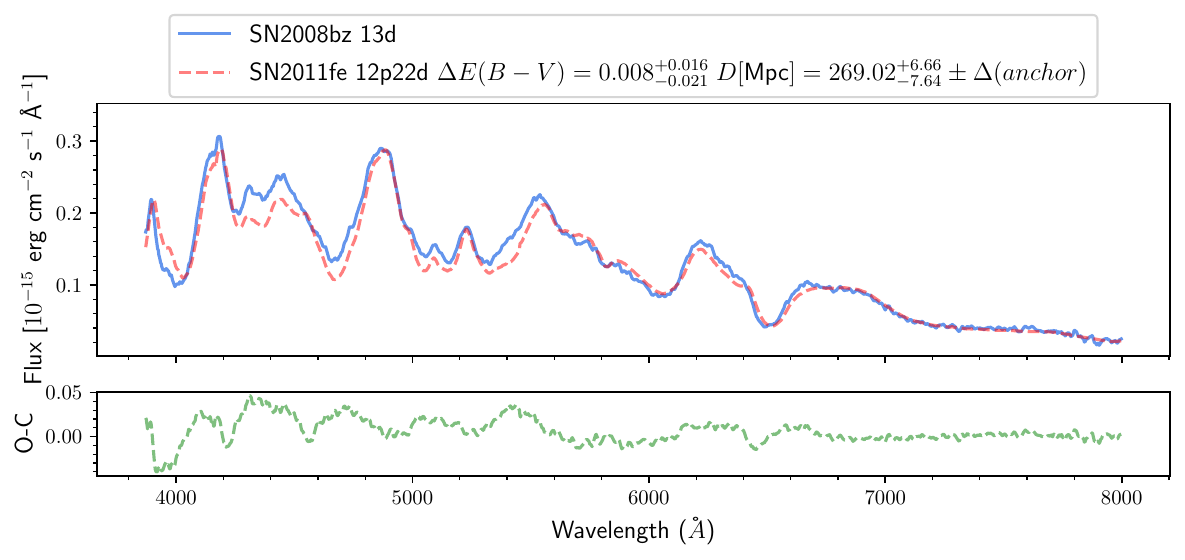}
\includegraphics[width=0.3\textwidth]{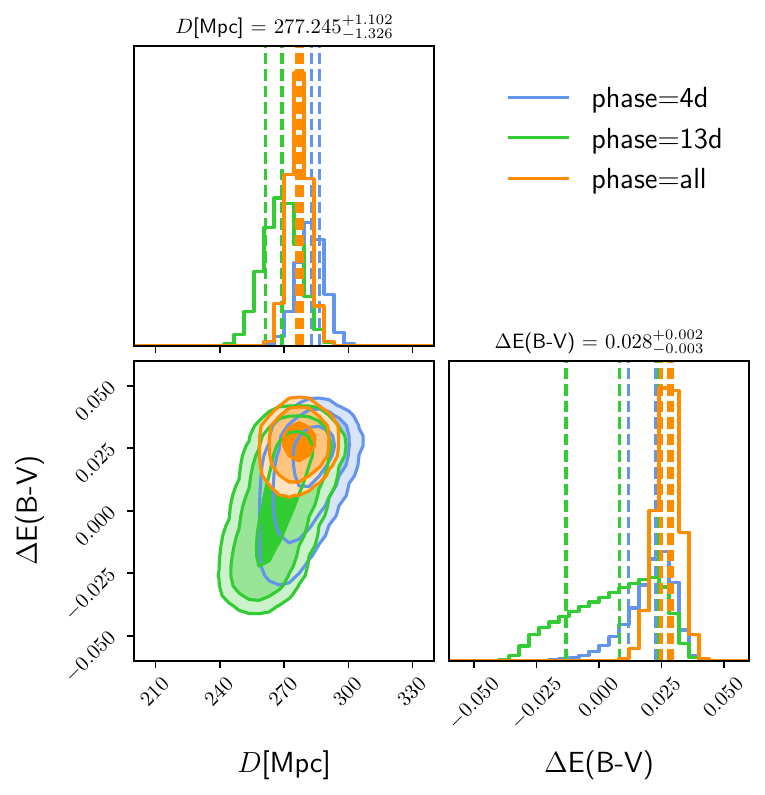}
\includegraphics[width=0.3\textwidth]{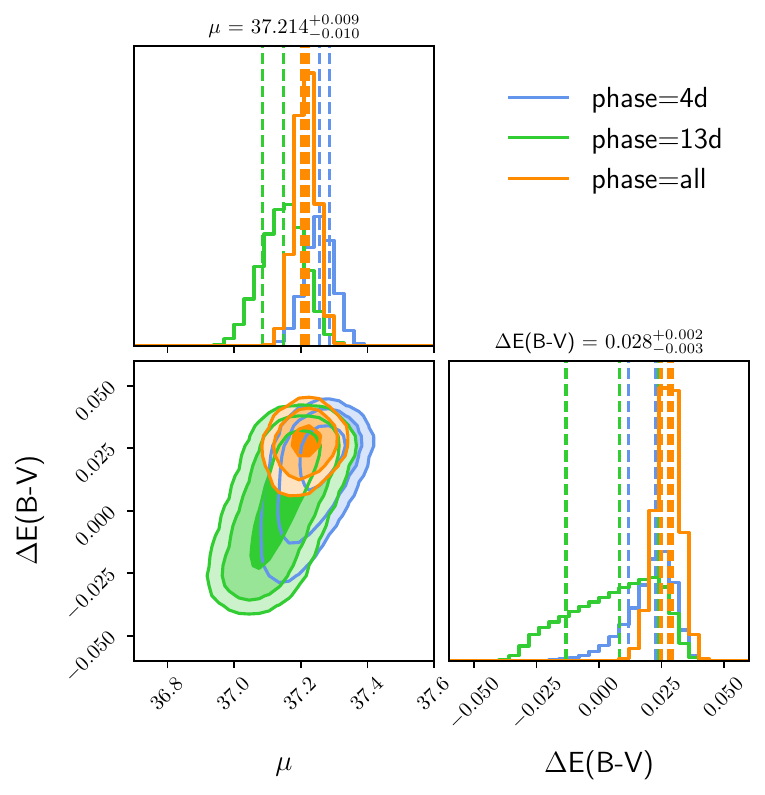}

\caption{ Top: Comparison of the early time spectrum of SN 2008bz
  with SN 2011fe  at
  4 days past  maximum light. Middle below top: Comparison of the spectrum of
  SN 2008bz at 13 days past maximum with the corresponding spectrum
  of SN 2011fe.
  Middle above bottom : Corner plot with the posterior probability at 1$\sigma$, 2$\sigma$,
  3$\sigma$ of distance and relative intrinsic reddening of 
 SN 2008bz in relation to SN 2011fe. Bottom: Corner plot with the posterior probability at 1$\sigma$, 2$\sigma$,
  3$\sigma$ of distance moduli and relative intrinsic reddening of
 SN 2008bz in relation to SN 2011fe.}
\end{figure}

\begin{figure}[H]
\includegraphics[width=0.45\textwidth]{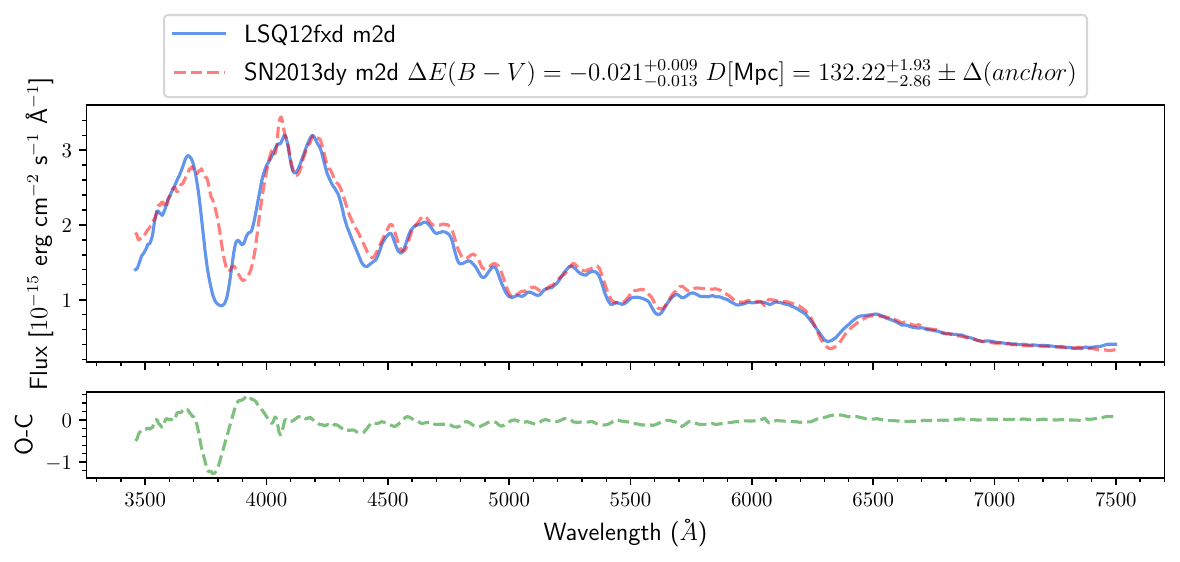}  
\includegraphics[width=0.45\textwidth]{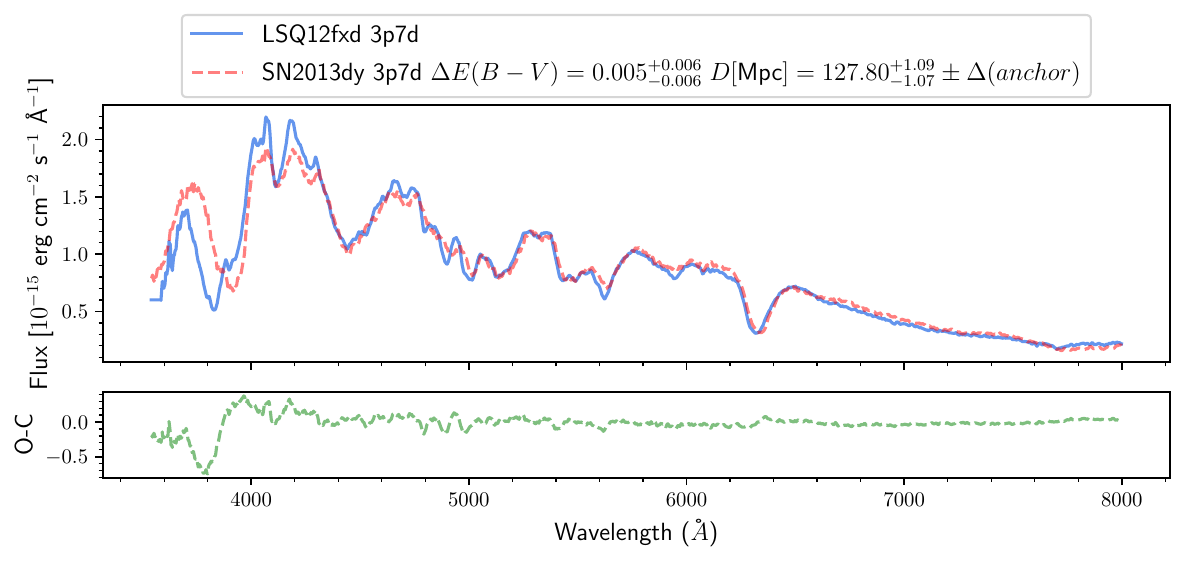}
\includegraphics[width=0.3\textwidth]{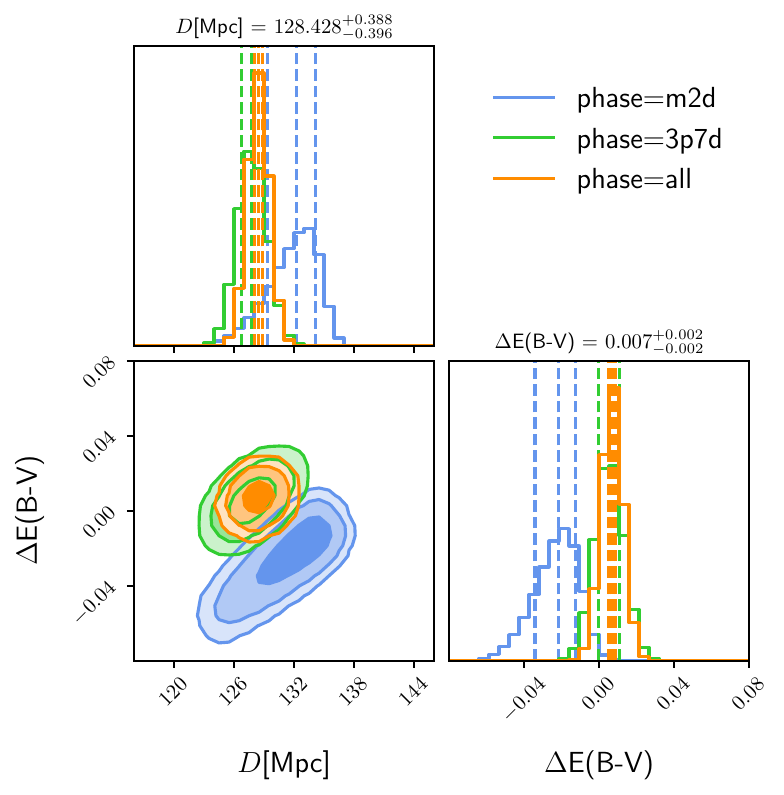}
\includegraphics[width=0.3\textwidth]{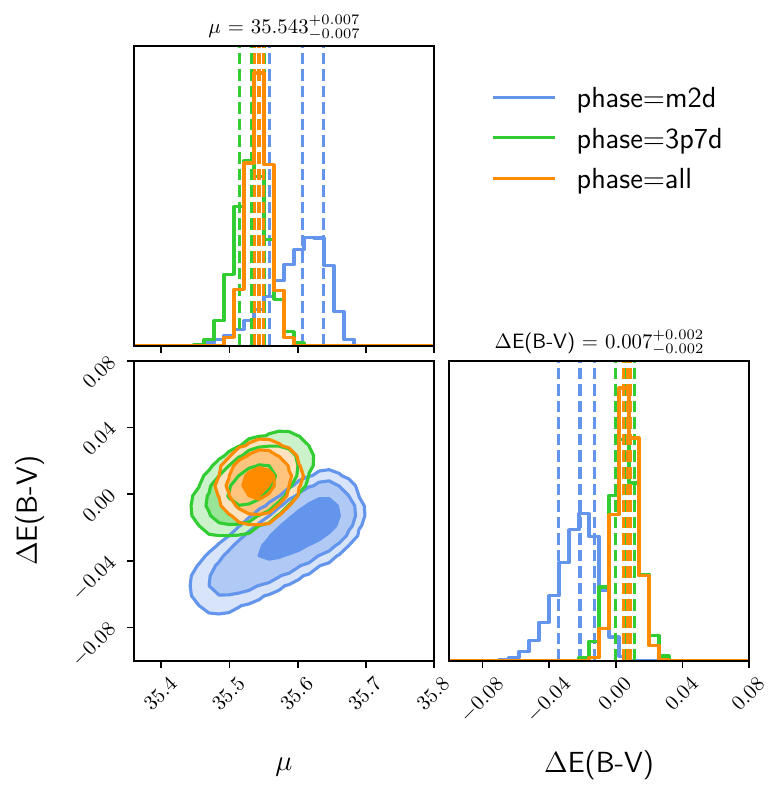}
\caption{ Top: Comparison of the
  spectrum of LSQ12fxd at 2 days before maximum
  with that of SN 2013dy in the same phase. Middle below top: Comparison
  of the spectrum of LSQ12fxd at 3.7 days past maximum with that of SN 2013dy.
  Middle above bottom:  Corner plot with the posterior probability at 1$\sigma$, 2$\sigma$,
  3$\sigma$ of distance and relative intrinsic reddening of 
 LSQ12fxd  in relation to SN 2013dy. Bottom: Corner plot with the posterior probability at 1$\sigma$, 2$\sigma$,
  3$\sigma$ of distance  moduli and relative intrinsic reddening of
 LSQ12fxd in relation to SN 2013dy.}
  \end{figure}

\begin{figure}[H]
  \centering
\includegraphics[width=0.45\textwidth]{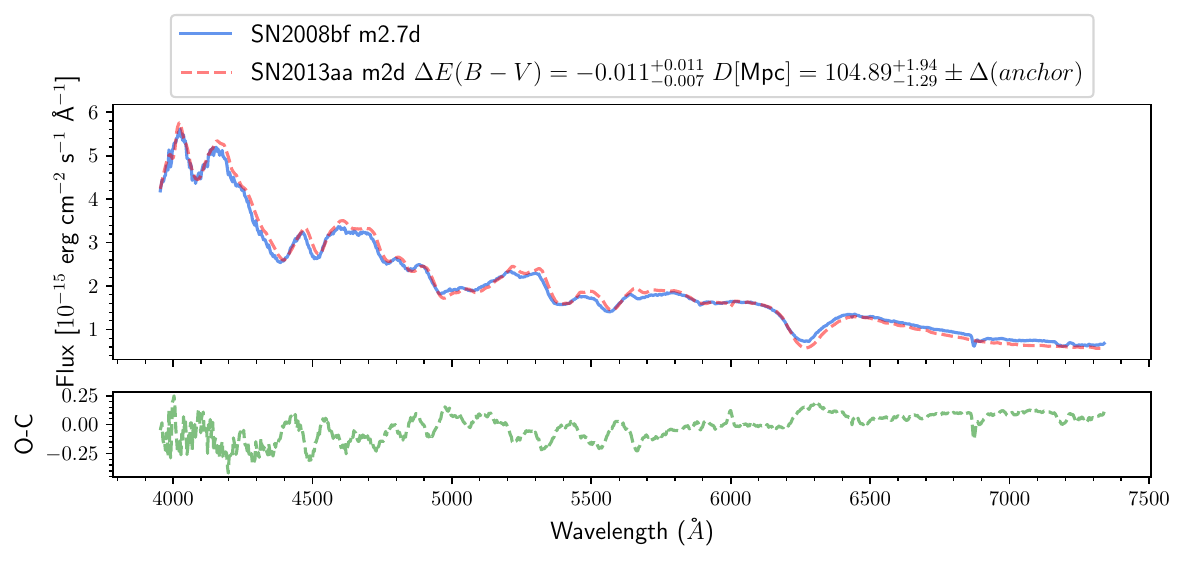}
\includegraphics[width=0.45\textwidth]{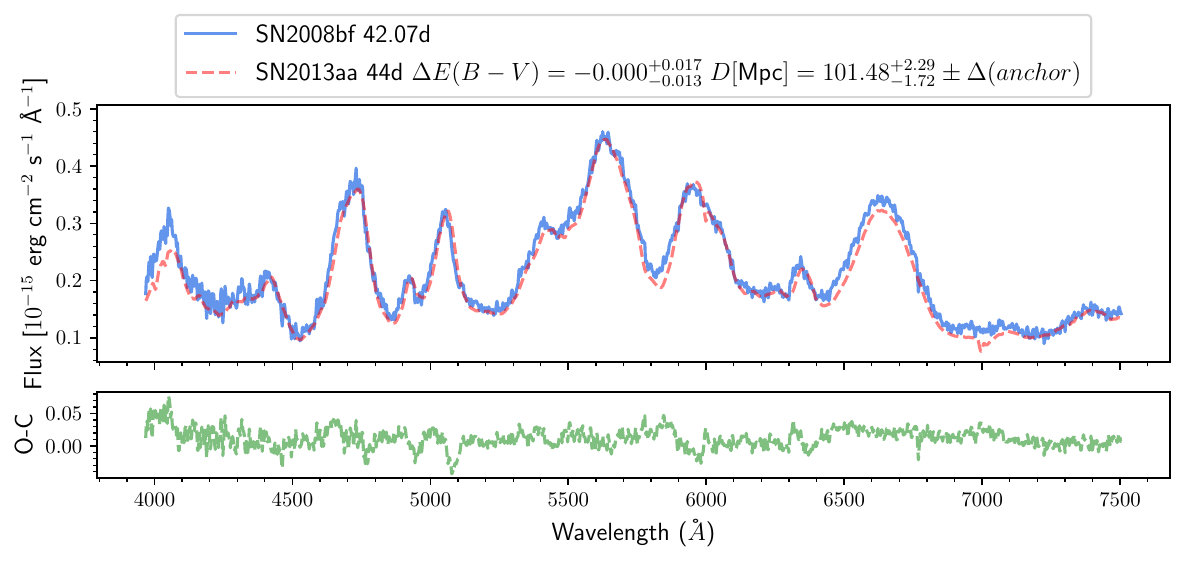}
\includegraphics[width=0.3\textwidth]{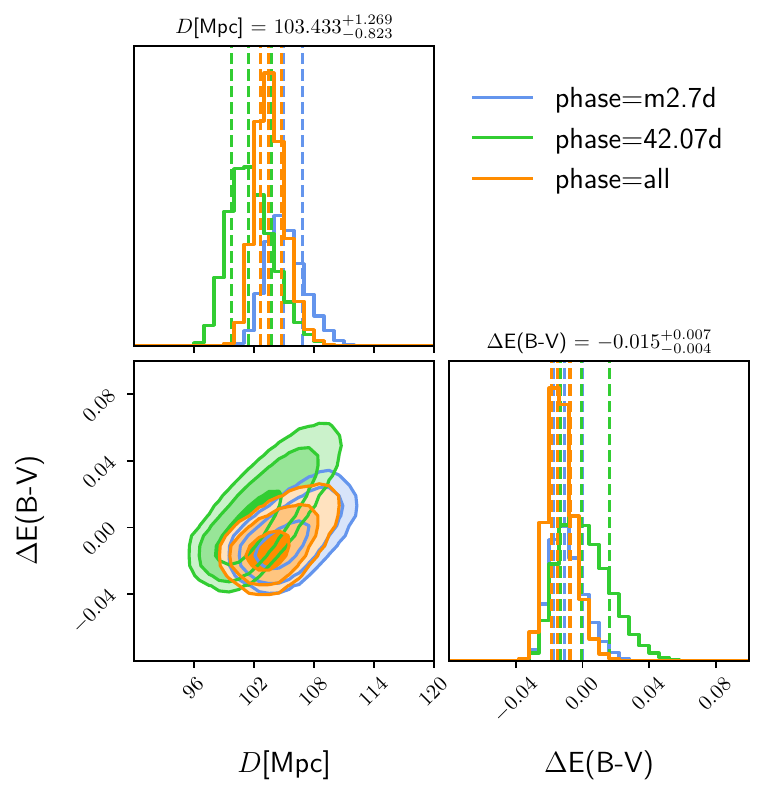}
\includegraphics[width=0.3\textwidth]{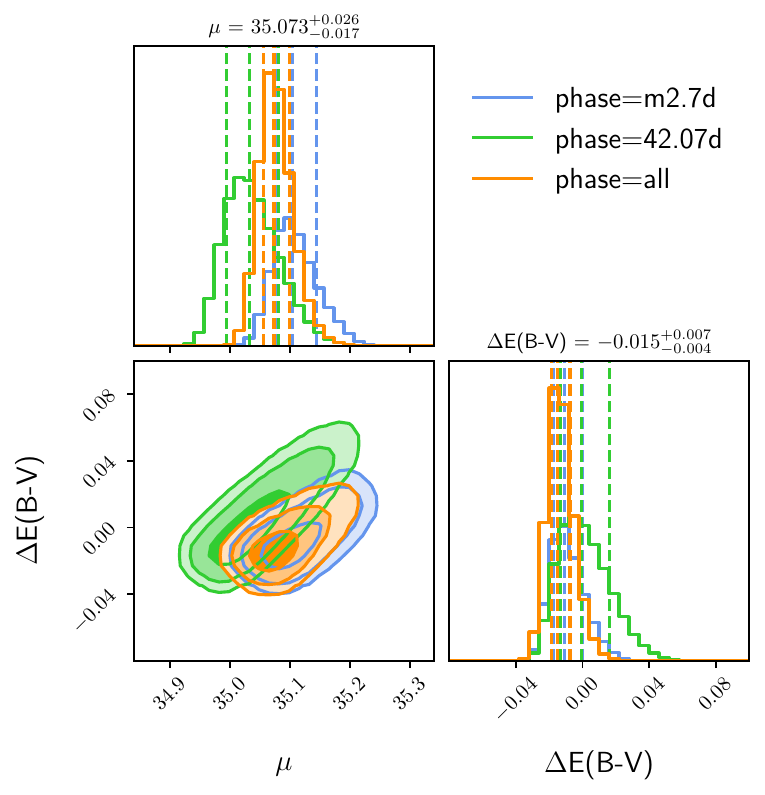}
\caption{Top: Comparison of the early time spectrum of SN 2008bf
  with SN 2013aa  at
  -2 days before maximum light. Middle below top: Comparison of the spectrum of
  SN 2008bf at 42 days past maximum with the corresponding spectrum
  of SN 2003aa. Middle above bottom: 
 Corner plot with the posterior probability at 1$\sigma$, 2$\sigma$,
  3$\sigma$ of distance and relative intrinsic reddening of 
  SN 2008bf in relation to SN 2013aa. Bottom: Corner plot with the posterior probability at 1$\sigma$, 2$\sigma$,
  3$\sigma$ of distance moduli
  and relative intrinsic reddening of 
  SN 2008bf in relation to SN 2013aa. 
 }
\end{figure}

\bigskip

\begin{figure}[H]
\includegraphics[width=0.45\textwidth]{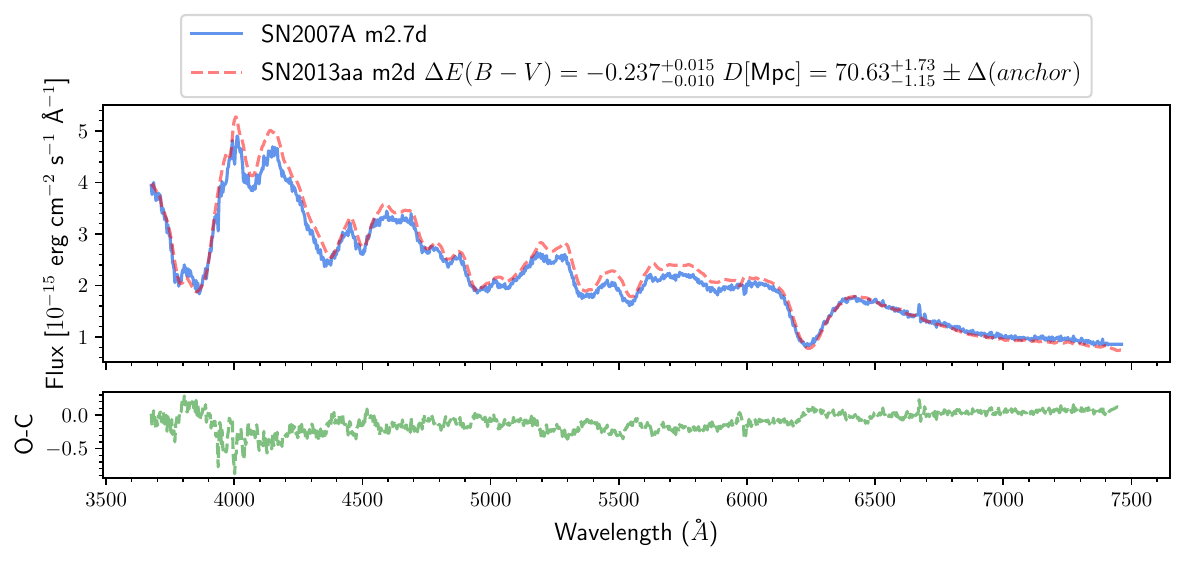}
\includegraphics[width=0.45\textwidth]{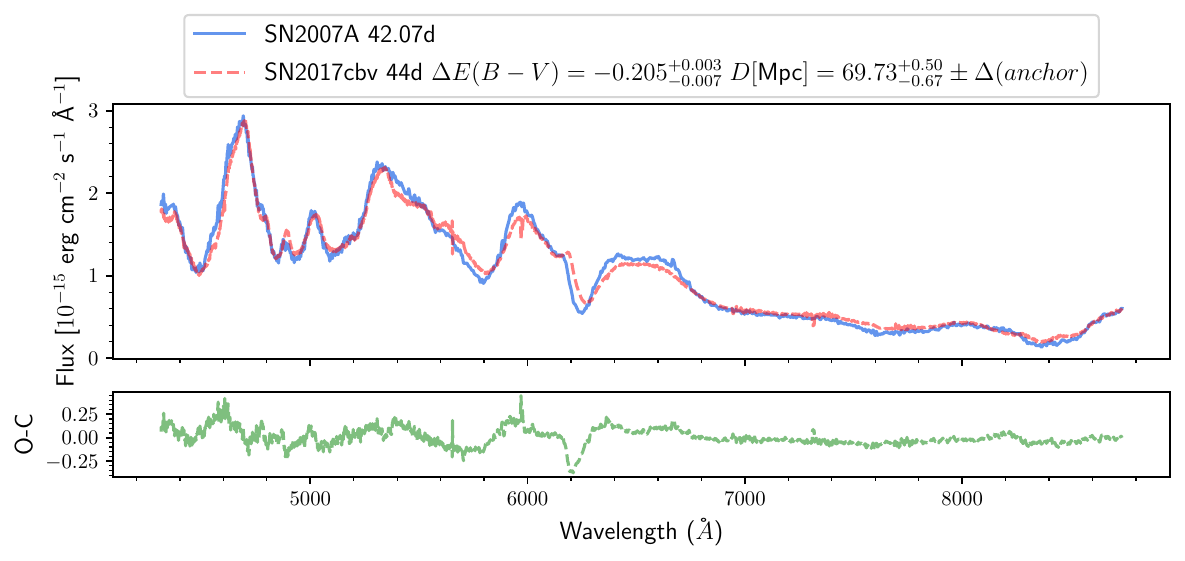}
\includegraphics[width=0.3\textwidth]{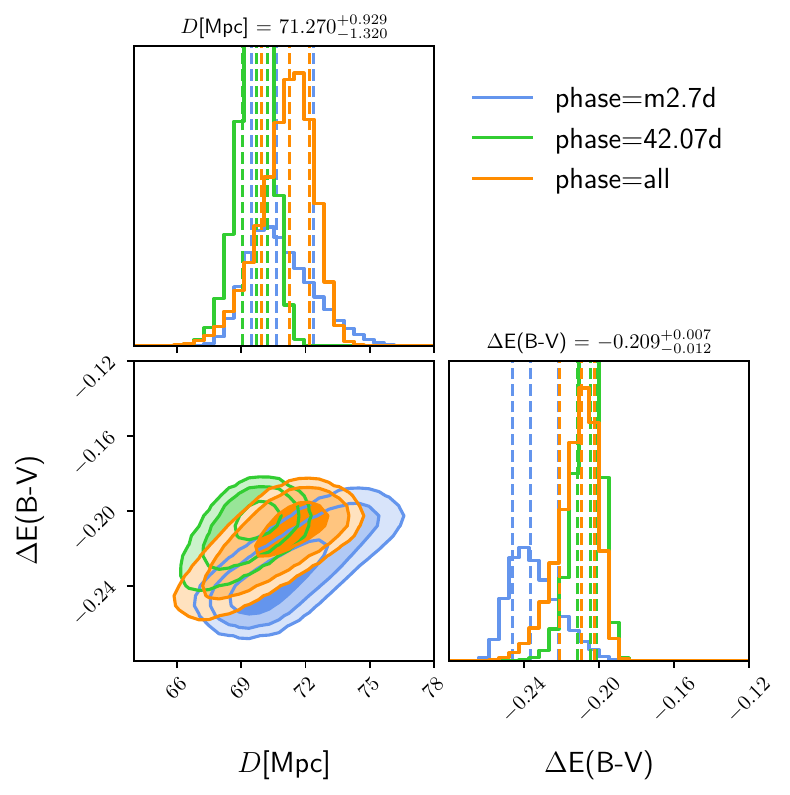}
\includegraphics[width=0.3\textwidth]{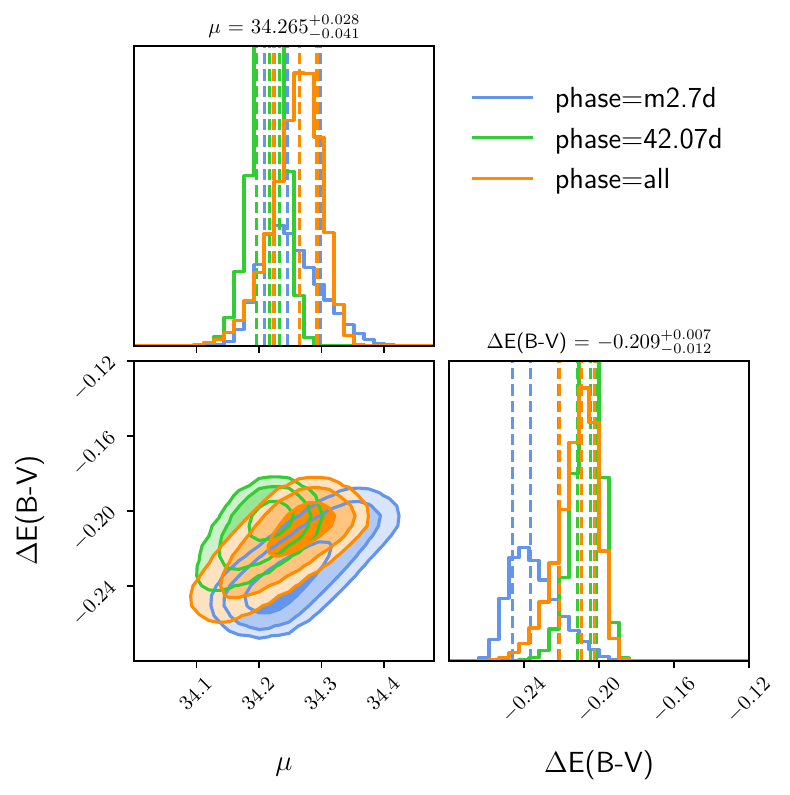}
\caption{Top: Comparison of the spectrum at 2 days before maximum
  of SN 2007A with that of SN 2013aa at the same phase.
  Middle below top: Comparison of the spectrum at 42 days past maximum of
  SN 2007A with that of SN 2013aa at the same phase.
  Middle above bottom: Corner plot with the posterior probability at 1$\sigma$,
  2$\sigma$,
  3$\sigma$ of distance and relative intrinsic reddening of 
  SN 2007A in relation to SN 2013aa.
  Bottom: Corner plot with the posterior probability at 1$\sigma$,
  2$\sigma$,
  3$\sigma$ of distance moduli and relative intrinsic reddening of
  SN 2007A in relation to SN 2013aa.}
\end{figure}

\begin{figure}[H]
\includegraphics[width=0.45\textwidth]{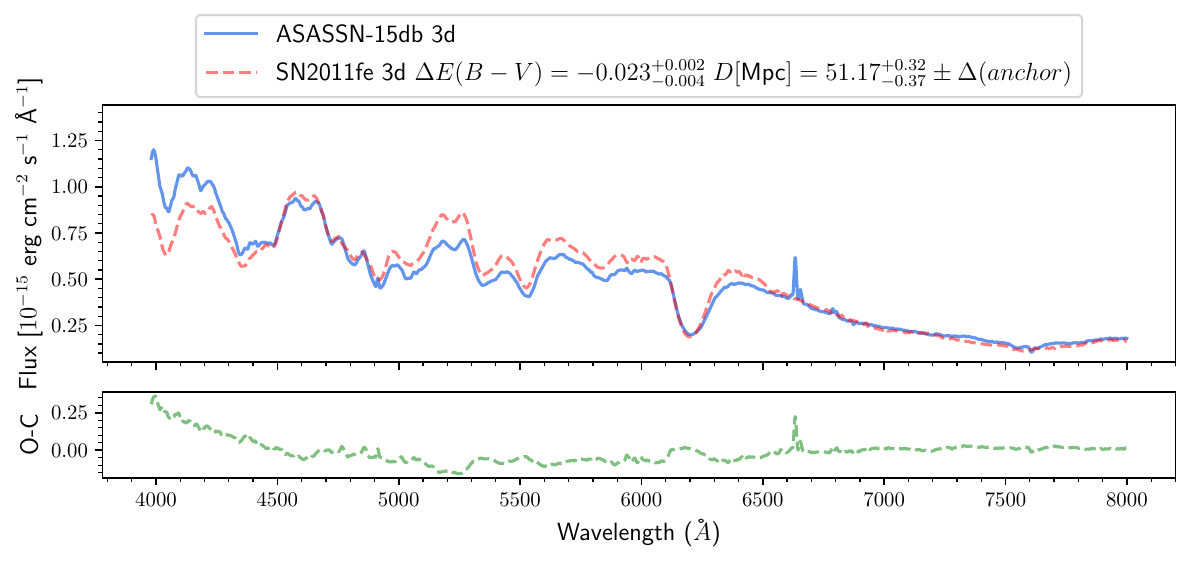}
\includegraphics[width=0.45\textwidth]{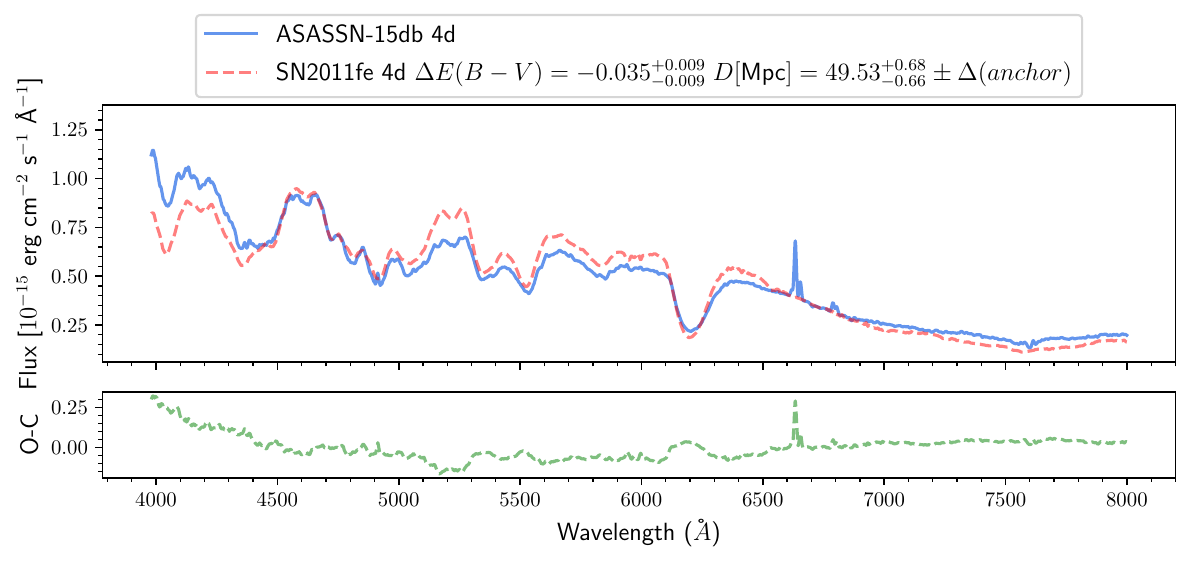}
\includegraphics[width=0.3\textwidth]{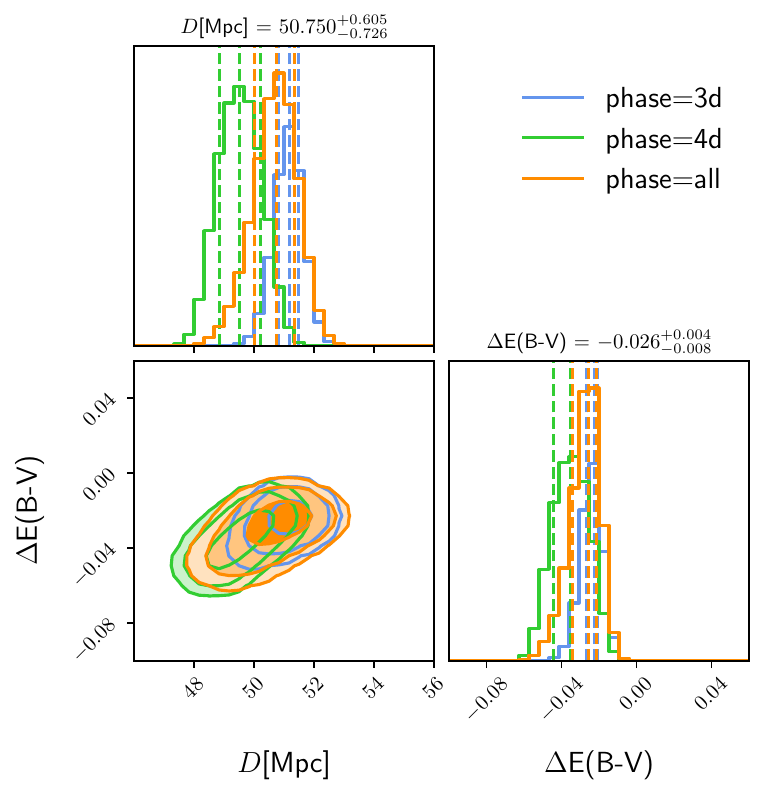}
\includegraphics[width=0.3\textwidth]{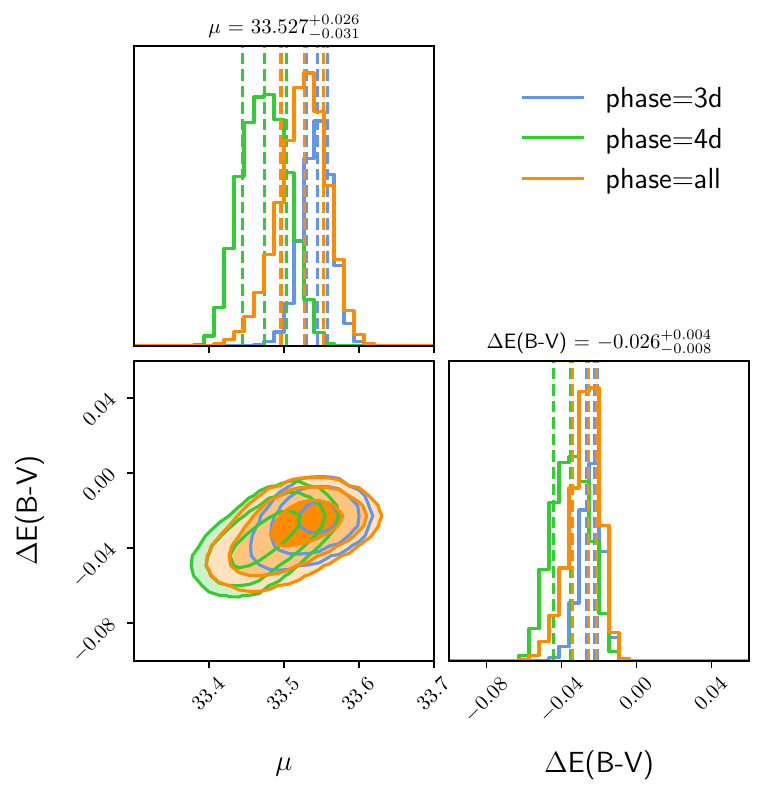}
\caption{Top: Comparison of the spectrum at 3 days past maximum
  of ASASSN-15db with that of SN 2011fe at the same phase.
  Middle below top: Comparison of the spectrum of ASASSN-15db at 4 days
   past maximum 
  with that of SN 2011fe at the same phase.
  Middle above bottom: Corner plot with the posterior probability at 1$\sigma$, 2$\sigma$,
  3$\sigma$ of distance and relative intrinsic reddening of
 ASASSN-15db in relation to SN 2011fe. Bottom: Corner plot with the posterior probability at 1$\sigma$, 2$\sigma$,
  3$\sigma$ of distance moduli and relative intrinsic reddening of 
 ASASSN-15db in relation to SN 2011fe.} 
\end{figure}

\begin{figure}[H]
  \includegraphics[width=0.45\textwidth]{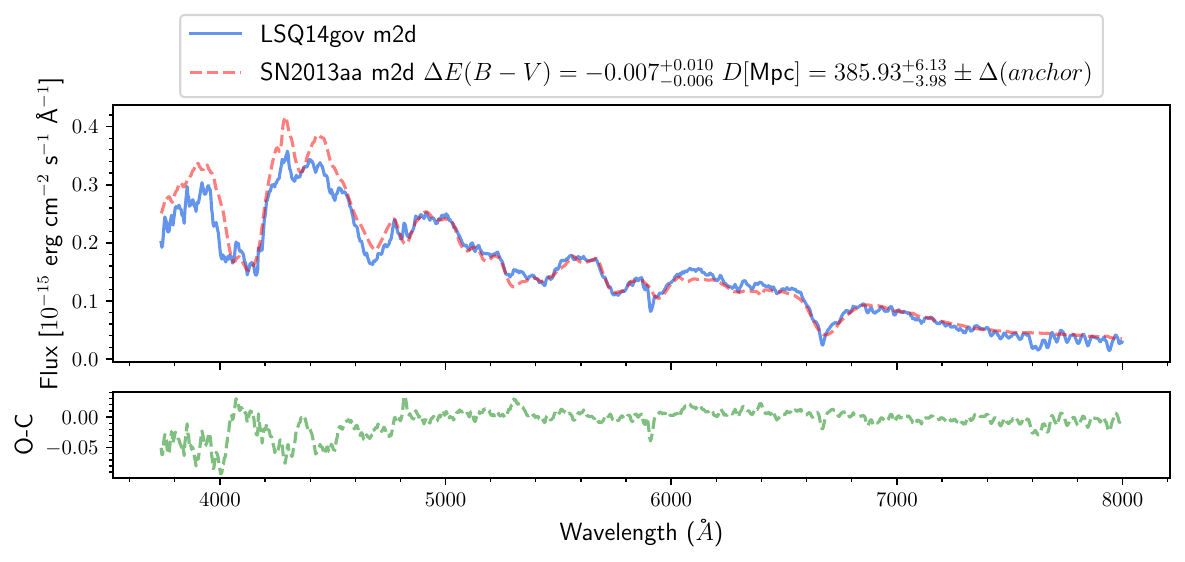}
  \includegraphics[width=0.45\textwidth]{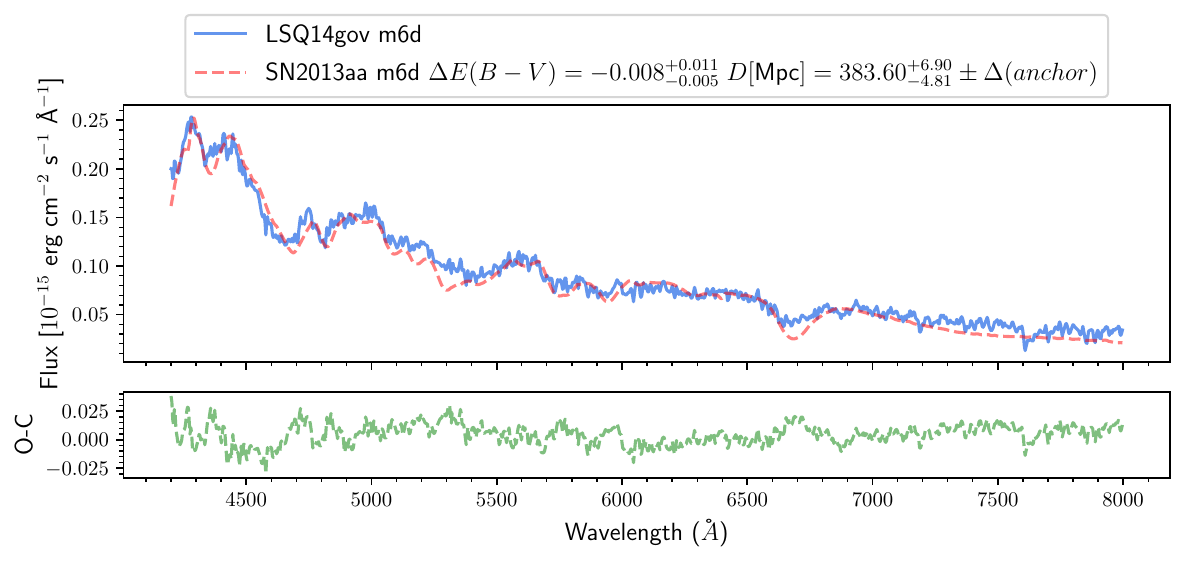}
  \includegraphics[width=0.3\textwidth]{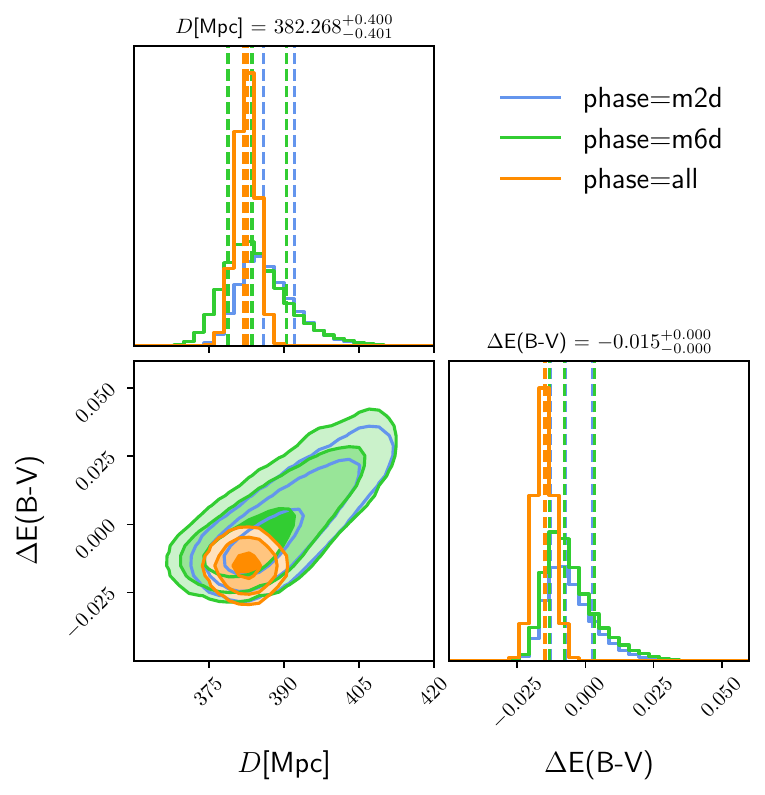}
\includegraphics[width=0.3\textwidth]{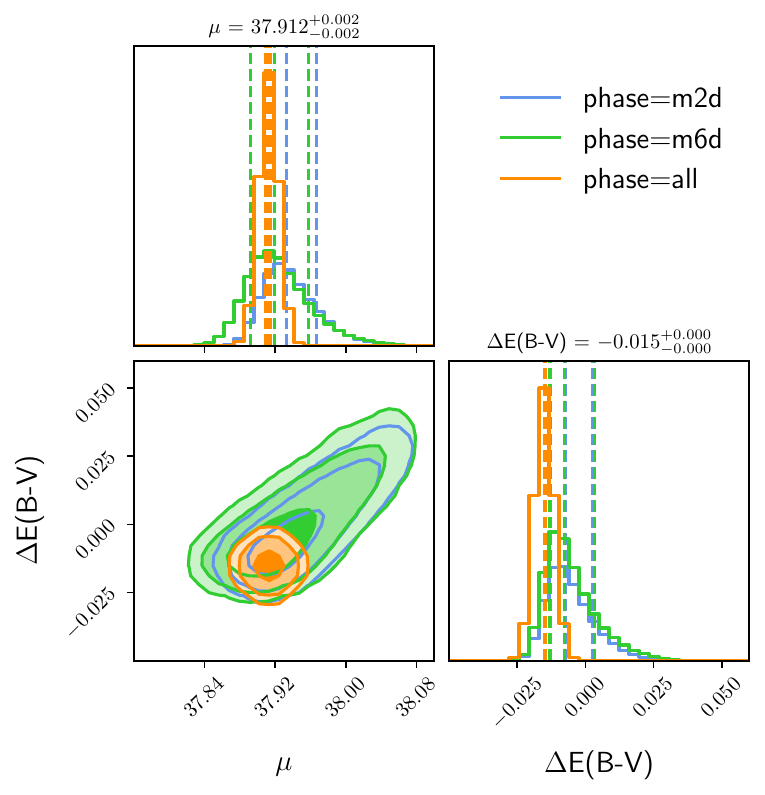}
\caption{Top: Comparison of the spectrum at 1 day past maximum
  of LSQ14gov with that of SN 2013aa at the same phase.
  Middle below top: Comparison of the spectrum at 6 days before m\'aximum
  with that of of SN 2013aa at a similar phase. 
  Middle above bottom: Corner plot with the posterior probability
  at 1$\sigma$, 2$\sigma$,
  3$\sigma$ of distance and relative intrinsic reddening of 
  LSQ14gov in relation to SN 2013aa. Bottom:
  Corner plot with the posterior probability at 1$\sigma$, 2$\sigma$,
  3$\sigma$ of distance moduli and relative intrinsic reddening of 
LSQ14gov in relation to SN 2013aa.   }
\end{figure}

\begin{figure}[H]
  \includegraphics[width=0.45\textwidth]{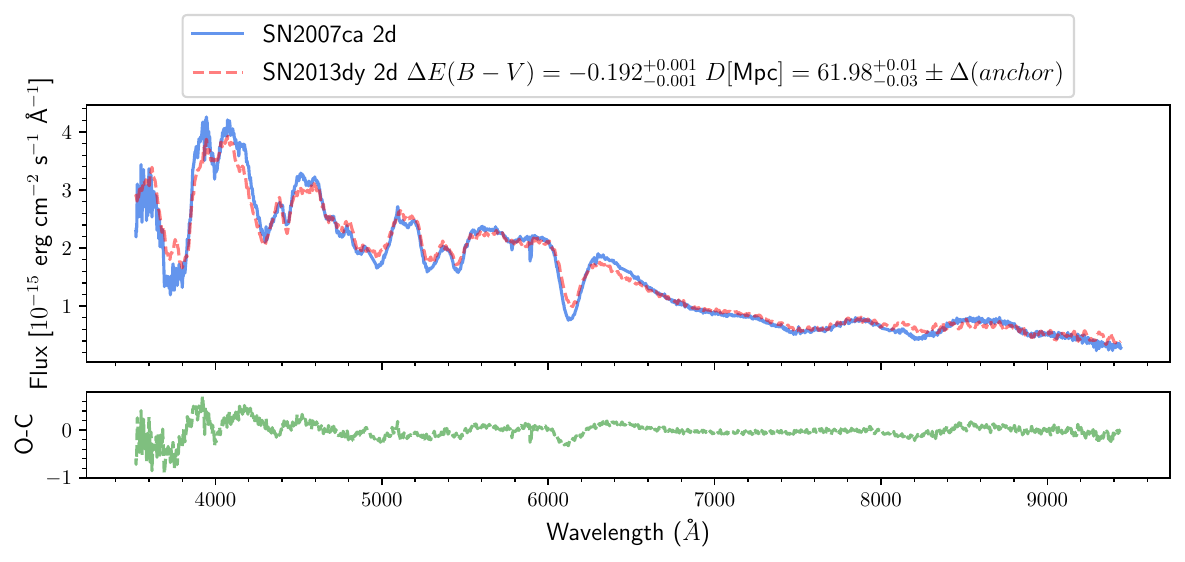}
  \includegraphics[width=0.45\textwidth]{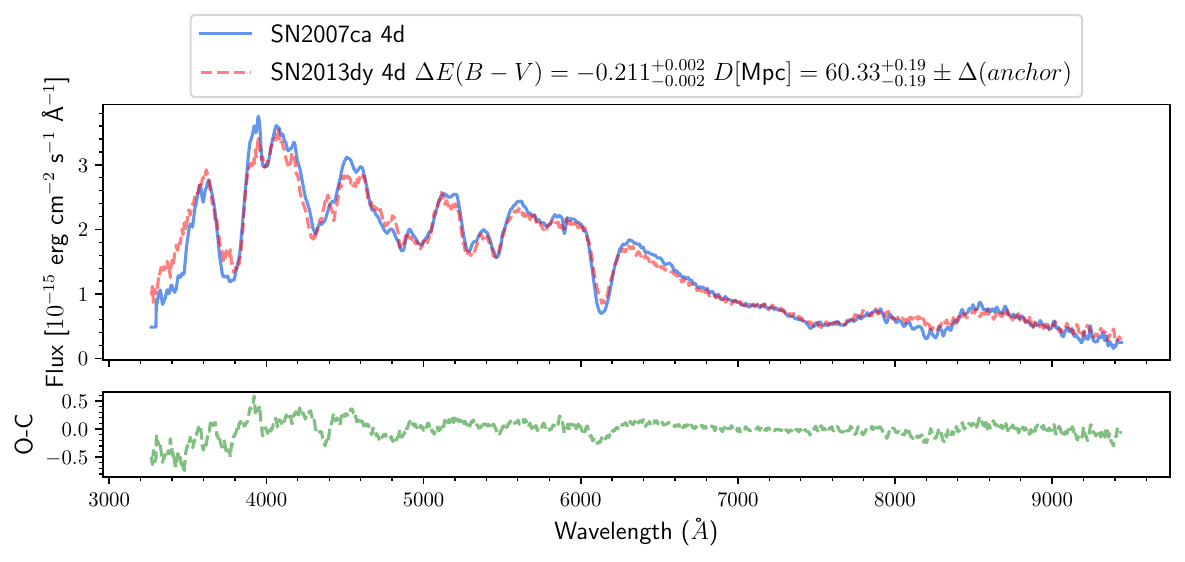}
  \includegraphics[width=0.3\textwidth]{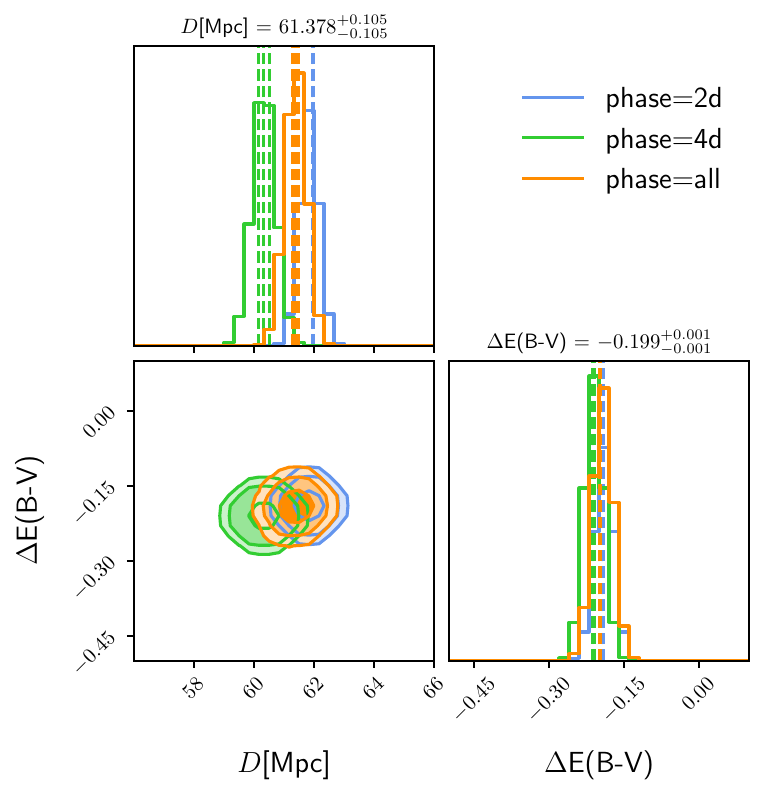}
\includegraphics[width=0.3\textwidth]{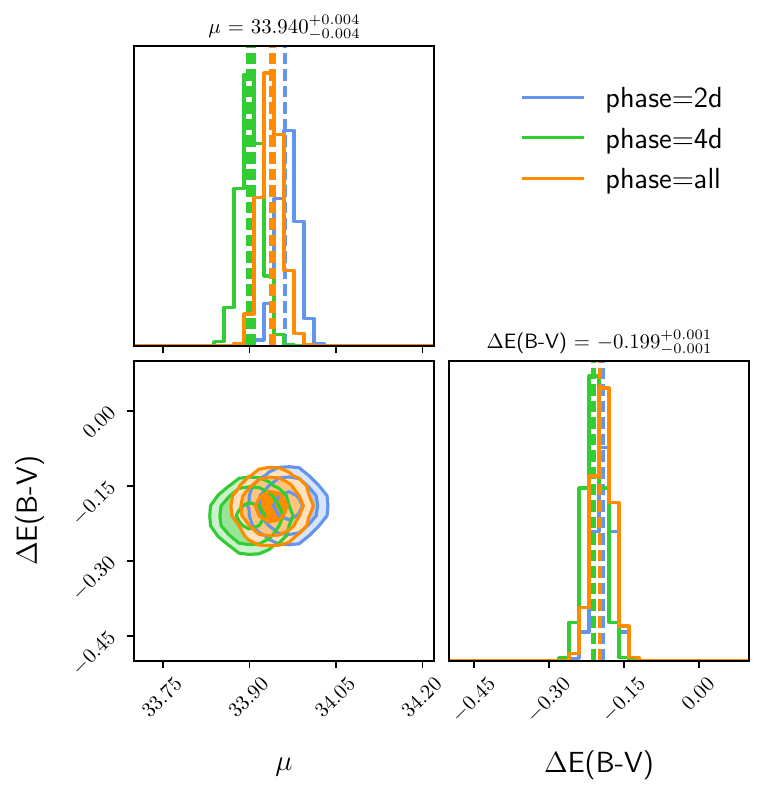}
\caption{Top: Comparison of the spectrum at 2 days past maximum
  of SN2007ca with that of SN 2013dy at the same phase.
  Middle below top: Comparison of the spectrum at 4 days past m\'aximum
  with that of of SN 2013aa at asimilar phase. 
  Middle above bottom: Corner plot with the posterior probability
  at 1$\sigma$, 2$\sigma$,
  3$\sigma$ of distance and relative intrinsic reddening of
  SN2007ca in relation to SN 2013dy. Bottom:
  Corner plot with the posterior probability at 1$\sigma$, 2$\sigma$,
  3$\sigma$ of distance moduli and relative intrinsic reddening of 
 SN2007ca in relation to SN 2013dy. }
\end{figure}

\begin{figure}[H]
  \includegraphics[width=0.45\textwidth]{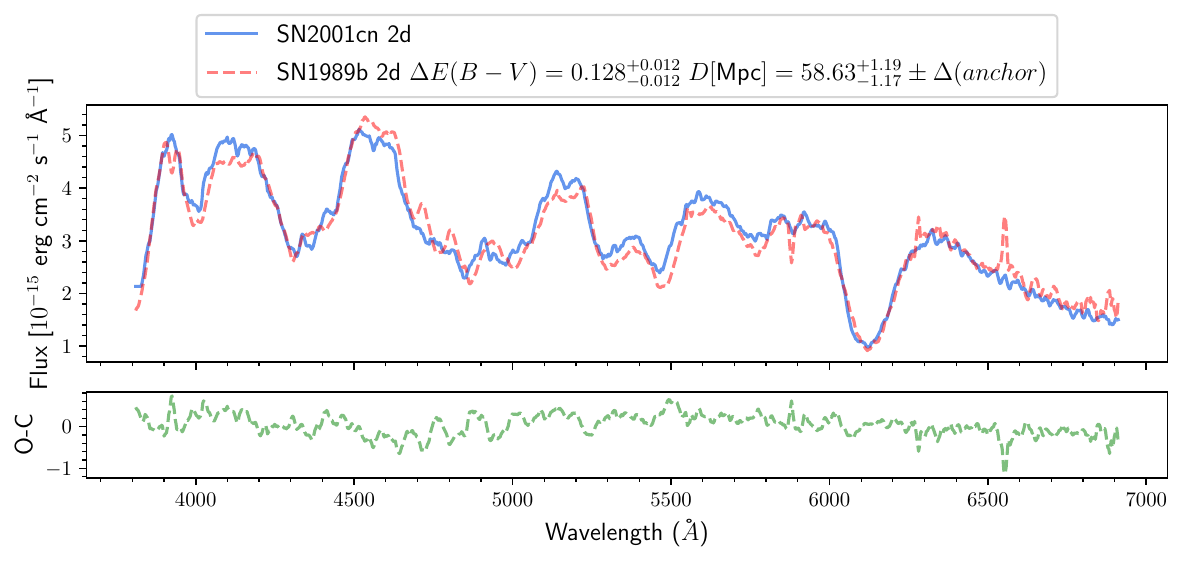}
  \includegraphics[width=0.45\textwidth]{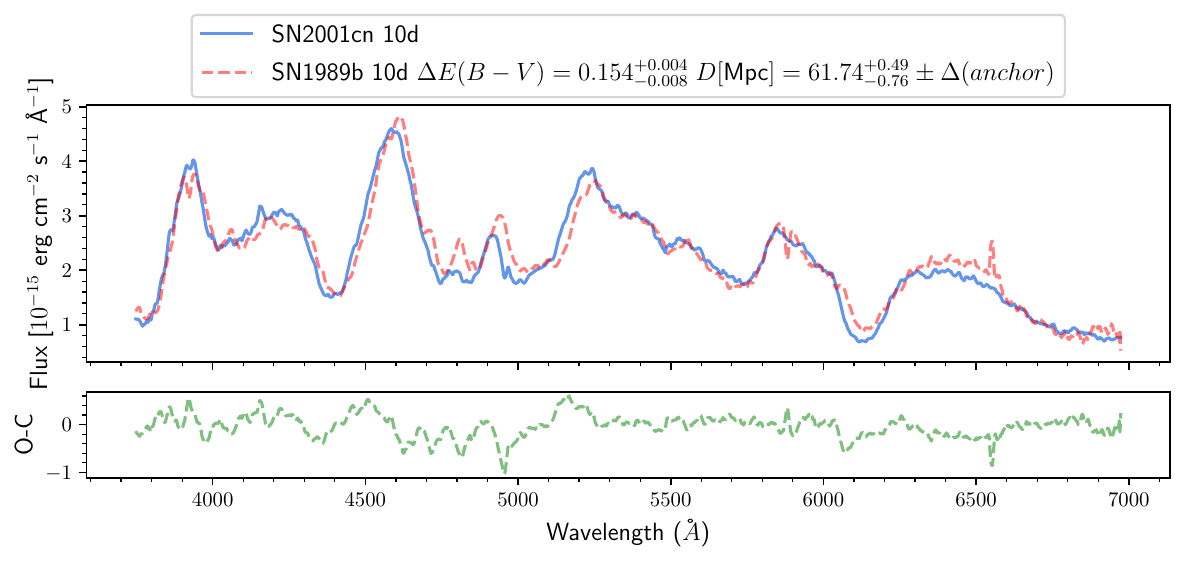}
  \includegraphics[width=0.3\textwidth]{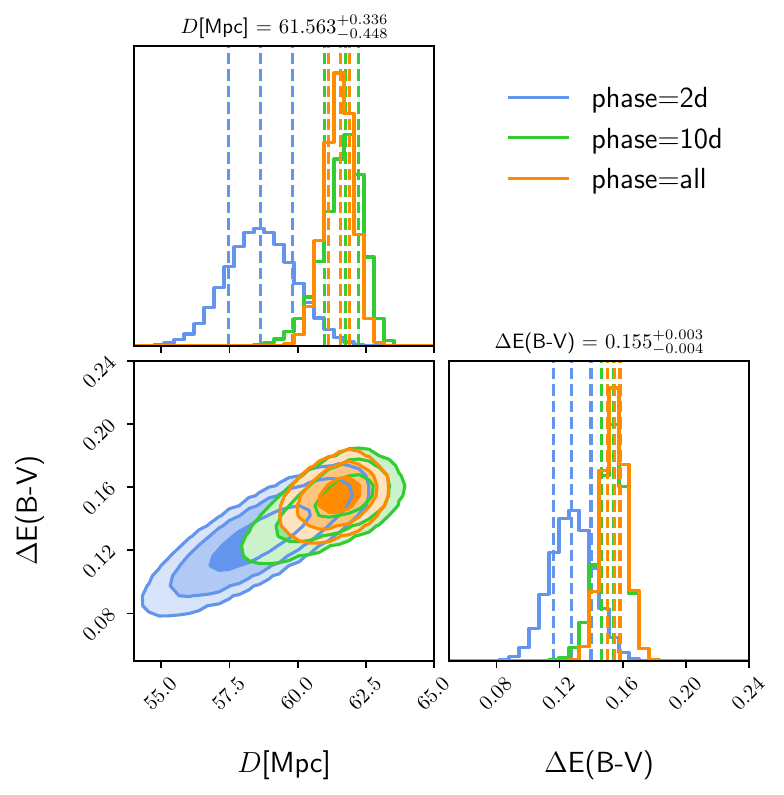}
\includegraphics[width=0.3\textwidth]{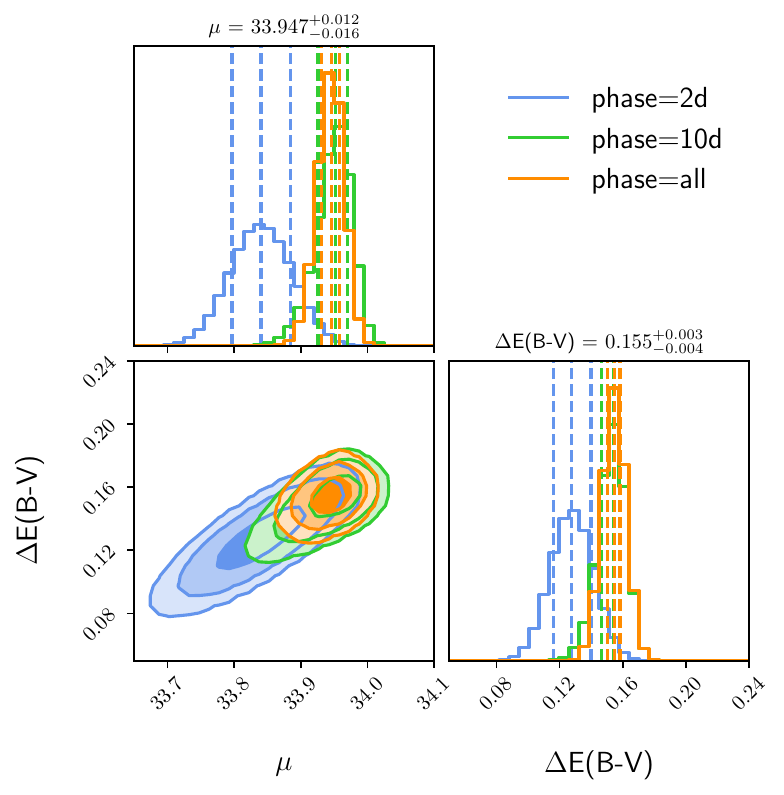}
\caption{ Top: Comparison of the spectrum at  2 days past maximum
  of SN 2001cn with that of SN 1989B at the same phase.
  Middle below top: Comparison of the spectrum at 10 days past maximum
  of SN 2001cn  with that of SN 1989B at about the same time.
  Middle above bottom: Corner plot with the posterior probability
  at 1$\sigma$, 2$\sigma$,
  3$\sigma$ of distance and relative intrinsic reddening of 
  SN 2001cn in relation to SN 1989B.
  Bottom:
  Corner plot with the posterior probability at 1$\sigma$, 2$\sigma$,
  3$\sigma$ of distance moduli  and relative intrinsic reddening of 
  SN 2001cn in relation to SN 1989B.  }
\end{figure}

\begin{figure}[H]
  \includegraphics[width=0.45\textwidth]{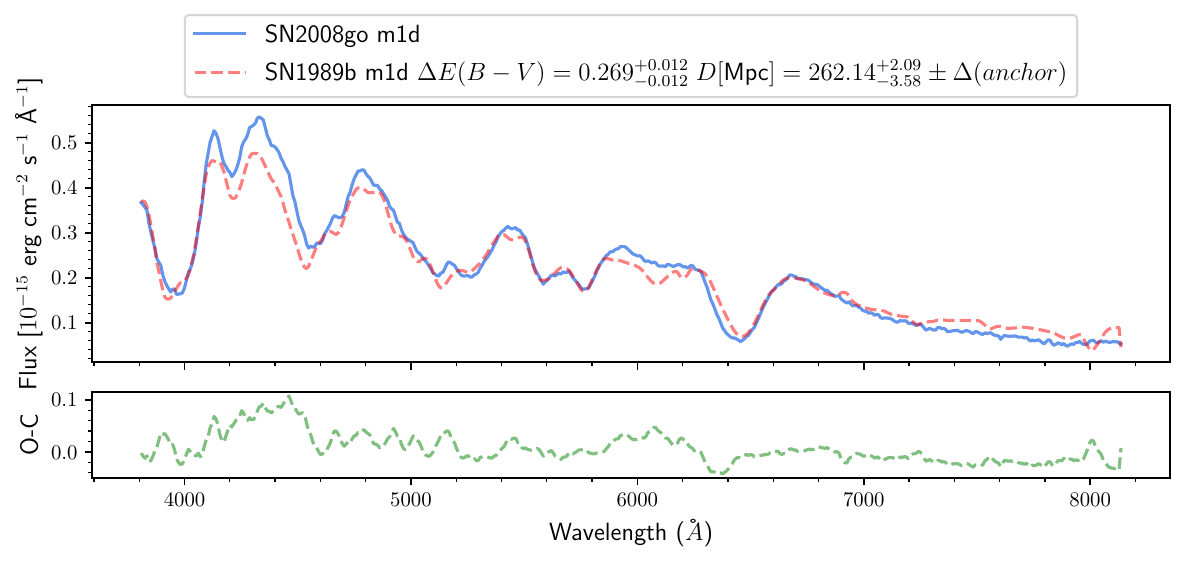}
  \includegraphics[width=0.45\textwidth]{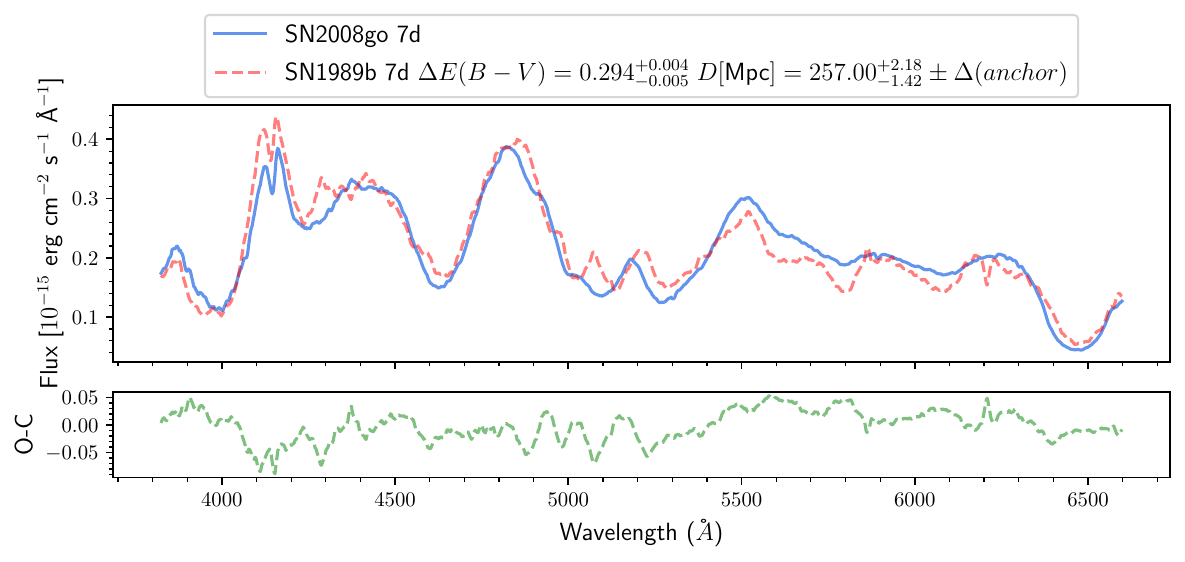}
\includegraphics[width=0.3\textwidth]{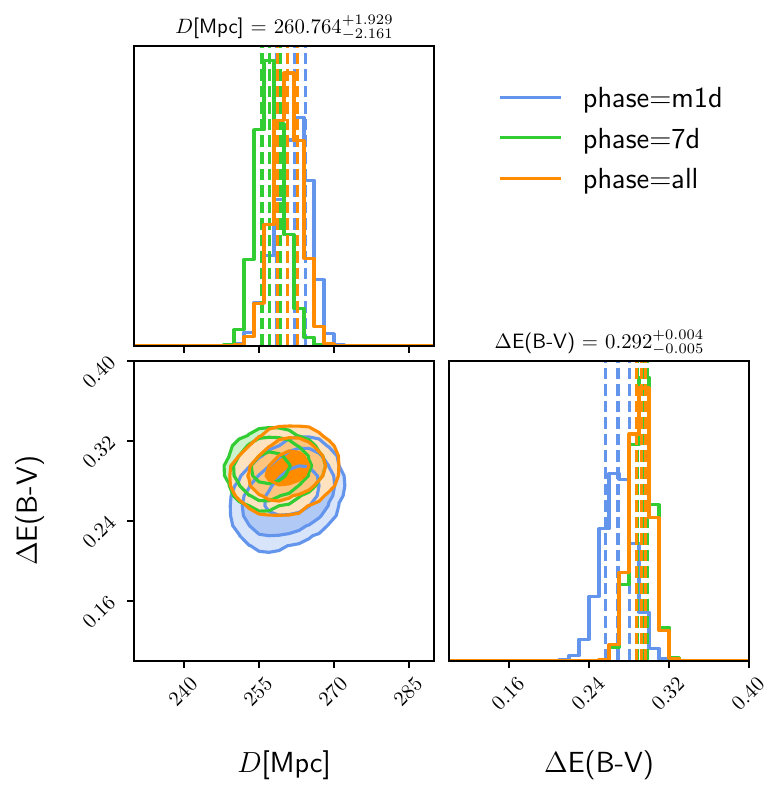}
  \includegraphics[width=0.3\textwidth]{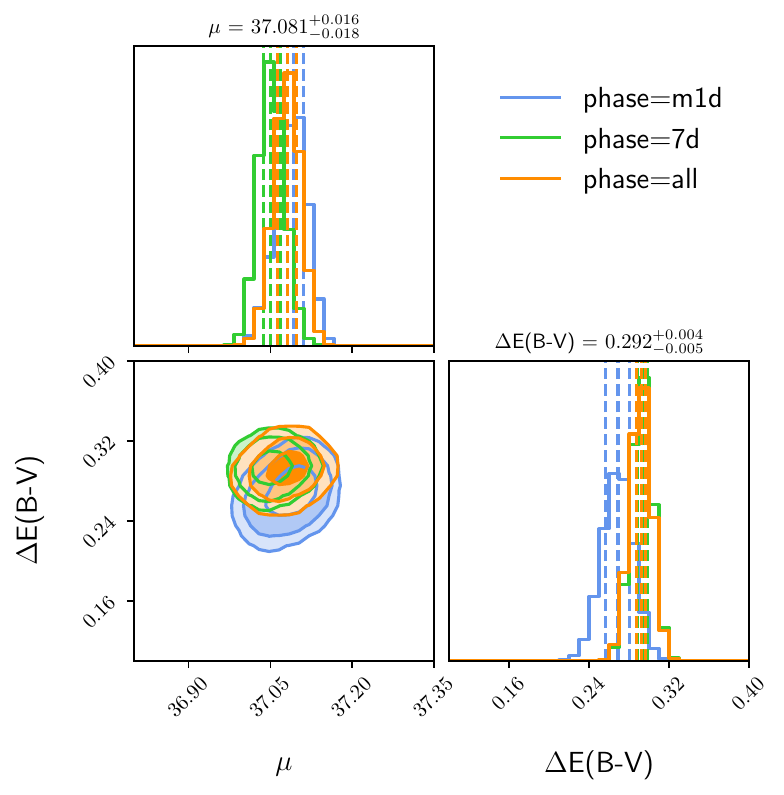}
\caption{ Top: Comparison of the spectrum at  1 day before maximum
  of SN 2008go with that of SN 1989B at the same phase.
  Middle below top: Comparison of the spectrum at 7 days past maximum
  of SN 2008go  with that of SN 1989B at about the same phase.
  Middle above bottom: Corner plot with the posterior probability
  at 1$\sigma$, 2$\sigma$,
  3$\sigma$ of distance and relative intrinsic reddening of
  SN 2008go relation to SN 1989B.
  Bottom:
  Corner plot with the posterior probability at 1$\sigma$, 2$\sigma$,
  3$\sigma$ of distance moduli  and relative intrinsic reddening of 
  SN 2008go in relation to SN 1989B.}
\end{figure}

\begin{figure}[H]
  \includegraphics[width=0.45\textwidth]{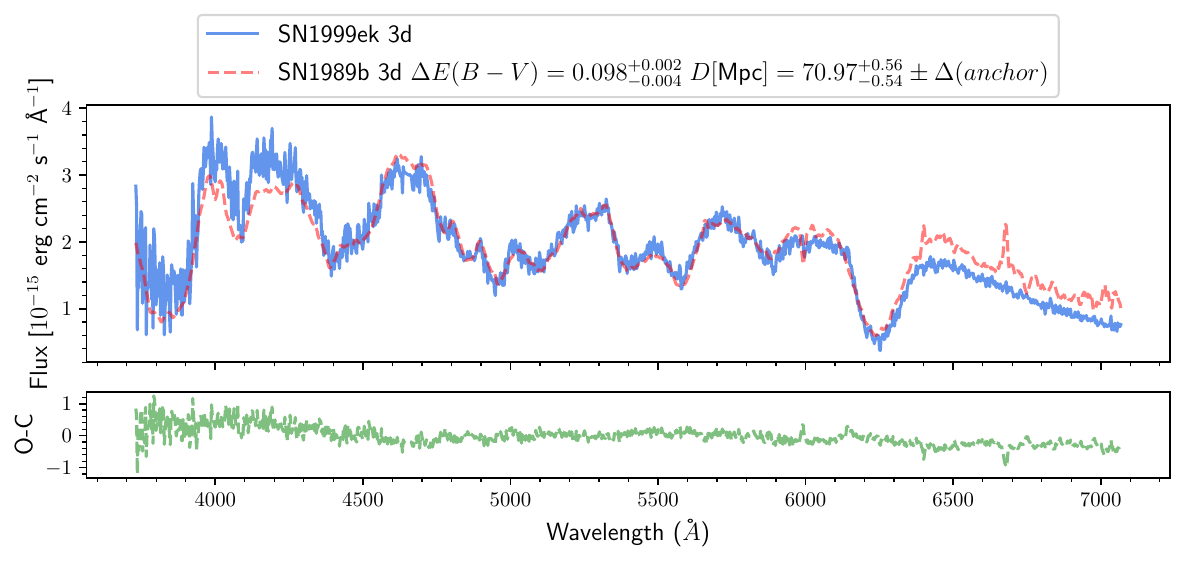}
  \includegraphics[width=0.45\textwidth]{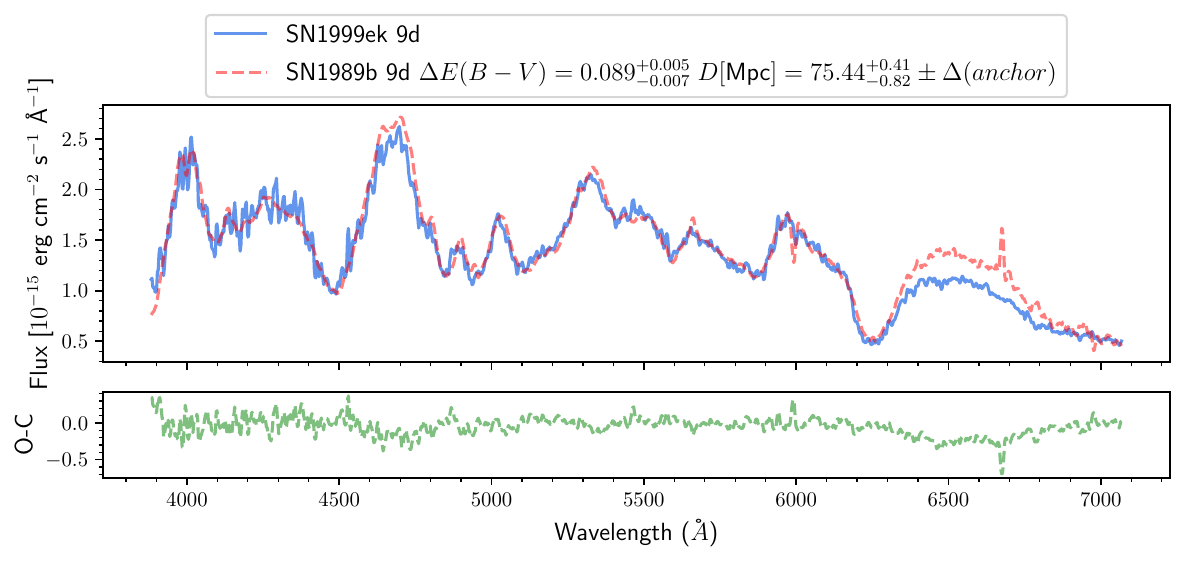}
  \includegraphics[width=0.3\textwidth]{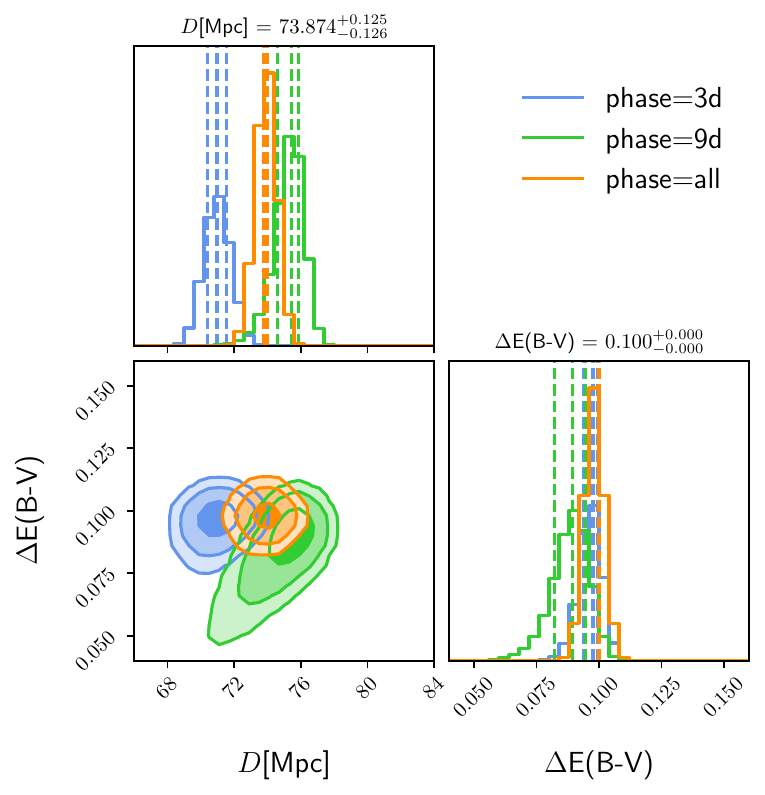}
\includegraphics[width=0.3\textwidth]{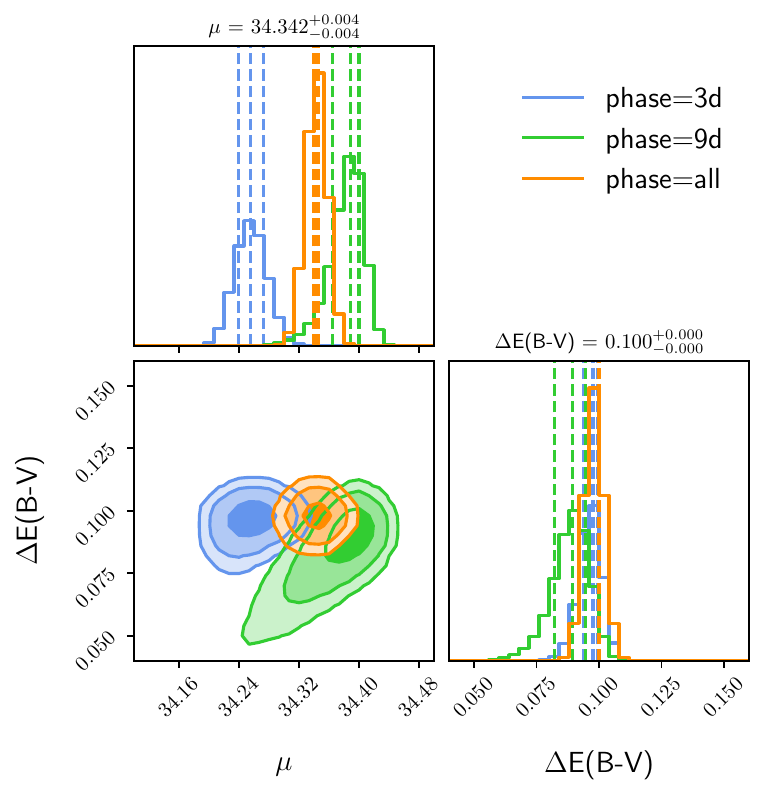}
\caption{ Top: Comparison of the spectrum at  3 days past maximum
  of SN 1999ek with that of SN 1989B at the same phase.
  Middle below top: Comparison of the spectrum at 9 days past maximum
  of SN 1999ek  with that of SN 1989B at about the same time.
  Middle above bottom:
  Middle above bottom: Corner plot with the posterior probability at 1$\sigma$, 2$\sigma$,
  3$\sigma$ of distance and relative intrinsic reddening of 
  SN 1999ek in relation to SN 1989B.
  Bottom:
  Corner plot with the posterior probability at 1$\sigma$, 2$\sigma$,
  3$\sigma$ of distance moduli  and relative intrinsic reddening of 
  SN 1999ek in relation to SN 1989B.}
\end{figure}

\begin{table*}[h!]
  \scriptsize
\centering
\caption{ Comparison of distance moduli and $H_{0}$ with Pantheon+ and CCHP
  results\footnote{This calibration uses the Cepheids distances to
the low--z anchor galaxies.}}

  \begin{tabular}{ccccccccccccccccccccccc}

    \hline

    && \ \ \ \ This work  &&&&& &&&&&  SH0ES &&&&&\   &&&&&   CCHP  \\

  \end{tabular}
  
  \begin{tabular}{llcccccc}

    SN & $z$_{hel}$$ &$\mu$ & $H_{0}$ & $\mu$  & $H_{0}$ & $\mu$ & $H_{0}$ \\

    \hline
    &&&&&&& \\
    2012 bo & 0.0265 &35.162$\pm$0.0453 & 75.291$\pm$3.243 & $\dots$ &
    $\dots$ & 35.446$\pm$0.217 & 66.040$\pm$6.599 \\
    &&&&&&& \\
    2008bq  & 0.0340 & 35.827$\pm$0.0739 & 72.419$\pm$3.240 &
    35.728$\pm$0.178 & 75.796$\pm$5.964 & 35.879$\pm$0.205 &
    70.705$\pm$8.132   \\
    &&&&&&& \\
    2008bz & 0.0603 & 37.214$\pm$0.0714 & 69.422$\pm$ 2.547 & 37.168$\pm$0.288
    & 71.072$\pm$3.945 & 37.155$\pm$0.280 & 71.334$\pm$9.198  \\
    &&&&&&& \\
    LSQ12fxd & 0.0312 & 35.543$\pm$0.0828 & 74.108$\pm$ 3.702& --- & --- &
-    35.615$\pm$0.207 & 72.497$\pm$6.911  \\
    &&&&&&& \\
    2008bf & 0.0235 & 35.073$\pm$0.112 & 74.085$\pm$4.820& 35.052$\pm$0.231 &
    74.716$\pm$7.494 & 35.011$\pm$0.222 & 76.192$\pm$7.789   \\
    &&&&&&& \\
    2007A & 0.0176 & 34.265$\pm$0.140 & 70.267$\pm$ 6.223 & 34.534$\pm$0.342 &
    62.043$\pm$9.041 & 34.317$\pm$0.312 & 68.604$\pm$9.857    \\
    &&&&&&& \\
    ASASSN-15db & 0.0109 & 33.527$\pm$0.0834 & 67.913$\pm$6.507 & --- & --- &
    33.532$\pm$0.330  & 67.757$\pm$10.297  \\
    &&&&&&& \\
    LSQ14gov & 0.0896 & 37.912$\pm$0.0539 & 73.594 $\pm$2.009  & --- & --- &
    37.940$\pm$0.185  & 73.732$\pm$6.336   \\
    &&&&&&& \\
    SN 2007ca & 0.0140 & 33.940$\pm$0.0524 & 74.440$\pm$5.256
    & 34.178$\pm$0.396 & 66.709$\pm$13.378 & 34.344$\pm$0.278 &
    61.802$\pm$7.912  \\
    &&&&&&& \\
    \hline

    & This work  &  & \ \ \ \ \ \ \ \  Cepheids & & \ \ \ \ \
    \  CCHP &   \\

    SN & $z$ & $\mu$ & $H_{0}$ & $\mu$  & $H_{0}$ & $\mu$ & $H_{0}$ \\

    \hline 
    &&&&&&& \\
    2001cn & 0.0156 & 33.947$\pm$0.0584 & 74.362$\pm$5.321
    & 33.887$\pm$0.2949 & 76.421 $\pm$ 11.119 &  ---    &  ---  \\
    &&&&&&& \\
    2008go & 0.0623 & 37.081$\pm$0.277& 72.284$\pm$9.299 & 36.96$\pm$0.19 
    & 77.955$\pm$7.128 & 37.004$\pm$0.193 & 76.392$\pm$6.790   \\
    &&&&&&& \\
    1999ek & 0.0177 &  34.342$\pm$0.0664 & 72.611$\pm$4.677 & --- & ---
     & --- & --- \\
    &&&&&&  \\
    \hline 

   \end{tabular}

\end{table*}

 \bigskip

 \bigskip

      \begin{figure}[h!]
       \includegraphics
      [width=0.45\textwidth]{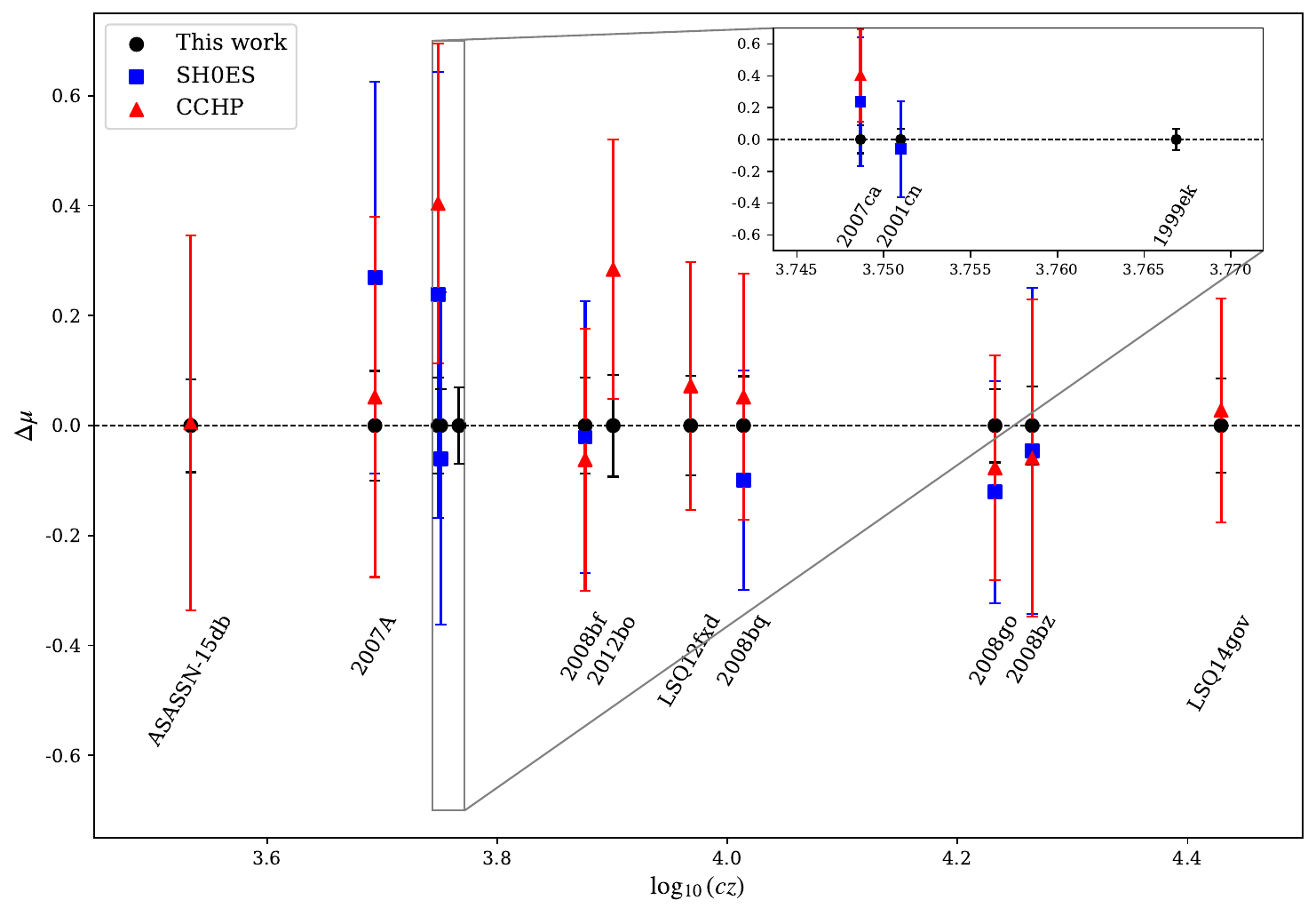}
      \caption{ Difference in moduli of distance between the Pantheon+ SH0ES
        and those obtained in this work and between the CCHP and
        those obtained in this work. The
        points in black are those obtained in this work anchored in Cepheids,
        points in blue are from
    Pantheon+ SH0ES using Cepheids and triangles in red are Cepheids distances
    from CCHP. Some results in blue are not
    aligned: those are the ones by Pantheon+ for distances where the
    color c of those SNe Ia is underestimated, in  particular
    for SN 2007A, SN 2007ca. The red triangles which  are not aligned,
    corresponding  to SNe Ia such as 
     SN 2007ca,  arise in an understimate of
    the reddening from the CCHP.
    There is also the misaligment of SN 2012bo linked to its parameterization.
    A view of the results by Pantheon+ and by the CCHP compared
    with the ``SNe Ia twins'' results can be seen in Figures 14 and 15
    for the sake of a better understanding.}

      \end{figure}

\bigskip

\noindent
We have modeled the statistical errors in the distance moduli for our method
adding the various contributions. We have added in quadrature (see 4) 
the error in the fit of  the distance
modulus obtained with  the MCMC,  $\sigma_{\rm fit}$,  and the covariance
between the modulus and
the $\Delta$(E(B-V)), cov ($\mu$, E(B-V)). We have 
as well the error
due to a difference of stretch between the Hubble flow SNeIa and their
nearby twins that goes into the distance modulus
$\sigma^{2}(\mu, \Delta s_{BV})$. We model it
taking into account the CSP impact of the error in stretch in the distance
moduli of the SNe Ia. 
We must say that, in our method, twins at low z and in the Hubble flow
have very close stretches: within $\sim$ 0.02 in s$_{BV}$. 
The light curves agree so well  between low z SNe Ia  and their
twins in the Hubble flow (once one makes the cosmological (1+z) time
broadening and magnitude dilution) 
that, in their  plot of one over the other, they are often
indistinguishable (see Appendix A). We add as well in quadrature
the estimated intrinsic error $\sigma_{int}$ (see Appendix B). 

\bigskip

\noindent
The error budget in $\mu$  is:

\begin{equation}
  \sigma_{\mu}^{2} = \sigma_{\rm fit}^{2} + 2 cov (\mu, E(B-V)) +
  \sigma^{2}(\mu, \Delta s_{BV}) + \sigma_{int}^{2}
\end{equation}\footnote{$\sigma^{2}(\mu, \Delta s_{BV}) = (P1+ 2P2
(\Delta s_{BV}-1))^{2}\sigma(\Delta s_{BV})^{2}$, see (7) for $P1, P2$.}

\noindent
where $\Delta s_{BV}$ is the difference of stretch between
the SNe Ia in the Hubble flow and in the low z twins. $\sigma_{int}$
is the intrinsic error due to intrinsic uncertainties in flux comparison
(see Methodology section).

\bigskip

\noindent
For the H$_{0}$ error calculation, we  incorporate the error
in peculiar velocity
affecting the cosmological z value.
This error in the peculiar velocity is not attributed to the distance
modulus because it enters into the H$_{0}$ calculation rather than in the
distance. We adopt an error of 300 km s$^{-1}$ in the calculation of
the cosmological redshift (the cosmological redshifts are adopted from
NASA/IPAC Extragalactic Database). 
  According to authors who model peculiar velocities for
  the fields of Pantheon+ (Carr et al. 2022), those
  peculiar velocities in galaxies
  can reach to 250 km s$^{-1}$, and 100 km s$^{-1}$
  for group of galaxies (Carrick et al. 2015). For this
  reason, 300 km s$^{-1}$ is a conservative estimate of the error.

\bigskip

\noindent
This error on H$_{0}$ and those from the terms that enter in (5, see  below)
are added 
to the error budget for H$_{0}$.

\bigskip

\noindent
We obtain the individual values of H$_{0}$ for each measured SN Ia.

\bigskip

\noindent
To calculate the value of H$_{0}$ we use the expression valid for a flat
cosmology, $\Omega_{k}$ $=$ 0:

\begin{dmath}
  H_{0}  = {cz\over D}\left({1 + z_{ hel}\over 1 + z}\right) \left[1 +
    {1\over 2}(1 - q_{0})z -{1\over 6}(1 - q_{0}
      -3q_{0}^{2} + j_{0})z^{2} + \mathcal{O}(z^{3})  \right]
\end{dmath}

\noindent
where H$_{0}$ is the Hubble constant, D is the distance to the SN Ia, c is the
speed of light, z is the cosmological redshift, z$_{hel}$ the heliocentric 
redshift. The first and second order terms multiplied by z in the  parenthesis
involve  the deceleration parameter, q$_{0}$, and the cosmic jerk, j$_{0}$.
q$_{0}$ $=$ $\Omega_{m}/2-\Omega_{de}$ which taking $\Omega_{m}$ $=$ 0.3 give
q$_{0}$ $=$ --0.55 ($\Omega_{de}$ $=$ 1 - $\Omega_{m}$ for a flat cosmology).
The cosmic jerk can be taken to be j$_{0}$ $=$ 1 for the range of  redshifts
that we are using. 

 \begin{table*}[h!]
  \centering
  \scriptsize
  \caption{{\bf Galactic and total reddenings of the SNe Ia in the Hubble flow}}
  \begin{tabular}{lccccc}
      \hline
      \hline
      Supernova &  $E(B-V)_{MW}$ & $\Delta E(B-V)_{Host}$ & $E(B-V)_{Tot}$ &
      $c$ & $B-V$ \\
      \hline
      SN 2012bo & 0.046  & 0.000 $\pm$ 0.004  & 0.046$\pm$0.004 & - & 0.012$\pm$0.006 \\
      SN 2008bq & 0.060 & 0.000 $\pm$ 0.003     &0.060$\pm$0.003 &  0.013$\pm$0.025$^{5}$ &
      0.045$\pm$0.009 \\
      SN 2008bz & 0.023 &  0.00$\pm$ 0.022   & 0.023$^{+0.022}_{-0.000}$ & -0.136$\pm$0.023$^{5}$ &
      -0.006$\pm$0.006 \\
     
      $\dots$ $\dots$ $\dots$ & $\dots$ & $\dots$ & $\dots$ &
      -0.137$\pm$0.036$^{65}$ & $\dots$ \\
      LSQ12fxd  & 0.022 & 0.000$\pm$0.002   & 0.022$\pm$0.002 & - & -0.006$\pm$0.006 \\
      
      SN 2008bf & 0.036 &  0.015$\pm$ 0.007  & 0.051$\pm$0.007 & -0.149$\pm$0.022$^{5}$ &
      -0.046$\pm$0.006 \\
      $\dots$ $\dots$ $\dots$ & $\dots$ & $\dots$ & $\dots$ &
      -0.149$\pm$0.026$^{57}$ & $\dots$ \\
      $\dots$ $\dots$ $\dots$ & $\dots$ & $\dots$ & $\dots$ &
      -0.056$\pm$0.026$^{64}$ & $\dots$ \\
      SN 2007A  & 0.074 & 0.209$\pm$ 0.012    & 0.283$\pm$0.012 & 0.117$\pm$0.025$^{5}$ &
      0.201$\pm$0.010 \\
      $\dots$ $\dots$ $\dots$ & $\dots$ & $\dots$ & $\dots$ &
      0.063$\pm$0.033$^{65}$ & $\dots$ \\

      ASASSN-15db & 0.03 & 0.026$\pm$0.008   & 0.056$\pm$0.008 & - & 0.029$\pm$0.003 \\
      LSQ14gov  & 0.0377 & 0.015$\pm$0.001  & 0.0527$\pm$0.001 & - & -0.104$\pm$0.008 \\
      SN 2007ca & 0.067  & 0.199 $\pm$ 0.001& 0.266$\pm$0.001 & 0.204$\pm$0.024$^{5}$ &
      0.260$\pm$0.010 \\
      $\dots$ $\dots$ $\dots$ & $\dots$ & $\dots$ & $\dots$ &
      0.151$\pm$0.023$^{57}$ & $\dots$ \\
      $\dots$ $\dots$ $\dots$ & $\dots$ & $\dots$ & $\dots$ &
      0.207$\pm$0.028$^{64}$ & $\dots$ \\
      SN 2001cn & 0.052 & 0.133 $\pm$ 0.03   & 0.183$\pm$0.03 & 0.118$\pm$0.027 & - \\
      SN 2008go & 0.033 &  0.020 $\pm$ 0.01  & 0.053$\pm$0.01 & -0.016$\pm$0.023$^{5}$ &
      0.042$\pm$0.010 \\
      $\dots$ $\dots$ $\dots$ & $\dots$ & $\dots$ & $\dots$ &
      -0.007$\pm$0.035$^{51}$ & $\dots$ \\
      SN 1999ek & 0.502 & 0.22 $\pm$ 0.006   &  0.722$\pm$0.006 & - & - \\
      \hline
      \hline
      \multicolumn{5}{c}{$^{5}$CSP \quad $^{50}$LOWZ/JRK07 \quad $^{51}$LOSS1 \quad $^{57}$LOSS2 \quad $^{64}$CfA3K \quad $^{65}$CFA4p2} \\
      \hline
\end{tabular}
\end{table*}

\bigskip

 \begin{table*}[h!]
    \scriptsize
  \centering
  \caption{Distance factors relative to anchors}

  \begin{tabular}{lcc}

    \hline

    SN & Distance factor & Anchor \\
    \hline
             &                       &      \\

    2012bo   &   5.180 $\pm$ 0.017  & 2013dy \\
             &                       &         \\
    2008bq   &   7.038 $\pm$ 0.033   & 2013dy \\
            &                       &         \\
    2008bz   &   40.291 $\pm$ 0.001   & 2011fe \\
            &                       &         \\
    LSQ12fxd &   6.175 $\pm$ 0.019    & 2013dy \\
            &                       &         \\
    2008bf   &    8.030 $\pm$ 0.098  & 2013aa \\
            &                       &         \\
    2007A    &   5.532 $\pm$ 0.102 & 2013aa \\
            &                       &         \\
    ASASSN-15db &   7.376 $\pm$0.105  & 2011fe \\
            &                       &         \\
    LSQ14gov &  29.674 $\pm$ 0.031 & 2013aa \\
            &                       &         \\
    SN2007ca &   2.966 $\pm$ 0.007  &  2013dy      \\
            &                       &         \\
    SN 2001cn  &    5.563 $\pm$ 0.040     &  1989B       \\
            &                       &         \\
    SN 2008go  &    23.564 $\pm$ 0.195      &  1989B     \\
            &                       &         \\
    SN 1999ek  & 6.675 $\pm$  0.011   &  1989B            \\

\hline

\end{tabular}

 \end{table*}

 \bigskip

\noindent
Our agreement  (Table 5) with the
distances using Cepheids in B by the CCHP is
consistent with the findings of Uddin et al. (2024).
Those authors obtained H$_{0}$  $=$ 72.37$\pm$ 0.71
km s$^{-1}$ Mpc$^{-1}$.
Concerning TRGB in B, Uddin et al. (2024)   gave  $H_{0}$ $=$
70.25  $\pm$ 0.71 km s$^{-1}$  Mpc$^{-1}$. And for {\it SBF},
(Surface Brightness Fluctuations) Uddin et al. (2024)
obtained $H_{0}$ $=$ 72.45 $\pm$ 0.94.

\bigskip

     \subsection{Limitations of the parameterizations}

     \bigskip

\noindent     
Pantheon+ obtains the SNe Ia distances
in the
Hubble flow through a process that implies describing  the
SNe Ia with SALT2. The CCHP also makes this three rung ladder parameterizing
the light curve-rate of decline relation but in a different way. 

\bigskip

     \noindent
     In this section, we  analyse some shortcomings that we have
     found in the  
     parameterization to obtain the distance moduli through the
     light curves using either SALT2 or the CSP new light curve-rate of decline relation.

     \begin{figure}[h!]
 \includegraphics[width=0.45\textwidth]{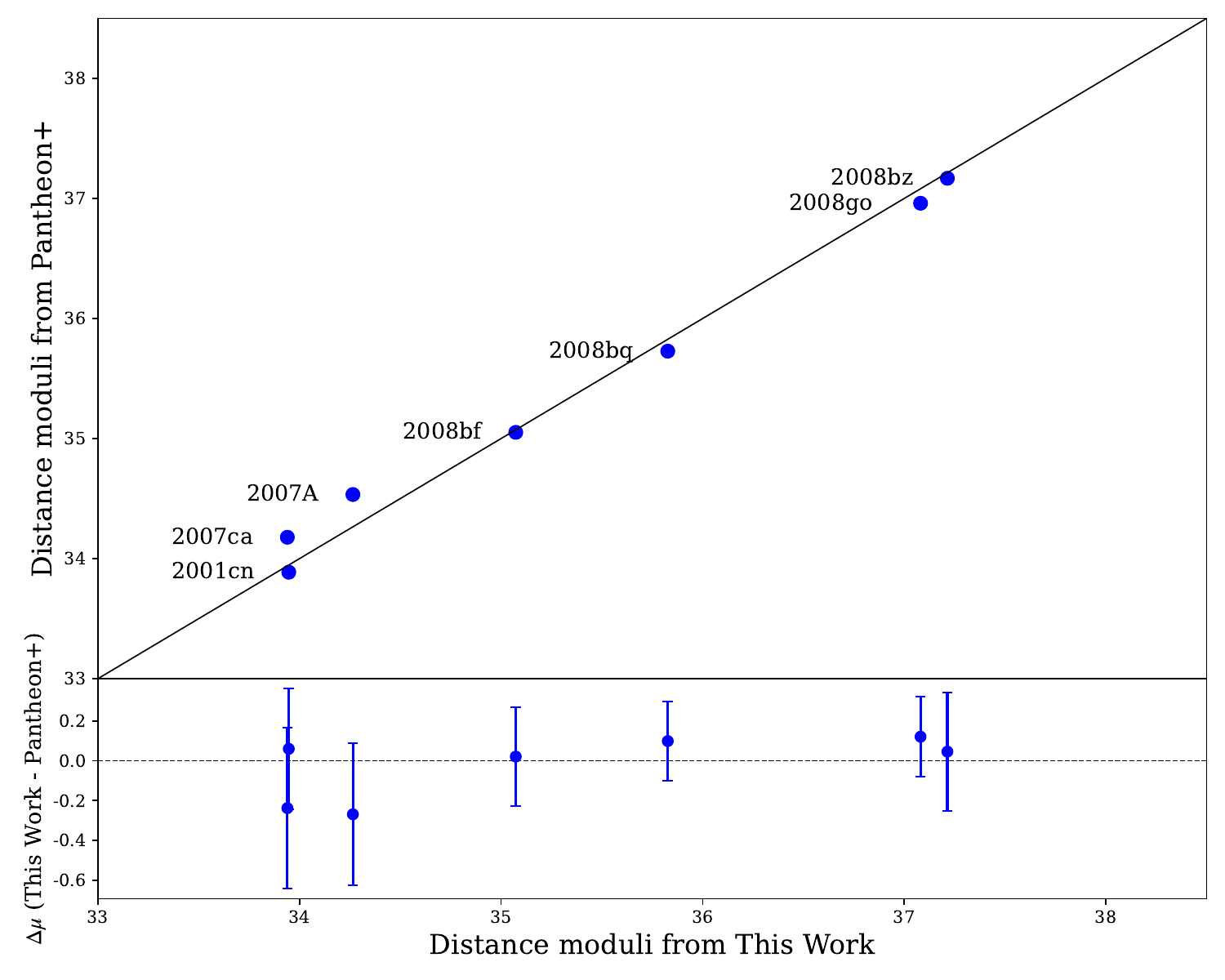}
  \caption{
    Top panel: Distance moduli in this work a versus 
     those of Pantheon+.
     The distances span a range from 60 Mpc to 400 Mpc. Here we
     can see how  for the Pantheon+ SN 2007A, SN 2007ca stand out due
     to a too low color 'c' estimation.}
  
 \end{figure}

 \begin{figure}[h!]
  \includegraphics[width=0.45\textwidth]{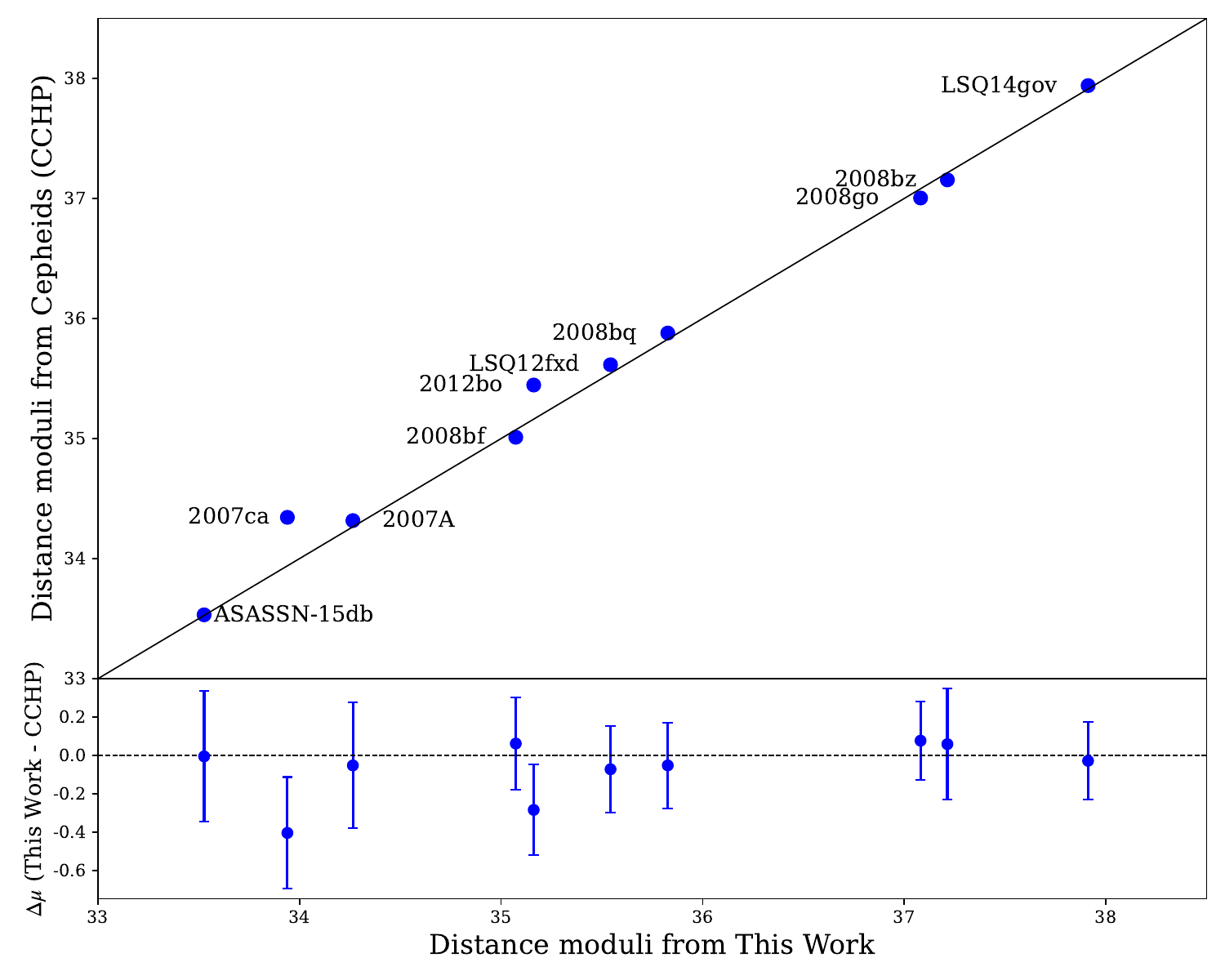}

\caption{Top panel: Distance moduli in this work a versus 
     those by the CCHP (Cepheids).
     The distances span a range from 60 Mpc to 400 Mpc. Here we
     can see how  SN 2007ca stands out due
     to a too low correction by BV in relation to E(B-V)
     estimation.  There is an error in the SN 2012bo parameters for the
      light curve--rate of decline correction.}
     \end{figure}

\bigskip

\noindent
The Pantheon+  light curves are parameterized through the SALT2 light--curve fitting
(Tripp 1998; Brout et al. 2022). In this approach
the distance modulus is defined as:

\begin{equation}
\mu= m_{B} + \alpha x_{1} + \beta c + M + \delta_{B,a} + \delta_{Host}
\end{equation}

\bigskip

\noindent
where $\alpha$ and $\beta$ are global nuisance parameters relating
stretch x$_{1}$ and the color c (respectively) to luminosity. M is the
fiducial absolute magnitude of an SN Ia, which can be calibrated
by setting an absolute distance scale with primary distance anchors
such as Cepheids. The $\delta_{B,a}$ bias is a correction
term to account for selection biases that is determined
from simulations following Popovic et al. (2021),
 $\delta_{Host}$ is the luminosity
correction (step) for residual correlations between the
standardized brightness of a SN Ia and the host-galaxy
mass.

\bigskip

\noindent
The relation used by the CSP and CCHP
instead of being 
linear  in stretch (like in Tripp 1998), is  quadratic.
The reddening correction
is simply a constant, $\beta$, multiplied by the observed color of the
SNe Ia, (B-V) where B and V are
the apparent, K-corrected peak magnitudes.
P1 is the linear coefﬁcient and P2 is the quadratic
coefﬁcient in (s$_{BV}$-1), which encapsulates the Phillips
relation in the s$_{BV}$ stretch computed by SNooPy;
$\alpha_{M}$  is the slope
of the correlation between peak luminosity and host stellar
mass M$_{*}$ (see for instance Freedman et al. 2019).

\bigskip

\begin{equation}
  \begin{split}
  \mu_{\rm obs} = m_{x} -P0 - P1 (s_{BV}-1) - P2 (s_{BV}-1)^{2} \\
  -\beta(B - V) - \alpha_{M}({\rm log_{10}}M_{*}/M_{\odot}-M_{0})
  \end{split}
  \end{equation}

\bigskip

\bigskip

\begin{figure*}[h!]
  \centering
  \includegraphics[width=0.75\textwidth]{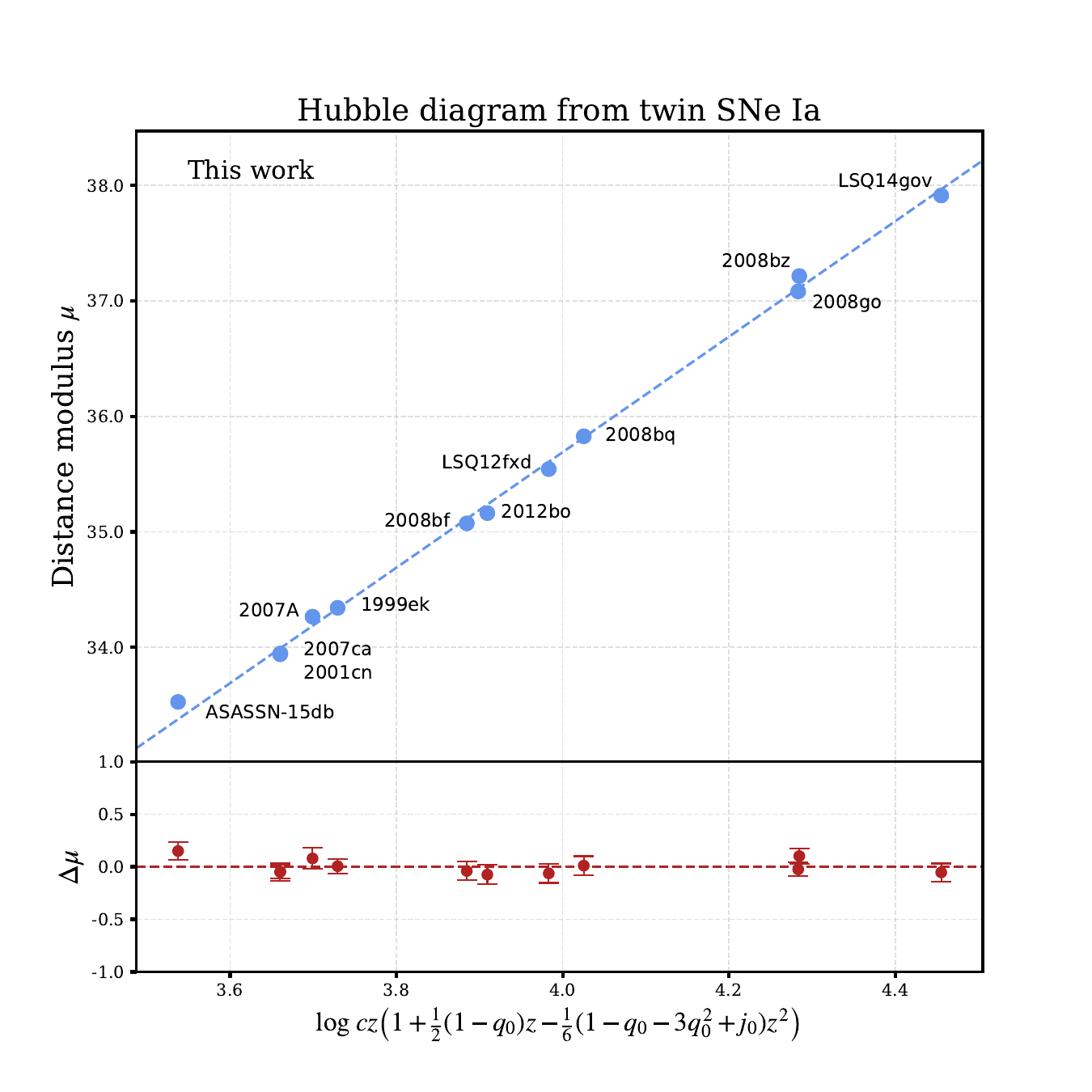}
  \caption{
    Hubble diagram with the distances obtained in this work.
    The intercept of  the straight line with the y-axis
    corresponds to H$_{0}$ $=$72.79 $\pm$ 0.78 
     km s$^{-1}$ Mpc$^{-1}$.}
\end{figure*}

\noindent
We have noticed that, for reddened SNe Ia, the two parameterizations of
the stretch fail to obtain good 'color'  values and this
impacts in the distance to SNe Ia.
For instance, within the SALT2 parametrization, the color 'c' in (5) obtained
by Pantheon+
tends to be low in the reddened SNe Ia in the Hubble flow (see Figure 14).
Such too low 'c' then reflects in a larger value for the distance
modulus and carries  into the Hubble flow values of the order of 60 to 65
km s$^{-1}$ Mpc$^ {-1}$ (Table 5).

\bigskip

\noindent
This happens as well with the parametrization by the CCHP  (see 6),
despite the fact that there is a term in (B-V). 
(B-V) acts as a color term that would include the presence of reddening.
In fact, the  BV color in the abovementioned CSP parameterization in (5) would
 be the combination of the intrinsic color of the SN Ia at maximum and 
 the E(B-V) reddening. At maximum the
 intrinsic (B$_{max}$ --V$_{max}$) of SNe Ia  can be slighlty negative, if
 $\Delta m(B)_{15}$ is lower than 1.1 (see expression 7 in Phillips et al. 1999)
 \footnote{$B_{max} -V_{max}$ $=$ $-0.070 (\pm 0.012) + 0.114(\pm 0.037) \times
( \Delta m_{15}(B)-1.1)$ , $\sigma$ $=$ 0.0042}.
 In Table 6, we include
the E(B-V)$_{MW}$ from Schlafly \& Finkbeiner (2011), the derived
E(B-V)$_{Tot}$ from the spectra, the 'c' parameter for this SN Ia
in the Pantheon+ compilation, and the 'BV' parameter in the CCHP
compilation.

\bigskip

\noindent
As mentioned in Freedman et al. (2019), there is an alternative
parameterization by Burns et al. (2014) where instead of the
 $\beta$(B--V) term there is a R$_{B}$ E(B--V) term to solve better
the extinction. However, that has not been implemented in the
present derivation of distance moduli for the CSP sample. 
So, this is the limitation we are seeing.  

\bigskip

\noindent
SN 2007A has a  wrong value of its color
 'c'. That causes  a too large distance
and individual point for H$_{0}$  with  low values.
It does not have a correct distance modulus in Pantheon+, as
it has a color parameter  in two diffferent entries of c $=$ 0.117 $\pm$
0.024  and c $=$ 0.067 $\pm$ 0.033. We have found and
E(B-V)$_{Tot}$ $=$ 0.283 $\pm$ 0.012. The CSP gives a value of
BV$=$ 0.201 $\pm$ 0.01.  
The CSP gets a consistent  distance modulus in this case as its 'BV' value
would coincide with an  E(B-V) from 0.283 $\pm$ 0.012 plus
a small negative value for the intrinsic color. The Pantheon+
  does not provide a consistent distant modulus.

\bigskip

\noindent
SN 2007ca requires a larger color correction than presented in 
Pantheon+. The color correction in Pantheon+, 
where it has 3 different entries for the SALT2 parameters is
c $=$ 0.151 $\pm$ 0.023, c $=$ 0.204 $\pm$ 0.023 and c $=$ 0.207
$\pm$ 0.028. The E(B-V) for this SN Ia is already 0.266  $\pm$ 0.001.
The CSP gives as its BV parameter 0.26 $\pm$ 0.01, and this time
it falls a little short.

\bigskip

\noindent
On the other hand, other SNe Ia have a consistent 'c' or 'BV' values. 
For instance, in  SN 2001cn, Pantheon+ gives a 
color with 'c' $=$ 0.118 $\pm$  0.027. 
The spectra  show that E(B-V) is 0.183 $\pm$ 0.005. Given that c should reflect
the intrinsic color of SNe Ia, which is (B$_{max,int}$ --V$_{max,int}$) $=$ --0.072 ($\pm$
0.049) mag
for this SN Ia, according to the expression (7) in Phillips et al (1999), if we add the reddening the final color 'c' is 0.111 $\pm$ 0.079. 
Here the final distances are similar, but the one by Pantheon+ has very large
errors, that might be linked to not have used the
light curves for this SN Ia (Krisciunas et
al. 2004) that would allow a very good x1 and c determination.

\bigskip

\bigskip

\begin{table}[ht]
\scriptsize
\centering
\caption{$H_{0}$ and distances calibrated with JAGB\footnote{The distances are
  anchored using JAGB by the CCHP for the
  9 SNe Ia in the upper part
  of the Table and using TRGB by the CCHP in the last three SNe Ia of the
  Table.}}
\setlength{\tabcolsep}{3pt}
  \begin{tabular}{lcc}
    \hline
    \hline \\

    SN & $\mu$ & $H_{0}$  \\
   
    \hline
        && \\
    2012bo & 35.162$\pm$0.0453 & 75.291$\pm$3.243\\
        && \\
    2008bq  &35.827$\pm$0.0739 & 72.419 $\pm$ 3.240 \\
        && \\
    2008bz & 37.236 $\pm$0.0714 & 68.722 $\pm$2.522 \\
        && \\
    LSQ12fxd &  35.543$\pm$0.0828 & 74.108 $\pm$3.702 \\
        && \\
    2008bf & 35.103 $\pm$0.112 & 73.069 $\pm$ 4.767  \\
        && \\
    2007A &    34.294 $\pm$0.140 & 69.335 $\pm$6.095  \\
        && \\
    ASASSN-15db &  33.549 $\pm$0.0834  & 67.228 $\pm$6.477 \\
        && \\
    LSQ14gov & 37.942 $\pm$0.0539 & 72.585 $\pm$2.009 \\
        && \\
    2007ca &   33.940  $\pm$0.0524 & 74.440 $\pm$5.256   \\
    && \\
   
    \hline

    SN & $\mu$ & $H_{0}$ \\
    \hline

    && \\
   
      2001cn & 33.947$\pm$0.0584 & 74.362 $\pm$ 5.321 \\
    && \\
     2008go & 37.081$\pm$0.277 &  72.284$\pm$9.299  \\
        && \\
    1999ek & 34.342$\pm$0.0664 & 72.611$\pm$4.677  \\

        && \\
    \hline
\end{tabular}
\end{table}

\begin{table}

\scriptsize
 
  \centering

  \caption{Comparison of JAGB distances (CCHP) with JAGB distances
    (SH0ES$^{2}$)}

\begin{tabular}{lcc}
 
  \hline
  \hline

  Galaxy & $\mu_{\rm JAGB}$ (CCHP)$^{1}$ & $\mu_{\rm JAGB}$ (SH0ES)$^{2}$ \\
  & (mag) & (mag) \\

  \hline

  M101 & 29.21$\pm$0.03 (stat)$\pm$0.03 (sys) & 29.12$\pm$0.06 \\
  NGC 1365 & 31.38$\pm$0.02 (stat)$\pm$0.03 (sys) & 31.34$\pm$0.04 \\
  NGC 2442 & 31.60$\pm$0.01 (stat)$\pm$0.04 (sys) & 31.56$\pm$0.05 \\
  NGC 4536 & 30.97$\pm$0.01 (stat)$\pm$0.03 (sys) & 30.93$\pm$0.04 \\
  NGC 4639 & 31.73$\pm$0.02 (stat)$\pm$0.03 (sys) & 31.69$\pm$0.04 \\
  NGC 5643 & 30.58$\pm$0.01 (stat)$\pm$0.04 (sys) & 30.54$\pm$0.04 \\
  NGC 7250 & 31.59$\pm$0.02 (stat)$\pm$0.04 (sys) & 31.59$\pm$0.04 \\
  NGC 3972 & 31.67$\pm$0.04 (stat)$\pm$0.03 (sys) & --- \\
  NGC 4038 & 31.53$\pm$0.06 (stat)$\pm$0.04 (sys) & --- \\
  NGC 4424 & 31.15$\pm$0.04 (stat)$\pm$0.03 (sys) & --- \\

  \hline

\end{tabular}
\begin{tabular}{lll}

$^{1}$Lee et al. (2025). & $^{2}$Li et al. (2025) &  \\

\end{tabular}

\end{table}

\bigskip     

\subsection{The value of H$_{0}$}

\bigskip

\noindent
Our results for individual SNe Ia, in Tables 5 and 8
clearly confirm that the Hubble tension is real, since they are
incompatible with the H$_{0}$ value derived from the CMB.
The results of the Hubble diagram, when the SNe Ia are anchored in JAGBs (CCHP)
(Table 8), are quite similar to the anchoring in Cepheids. 

\bigskip

\noindent
The 
value, for the 12 SNe Ia, is H$_{0}$ =  72.56 $\pm$ 1.54 (stat) $\pm$
1.33 (syst) km s$^{-1}$ Mpc$^{-1}$ when calibrating the anchor distances with
Cepheids (by SH0ES)  and  72.20 $\pm$ 1.53 (stat) $\pm$ 1.33 (sys)  km
s$^{-1}$ Mpc$^{-1}$ when calibrating with JAGB (by the CCHP).
The average is H$_{0}$ = 72.38 $\pm$ 1.54 (stat) $\pm$ 1.33
(sys) km s$^{-1}$ Mpc$^{-1}$, far from the
67.4 $\pm$ 0.5 km s$^{-1}$ Mpc$^{-1}$ of the Planck Collaboration.

\bigskip

\noindent
Figure 16 is the Hubble diagram of the SNe Ia in this work anchored
in Cepheids (Table 5). The intercept corresponds to H$_{0}$ $=$  72.79 $\pm$ 0.78 km s$^{-1}$ Mpc$^{-1}$.

\bigskip

\noindent
 If we take the SNe Ia of the BL class only, for which the anchor distance
is known with the greatest precision (coincidence between CCHP distances
 and Saha et al. 1999), the simple mean is
H$_{0}$ = 73.00 $\pm$ 3.98 (stat)  km s$^{-1}$ Mpc$^{-1}$
  and the weighted average from the three SN Ia 73.24 $\pm$ 3.29 (stat) km s$^{-1}$ Mpc$^{-1}$.

\bigskip

\noindent
Our results from comparing the spectra of very nearby SNe Ia with those of
their twins in the Hubble flow also validates the three rung procedures used
by the Pantheon+ and CCHP collaborations, even at the level of individual
SNe Ia, since the discrepancies can easily be explained by differences in
the evaluation of the total reddenings or  the different light-curve
parameters adopted, as we have seen in the previous Section.

\bigskip

\noindent
To leave our results stable against future changes in the distances of the
anchors in the nearby galaxies,
we publish the distance factors between the anchors and the SNe Ia
distances. The factors should be multiplied by the distances of the anchors
to obtain the distances to the SNe Ia in the Hubble flow (Table 7).

\bigskip

\noindent
Concerning the different calibrations used for setting our anchor distances,
we see that the distance moduli from  TRGB by
the CCHP have fluctuated during the last year. In addition, their distances
are very different when using the same data but analysed by
other groups (Anand et al. 2022).

\bigskip

\noindent
Consequently, we have obtained the two H$_{0}$ values given above, for the
two different calibrations of our anchors using Cepheids and JAGBs (Table
5 and Table 8). 

\bigskip

\noindent
It is worth to remark the distance to M101 measured from SH0ES with
Cepheids and with the JAGB stars from both the CCHP and SH0ES.
Those distance values clearly favor a value below 70 km s$^{-1}$ Mpc $^{-1}$
for H$_{0}$ from 
SNe Ia twins of SN 2011fe. Table 9  illustrates the
difference in distance moduli to M101 obtained by the two collaborations
with the same method: it is a difference of $\Delta \mu_{M101}$ $=$ 0.09 mag,
which  translates in $\sim$ 4 km s$^{-1}$ Mpc$^{-1}$ for $\Delta$H$_{0}$.
If our sample would
be dominated by 2011fe--like SNe Ia, the total H$_{0}$ would be lower
than what it is. However, the other values of distances using JAGB stars
for the rest of galaxies differ only  by 0.04 mag between the larger
distance value by CCHP and the shorter  one from  SH0ES. As the sample contains
SNe Ia twins from SN 2013aa and SN 2013dy, the values above
70 km s$^{-1}$ Mpc $^{-1}$ dominate. We pointed, in RLGH24, a preference
for the distance to SN 2011fe from the spectra of $\sim$ 6.5 Mpc, in agreement
with the value obtained from Blue Supergiants (Bresolin et al. 2022)
of 6.5 $\pm$ 0.2 Mpc. 
Inevitable,  using the 10 galaxy calibrators by the JAGB in rung 2,
as it is done  by the CCHP, could  bring H$_{0}$ to lower values than 70
km s$^{-1}$ Mpc$^{-1}$ given the calibration of M101. That would depend
on the weight of M101 in the whole sample.  However, it is
surprising such low  mean value of
H$_{0}$ $\sim$ 67 km s$^{-1}$ Mpc$^{-1}$ published by Freedman et al. (2025).
The case deserves further
exploration.

\bigskip

\section{Summary and conclusions}

\bigskip

\noindent
The use of the ''SNe Ia twins'' to obtain H$_{0}$  has revealed that
the Hubble tension is real. The distances obtained through this direct path
go from   ~6 Mpc to  around 300 Mpc in a single step.

\bigskip

\noindent
Comparing our distances with those in the
Pantheon+ and CCHP samples using Cepheids, we find a general agreement,
with a few exceptions, mostly  related with the determination of the reddening of some SNe Ia. The results in the Hubble diagram coincide and the
 values of  H$_{0}$ are very close. 

\bigskip

\noindent
This implies that the SNe Ia calibrators in rung 2 are a good 
sample that achieves a consistent final value of H$_{0}$  in rung 3.
The three--rung method seems to work.
It could fail if  a biased sample in rung 2 were taken. Then,
the final value of H$_{0}$
would be distorted, and the results of the Hubble diagram changed.
The large samples in Pantheon+ or CCHP using Cepheids as calibrators 
have values of individual H$_{0}$ 
whose average give close values for H$_{0}$. 
Using the TRGB obtained by the CCHP, the distances of the galaxies in rung
2 would be larger and the H$_{0}$ smaller, in contrast to the 
TRGB from SH0ES. A measurement
that  uses Cepheids from SH0ES and JAGB from the CCHP with the twins method
crosses the aisle between the two collaborations. The opportunity comes from
similar distance values for a few galaxies below 20 Mpc.

\bigskip

\noindent
At the individual level,
the distances to each SN Ia are affected by the limitations of the
parameterization of the light curves used by the two collaborations. 
The method of twin SNe Ia, with its use of spectra, is very powerful 
in getting the reddening of the SNe Ia in the Hubble
flow.
An approach as that developed here  can point to the failures in the distance
moduli for these supernovae that give H$_{0}$ in the extremes of
the distribution (very low or very large H$_{0}$).
SNe Ia with color larger than 'c' $=$ 0.1 
in the Pantheon+ sample give rise to low reddening correction and larger
distance moduli. This leads  to 
individual  points of H$_{0}$  falling out of the range of discussion,
i.e., in the tail 
of H$_{0}$ below 65 km s$^{-1}$ Mpc$^{-1}$. A similar effect is seen with
the BV color. 
Those tails come from SNe Ia where the color terms 'c', or '(BV)' are not well
estimated or even  the rates of decline of the lightcurves
parameterized in the form of 'x1' in SALT2 or s$_{BV}$ in CCHP
have been wrongly estimated. Sometimes large uncertainties are  quoted to accompany those values, but in some cases they are not.
Those extreme values do not alter significantly the mean for H$_{0}$.
However, for those particular SNe Ia in the Hubble flow the distance remains unknown.

\bigskip

\noindent
Inaccurate  point values in distances and therefore in
 H$_{0}$ individual points (even when enclosing these points within a 
a large error bar)  would interfere with
the proper determination  of possible cosmic  bulk flow in the z
range 0.015 $<$ z $<$ 0.1. 
An important task is to improve the Pantheon+ and CSP distances. 
In this sense, our approach can help a lot  to give the right
distances and right individual H$_{0}$ values. 

\bigskip

\noindent
The use of SN 1989B  in M66 as anchor has
been very fruitful. It has allowed to get accurate distances
to three SNe Ia: SN 2001cn, SN 2008go and SN 1999ek.
As said before, the distance to M66 with TRGB  by Freedman et al. (2019)
(obtained by  Hoyt et al. 2019) 
coincides with the Cepheids value from Saha et al. (1999).

\bigskip

\noindent
We finally stress that the present use of two different calibrations, that with Cepheids from Pantheon+ and 
the one with JAGB from CCHP, do converge to the same result.
This concordance found  using  anchors from the two collaborations 
rarely happens and looks very promising. 
 There is  a very good agreement on the distances to NGC 7250 and
  NGC 5643 derived with Cepheids by SH0ES and
   derived with the use of J-Asymptotic Giant Branch stars
  (JAGB stars) by the CCHP.  
  The difference for SNe Ia in the Hubble flow anchored in one or the other way
  is negligible:  our SNe Ia  sample in the Hubble flow 
  anchored in Cepheids gives H$_{0}$ $=$
  72.56 $\pm$ 1.54 (stat) $\pm$ 1.33 (sys) and  
  anchored in JAGB stars gives
H$_{0}$ $=$  72.20 $\pm$ 1.53 (stat) $\pm$ 1.33 (sys). 
The mean would be H$_{0}$ $=$  72.38$\pm$ 1.54 (stat) $\pm$ 1.33 (sys).

\bigskip
\bigskip

\noindent
This work has made use of spectra from the {\it Weizmann Interactive
  Supernova data REPository} ({\it WISeREP}), spectra and photometry of the
Carnegie Supernova Project (CSP). The authors acknowledge Pantheon+ and the CSP
for opening their data releases and software to the community.
We are very grateful to Nidia Morrell (from the CSP)
for exchanges in relation to the photometry of CSP SNe Ia.
We thank the referee for his careful and very helpful report.
PR-L and A.Q-E acknowledge support from grant PID2021-123528NB-I00, from the
the Spanish Ministry of Science and Innovation
(MICINN).
JIGH acknowledges financial
support from MICIU grant PID2023-149982NB-I00.
A.P. acknowledges support from the PRIN-INAF 2022 project “Shedding light on the nature of gap transients: from the observations to the model”.

\bigskip

\noindent
    {\it Software}: AstroPy (Price-Whelan  et al. 2018);
    H0CSP.ipynb (Uddin et al. 2024);
    Matplotlib (version 3.2.1, Hunter 2007);
    NumPy (version 1.18.2, Oliphant 2006;
    van der Walt et al. 2011). 
    SNooPy (Burns et al.2011);

    \bigskip


\noindent{REFERENCES}

\bigskip
    
\noindent
  Anand, G.S., Tully, R.B., Rizzi, L., Riess, A.G., \& Yuang, W. 2022, ApJ, 932,
  15
 
  \noindent
  Boone, K., Fakhouri, H., Aldering, G.A., et al. 2016, AAS Meeting, 227,
  237.101

\noindent
  Branch, D., Dang, L.C., Hall, N., et al. 2006, PASP, 118, 560

\noindent
  Bresolin, F., Kudritzki, R.P., Urbaneja, M.A., Sextl, E., \& Riess, A.G.
  2025, ApJ, 991, 151

\noindent
  Brout, D., Taylor, G., Scolnic, D., et al. 2022, ApJ, 938, 111

  \noindent
Burns, C.R., Stritzinger, M., \& Phillips, M.M. 2011, AJ, 141, 19
  
 \noindent
Burns, C.R., Stritzinger, M., Phillips, M.M., et al. 2014, ApJ, 789, 32 

\noindent
  Burns, C.R., Ashall, C., Contreras, C., et al. 2020, ApJ, 895, 11J8

\noindent
  Burrow, A., Baron,, E., Ashall, et al. 2020, ApJ, 901, 154

\noindent
 Cardelli, J. A., Clayton, G. C., \& Mathis, J. S. 1989, ApJ, 345, 245

\noindent
  Carr, A., Davis, T.M., Scolnic, D, et al. 2022, PASA, 39, e046

\noindent
   Carrick, J., Turnbull, S.J., Lavaux, G., \& Hudson, M.J. 2015,
    MNRAS, 450, 317

\noindent
  Casper, C., Zheng, W., Li, W., et al. 2013, CBET, 3588
 
\noindent
  Cenko, S.B., Thomas, R.C., Nugent, P.E., et al. 2011, ATel, 3583, 1

\noindent
  Chassagne, R., \& Santallo, R. 2001, IAUC, 7643

\noindent
  Di Valentino, E., Mena, O., Pan, S., et al. 2021a, CQGra, 38, 153001

\noindent
  Di Valentino, E., Melchiorri, A., \& Silk, J. 2021b, ApJL, 908, L9

\noindent
  Di Valentino, E., Levi Said, J., \& Sandakis, E.N. 2025a, arXiv:2509.25288

\noindent
Di Valentino, E.,  Said, J.L., Riess, A.G., et al.. 2025b, PDU, 49, 101965

\noindent
  Ducroft, M., Leget, P.F., Chotard, N., et al. 2014, ATel, 6859

\noindent
  Fakhouri, H.K., Boone, K., Aldering, G., et al. 2015, ApJ, 815, 58

\noindent
  Folatelli, G., Morrell, N., Phillips, M.M., et al. 2013, ApJ, 773, 53

    \noindent
  Foreman-Mackey, D., Hoog, D.W., Lang, D., \& Gooodman, J. 2013, PASP, 125,
  300

 \noindent
  Freedman, W.L., Madore, B.F., Hatt, D., et al. 2019, ApJ, 882, 34

\noindent
  Freedman, W.L., Madore, B.F., Hoyt, T.J., Jang, I.S., Lee, A.J., \&
  Owens, K.A. 2024, arXiv:2408.06153

\noindent
  Freedman, W.L., Madore, B.F., Hoyt, T.J., Jang, I.S., Lee, A.J., \&
  Owens, K.A. 2025, ApJ, 985, 203

  \noindent
  Galbany, L., de Jaeger, T., Riess, A.G., et al. 2023, A\&A, 679, A95

\noindent
  Garnavich, P., Wood, C.M., Milne, P., et al. 2023, ApJ, 953, 35

  \noindent
  Gelman, A., Roberts, G.O., \& Gilks, W. 1996, Bayesian Statistics, 

\noindent
  Goodman, J., \& Weare, J. 2010, Communications in Applied Mathematics and
  Computational Science, 5, 65

\noindent
  Griffith, C., Li, W., Cenko, S.B., \& Filippenko, A.V. 2008, CBET, 1553

\noindent
  Holoien, T.W.S., Stanek, K.Z., Kochanek, C.S., et al. 2015, ATel, 7078

\noindent
 Hoyt, T.J., Freedman, W.L., \& Madore, G.F. 2019, ApJ, 882, 150
  
\noindent
 Hoyt, T.,J., Beaton, R.L., Freedman, W.L., et al. 2021, ApJ, 915, 34

\noindent
 Hoyt, T.J., Jang, I.S., Freedman, W.L., Madore, B.F., Owens, K.A., \& Lee, A.J. 2025, arXiv:2503.11769
  
\noindent
  Huang, C.D., Yuan, W., Riess, A.G., et al. 2024, ApJ, 963, 83

\noindent
  Hunter, J.D. 2007, CSE, 9, 3

 \noindent
  Itagaki, K., Nakano, S., Quimby, R., et al. 2007, IAUC, 8843

\noindent
  Itagaki, K., Yusa, T., Noguchi, T., et al. 2012, CBET, 3079

\noindent
Jacobson-Galan, V.W., Dimitriadis, G., Foley, R.J., \& Kirpatrick, C.D.
2018, ApJ, 857, 88

\noindent
  Jensen, J.B., Blakeslee, J.P., Cantiello, M., et al. 2025, ApJ, 987, 87

\noindent
  Johnson, R., \& Li, W. 1999, IAUC, 7286

\noindent
  Kamionkowski, M,, \& Riess, A.G. 2023, ARNPS, 73, 153

\noindent
Krisciunas, K., Suntzeff, N.B., Phillips, M.M., et al. 2004, AJ, 128, 3034
  
\noindent
  Krisciunas, K., Contreras, C., Burns, C.R., et al. 2017, AJ, 154, 211

\noindent
Lee, A.J., Freedman, W.L., Madore, B.F., Jang, I.S., Owens, K.A. \& Hoyt, T.J. 2025, ApJ, 985, 182
  
\noindent
  Li, S., Riess, A.G., Casertano, S., et al. 2025, ApJ, 988, 97

\noindent
  Luckas, P., Trondal, O., \& Schwartz, M. 2008, CBET, 1328

  \noindent
    Morrell, N., Phillips, M.M., Folatelli, G., et al. 2024, ApJ, 967, 20

    \noindent
    M\"ortsell, E., \& Dhawan, S. 2018, JCAP, 1809, 025

\noindent
  Murakami, Y.S., Riess, A.G., Stahl, B.E., et al. 2023, JCAP, 11, 046

\noindent
Nugent, P.E., Sullivan, M., Cenko, S.B., et al. 2011, Natur, 480, 344

\noindent
  Oliphant, T.E. 2007, CSE, 9, 10

\noindent
  Pan, Y.C., Foley, R.J., Kromer, M., et al. 2015, MNRAS, 452, 4307

\noindent
  Pansky, X., Li, W., \& Filippenko, A.V. 2008, CBET, 1307

\noindent
  Pantos, I., \& Perivolaropoulos, L. 2026, arXiv:2601.00650v2
  
\noindent
  Parrent, J.T., Sand, D., Valento, M., Graham, D.A., \& Howell, D.A. 2013,
  ATel, 4817, 1
  
\noindent
  Perlmutter, S., Aldering, G., Goldhaber, G., et al. 1999, ApJ, 517, 565

\noindent
  Phillips, M.M. 1993, ApJL, 413, L105

\noindent
  Phillips, M. M., Lira, P., Suntzeff, N.B., Schommer, R. A., Hamuy, M.,
  \& Maza, J. 1999, AJ, 118, 1766
  
  \noindent
    Planck Collaboration 2020, A\&A, 641, A6

  \noindent
    Popovic, B., Brout, D., Kessler, R., Scolnic, D., \& Lu, L. 2021,
    ApJ, 913, 49

\noindent
Poulin, V., Smith, T.L., Calder\'on, R., \& Simon, T. 2025, PhRvD, 111, 083552
    
\noindent
  Price-Whelan, A.M., Sip\~ocz, B.M., G\"unther, H.M., et al. 2018, AJ, 156,
  123
    
  \noindent
    Puckett, T., Orff, T., Newton, J., Madison, D., \& Li, W. 2007, CBET, 795

  \noindent
    Richmond, M.W., \& Smith, H.A. 2012, JAVSO, 40, 872

\noindent
  Riess, A.G., Filippenko, A.V., Challis, P., et al. 1998, AJ, 116, 1009

\noindent
  Riess, A.G., Yuan, W., Macri, L.M., et al. 2022, ApJ, 934, L7
  
\noindent
  Riess, A.G., Li, S., Anand, G. et al. 2025, ApJL, 922, L34

\noindent
  Riess, A.G., Scolnic, D., Gagandeep, S., et al. 2025, ApJ, 977, 120

  \noindent
  Rubin, D., Aldering, G., Betoule, M., et al. (2023, arXiv:2311.12098)
  2025, ApJ, 986, 231

\noindent
  Ruiz-Lapuente, P., \& Gonz\'alez Hern\'andez, J.I. 2024, ApJ, 977, 180

\noindent
  Ryan, S., \& Visvanathan, R. 1989, IAUC, 4730

\noindent
  Saha, A., Sandage, A., Tammann, G.A., Labhardt, L., Machetto, F.D.,
  \& Panagia, N. 1999, ApJ, 522, 802

\noindent
  Sch\"oneberg, N., Franco Abell\'an, G., P\'erez S\'anchez, A., Witte, S.J.,
  Poulin, V., \& Lesgourges, J. 2022, PhR, 984, 1 
  
  \noindent
    Scolnic, D., Brout, D., Carr, A., et al. 2022, ApJ, 938, 113
 
  \noindent
Schlafly, E.F., \& Finkbeiner, D.P. 2011, ApJ, 737, 103

\noindent
    Schneider, S.E., Thuan, T.X., Mangum, J.G., \& Miller, J. 1992, ApJS, 81, 5

\noindent
  Stahl, B.E., Zheng, W., de Jaeger, T., et al. 2020, MNRAS, 492, 4325

\noindent
  Tartaglia, R., Sand, D., Wyatt, S., et al. 2017, ATel, 4158, 1

\noindent
  Tripp, R. 1998, A\&A, 331, 815

\noindent
  Uddin, S.A., Burns, C.R., Phillips, M.M., et al. 2020, ApJ, 901, 143
  
\noindent
  Uddin, S.A., Burns, C.R., Phillips, M.M., et al. 2024, ApJ, 970, 72

\noindent
  Van Der Walt, S., Colbert, S.C., \& Varoquaux, G. 2011, CSE, 13, 22

  \noindent
  Vogl, C., Taubenberger, S., Cs\'omier, G., et al. 2025, A\&A, 702, A41

  \noindent
    Waagen, E.O. 2013, AAN, 479, 1

  \noindent
    Wells, L.A., Phillips, M.M., Suntzeff, B., et al. 1994, AJ, 108, 2233

\noindent
  Yang, Y., Hoeflich, P., Baade, D., et al. 2020, ApJ, 902, 46

\noindent
  Yuan, F., Quimby, R., Chamarro, D., et al. 2008, CBET, 1353

\noindent
    Zheng, W., Silverman, J.M., Filippenko, A.V., et al. 2013, ApJL, 788, L15

\section{APPENDIX A: Light curves of SNe Ia in the Hubble flow, compared with
  those of their low-redshift twins}

\bigskip

\noindent
In this Appendix, we show how the light curves of the SNe Ia in the
Hubble flow from the CSP I coincide with those of our twins of reference,
once these last ones are placed in the corresponding  cosmological frame.

\begin{figure}[h!]
  \includegraphics[width=0.35\textwidth]{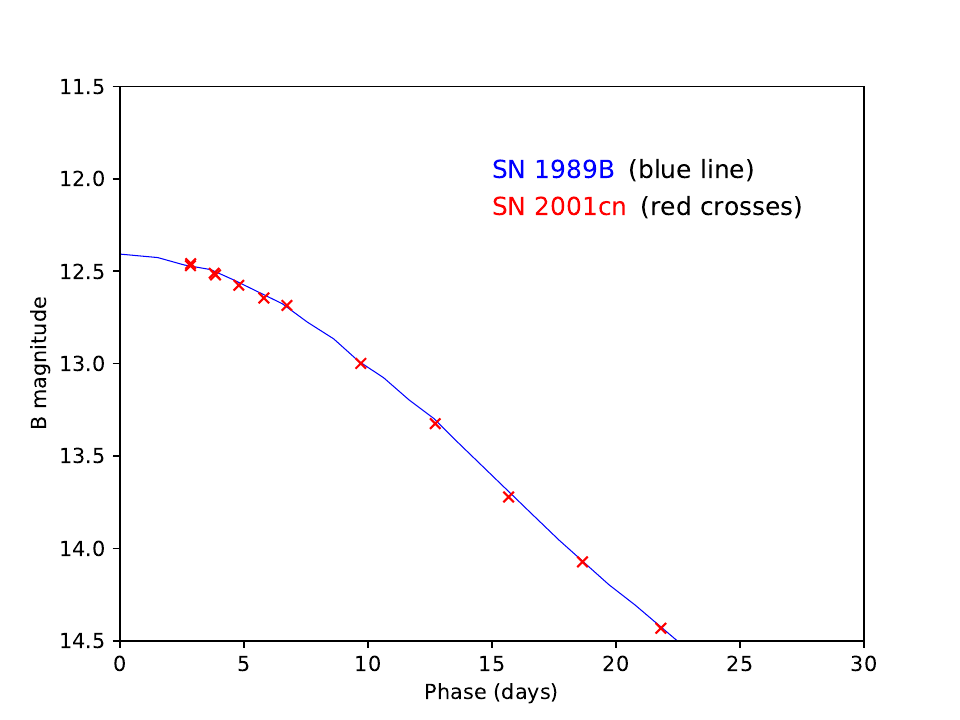}
  \includegraphics[width=0.35\textwidth]{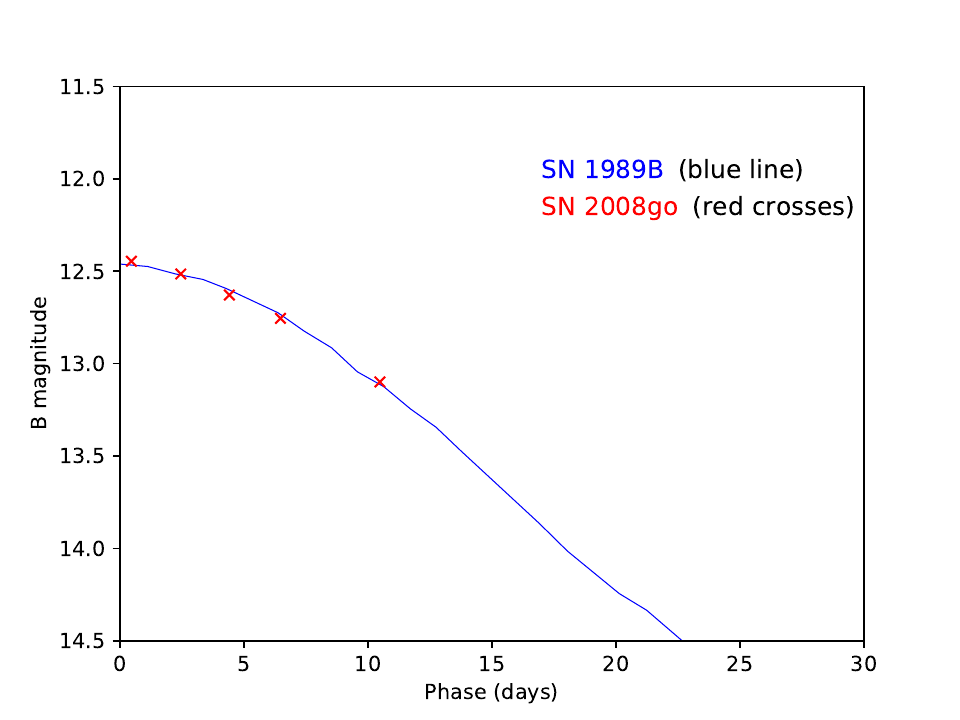}
  \includegraphics[width=0.35\textwidth]{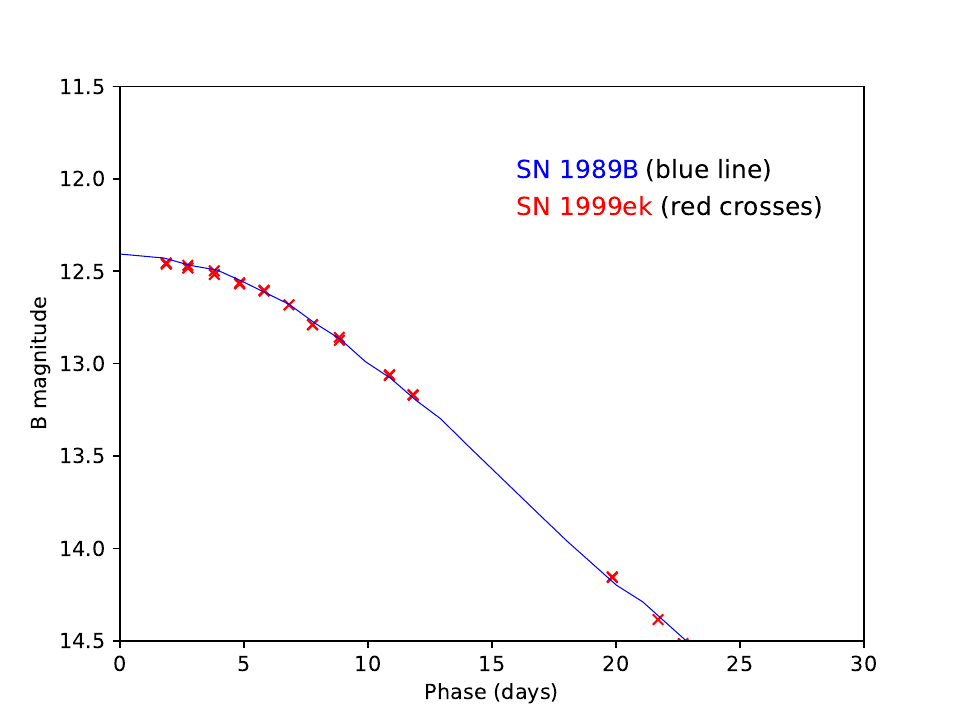}
  \caption{B light curves of the broad-line SNe Ia twins SN 2001cn/SN 1989B
    (left), SN 2008go/SN 1989B (middle) and SN 1999ek/SN 1989B
  (right).}
\end{figure}

\begin{figure}[h!]
  \centering
  \includegraphics[width=0.35\textwidth]{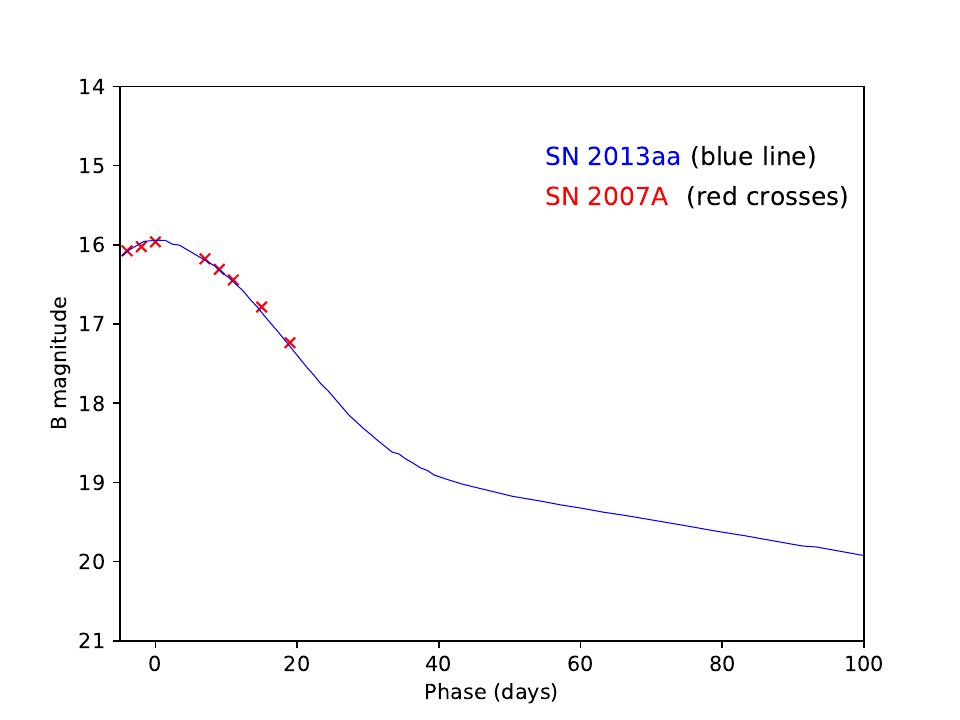}
  \includegraphics[width=0.35\textwidth]{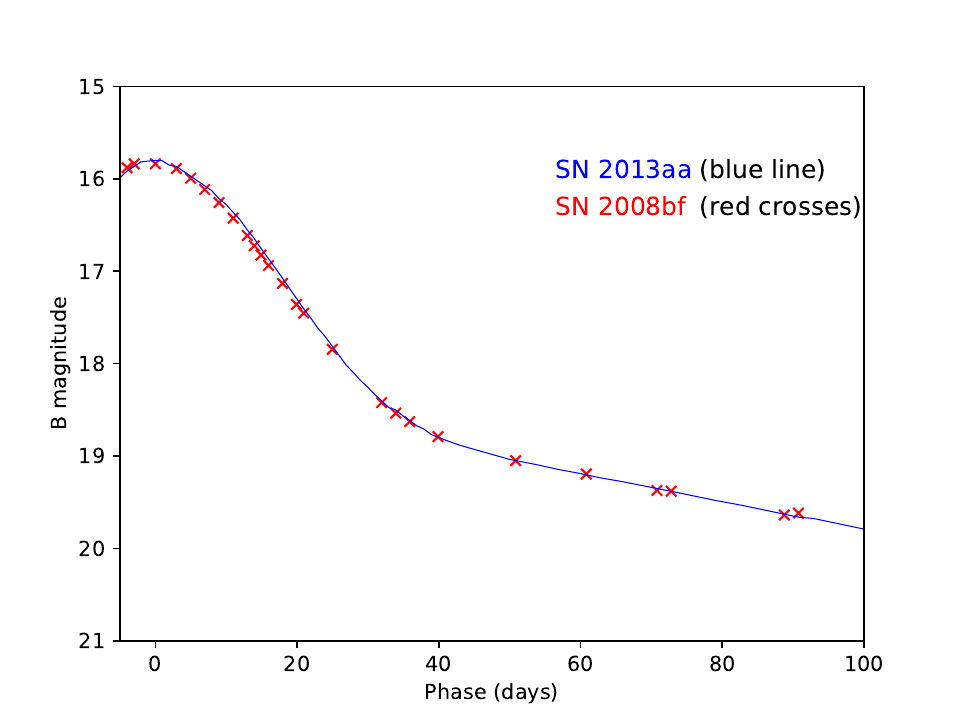}
  \caption{B ligh curves of the core-normal SNe Ia twins SN 2007A/SN 2013aa
  (left) and SN 2008bf/SN 2013aa (right).}
  \end{figure}

\begin{figure}[h!]
  \includegraphics[width=0.35\textwidth]{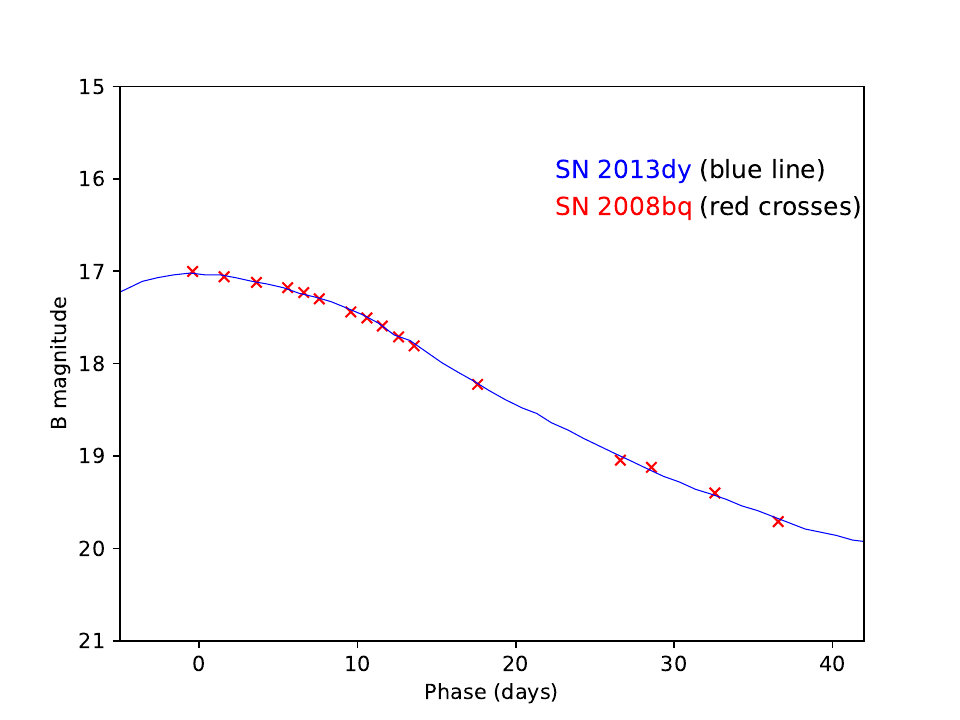}
  \includegraphics[width=0.35\textwidth]{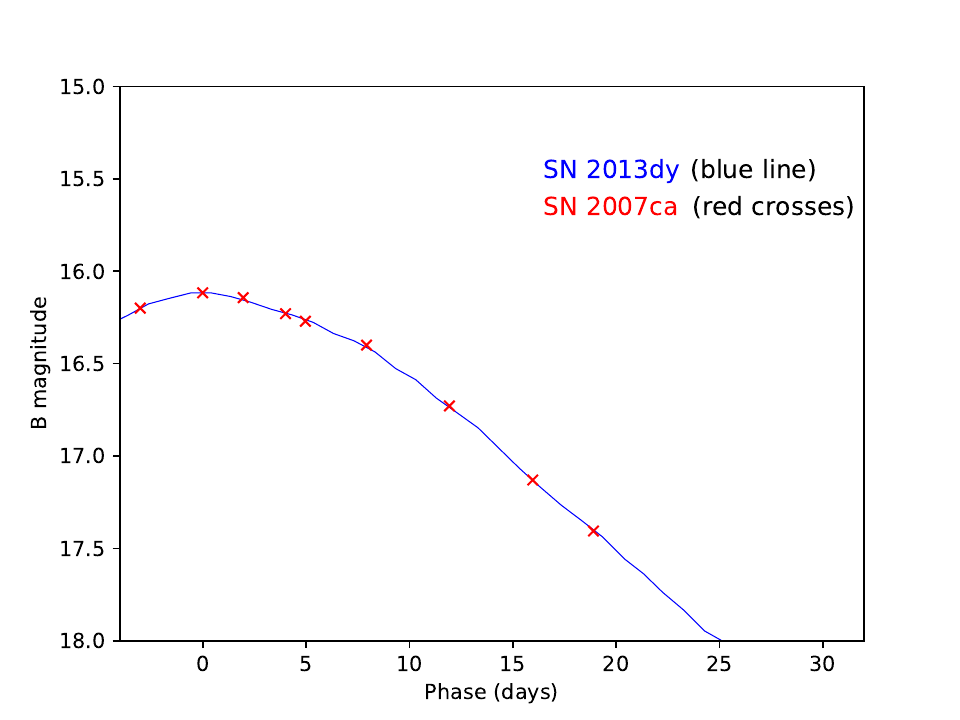}
    \includegraphics[width=0.35\textwidth]{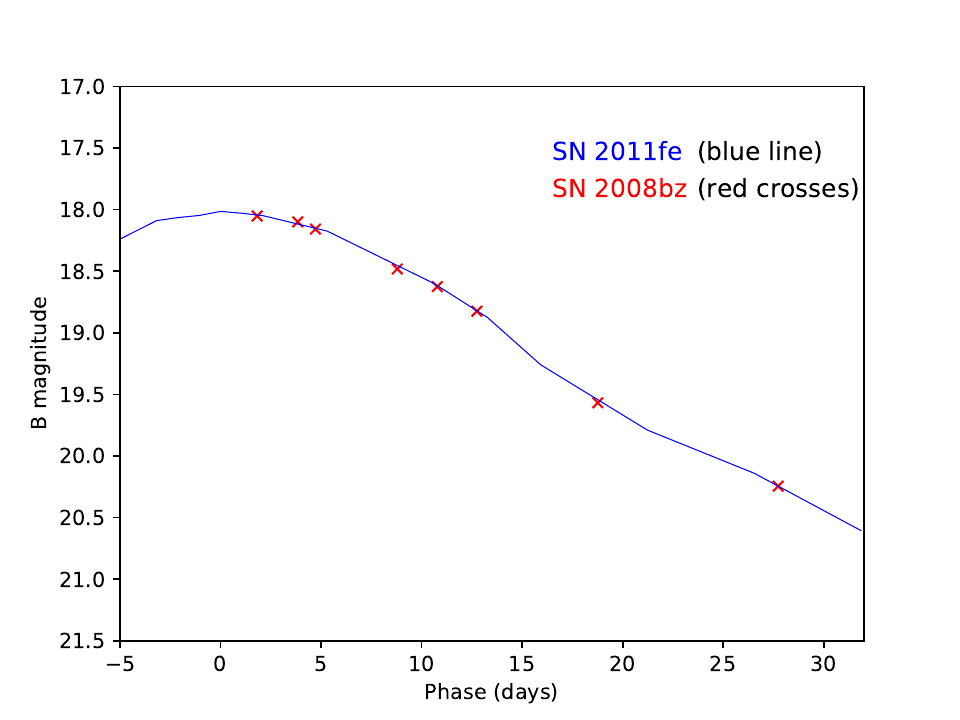}
   \caption{B light curves of the core-normal SNe Ia twins
     2008bq/SN 2013dy (left), SN 2007ca/SN 2013dy (middle) and
    SN 2008bz/SN 2011fe (right) .}
\end{figure}

\section{APPENDIX B:  Intrinsic error with the ''SNe Ia twins'' method}

\bigskip

\noindent
In addition to the approach used in RLGH24 to estimate the error within the “twin SNe Ia” method, we have used in this work a refined estimation by introducing an additional term in the total error budget, namely the intrinsic scatter, $\sigma_{int}$, treated as a free parameter within the MCMC fit. This parameter is included in all fits and marginalized over. We present here the corner plots showing $\sigma_{int}$ for two SNe Ia, SN 2001cn (Figure 20) and SN 1999ek (Figure 21).

\bigskip

\begin{figure*}[h!]
  \centering
  \includegraphics[width=0.8\textwidth]{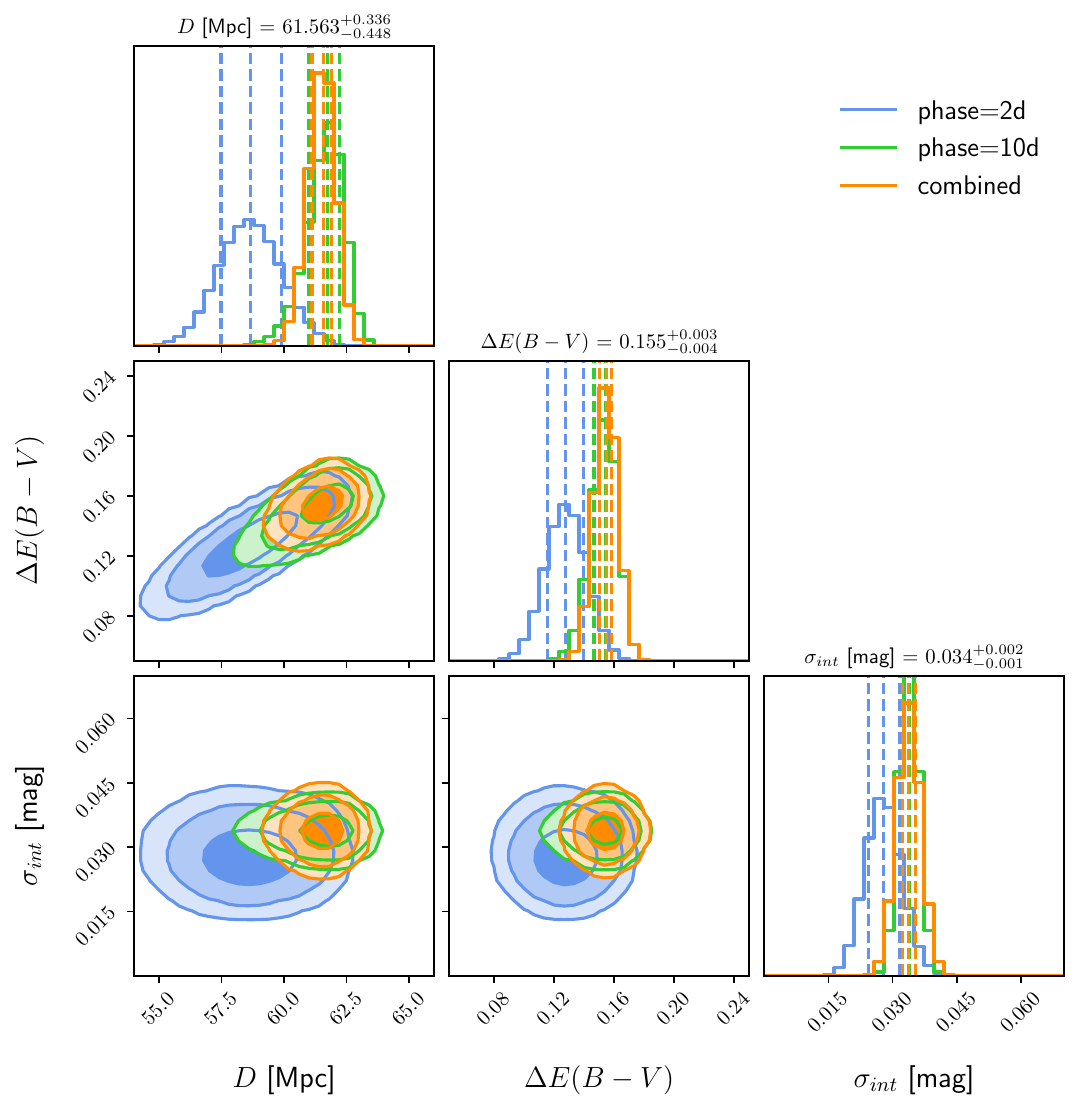}
\caption{
 Corner plot with the posterior probability at 1$\sigma$, 2$\sigma$,
  3$\sigma$ of distance, relative intrinsic reddening of SN 2001cn in relation to SN 1989B and the intrinsic scatter magnitude $\sigma_{int}$ of the method
  which results, in SN 2001cn, of 0.034$^{+0.002}_{-0.001}$ mag.  }  
\end{figure*}

\begin{figure*}[h!]
  \centering
  \includegraphics[width=0.8\textwidth]{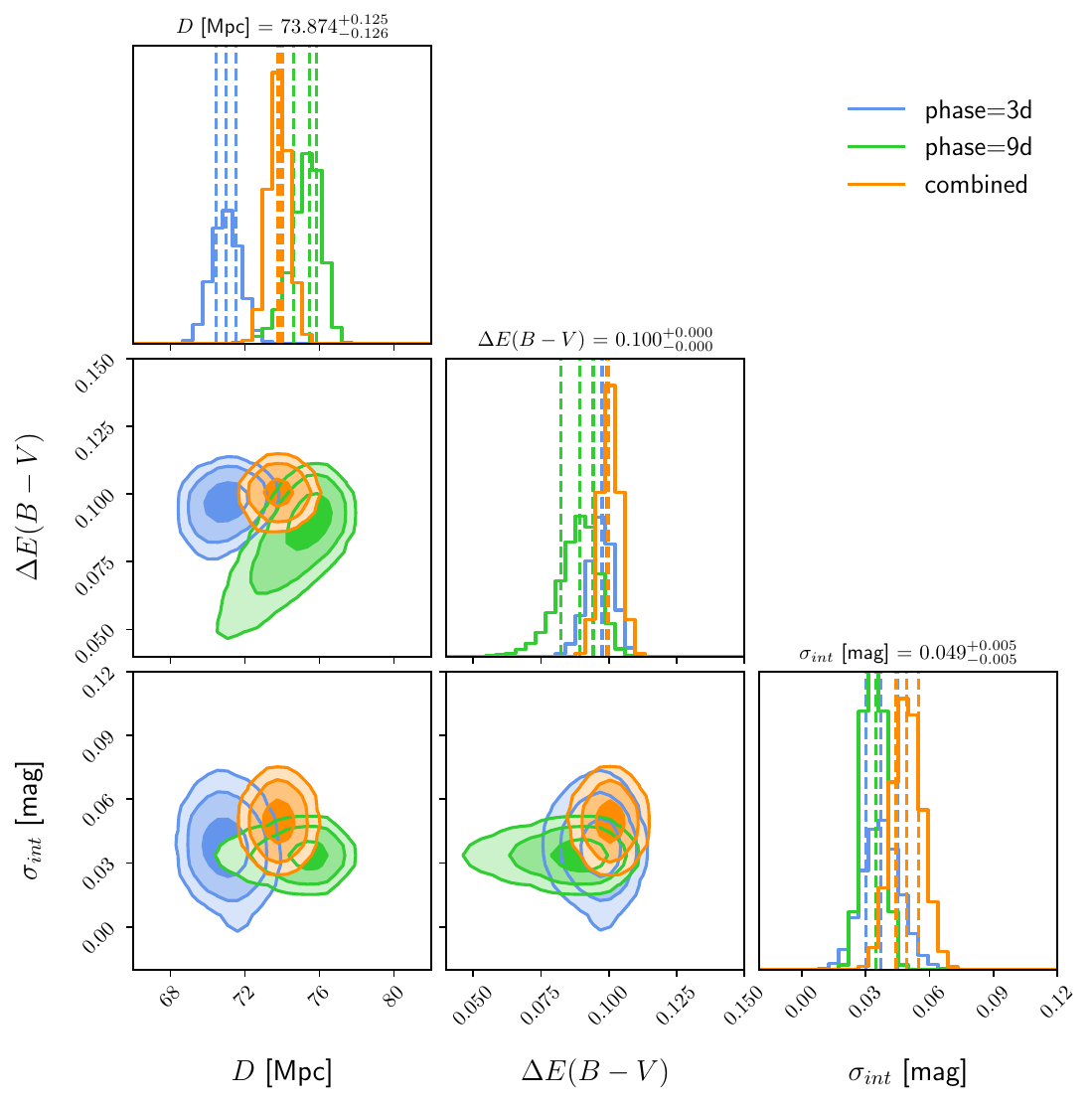}
  \caption{ Corner plot with the posterior probability at 1$\sigma$,
    2$\sigma$, 3$\sigma$ of distance,  relative intrinsic reddening of
    SN 1999ek in relation to SN 1989B and the intrinsic scatter
      magnitude $\sigma_{int}$ of the method which results, in SN 1999ek, of
      0.049$\pm$ 0.005 mag.}  
\end{figure*}

\end{document}